\documentclass[onecolumn,12pt]{IEEEtran}
\usepackage{latexsym}
\usepackage{amssymb}
\usepackage{amsmath}
\usepackage{citesort}
\usepackage{xspace}
\usepackage{subfigure}
\usepackage{fancyhdr}
\usepackage{psfrag}
\renewcommand{\baselinestretch}{1.5}
\normalsize
\usepackage[a4paper, textwidth=6.5in, lines=33, top=1.25in ]{geometry}
\usepackage{mat_eqn}
\ifpdf
    \usepackage[pdftex]{graphicx}
    \usepackage[pdftex]{hyperref}
    \usepackage[usenames,dvipsnames]{color}
\else
    \usepackage{graphicx}
    \usepackage[usenames,dvips]{color}
\fi

\ifpdf
\hypersetup{    backref=section,
                colorlinks=true,
                plainpages=false,
                breaklinks=true,
                colorlinks=true,
                urlcolor=blue,
                citecolor=blue,
                linkcolor=blue,
                bookmarks=true,
                pdfpagemode=none,
                pdfstartview=FitH}
\fi

\def\figw{0.56\columnwidth}
\def\subfigw{0.45\columnwidth}

\centerfigcaptionstrue

\begin{document}
\title{\Large\bfseries Unified Large System Analysis of MMSE and Adaptive Least Squares Receivers for a class of Random Matrix Channels}

\author{ %
\large Matthew J.\  M.\  Peacock$^\dagger$\thanks{$^\dagger$ Matthew Peacock is supported in part
by the Australian CSIRO}
{~~ and ~~} Iain B.\  Collings\\
\normalsize School of Electrical \& Information Engineering \\
\normalsize The University of Sydney, Sydney, NSW 2006, Australia.\\
\normalsize \{mpeac,iain\}@ee.usyd.edu.au \\
\and \vspace*{5mm}%
\large Michael L.\ Honig$^\ddag$ \thanks{$^\ddag$ Supported by the U.S. Army Research
Office under DAAD19-99-1-0288 and the National Science Foundation under grant CCR-0310809} \\
\normalsize Department of Electrical \& Computer Engineering\\
\normalsize Northwestern University, Evanston, IL 60208, USA.\\
\normalsize mh@ece.northwestern.edu }

\maketitle %

\thispagestyle{empty} %
\vspace*{-15mm}
\begin{abstract}
We present a unified large system analysis of linear receivers for a class of random matrix channels. The technique unifies the analysis of both the minimum-mean-squared-error (MMSE) receiver and the adaptive least-squares (ALS) receiver, and also uses a common approach for both random \iid and random orthogonal precoding.  We derive expressions for the asymptotic signal-to-interference-plus-noise (SINR) of the MMSE receiver, and both the transient and steady-state SINR of the ALS receiver, trained using either \iid data sequences or orthogonal training sequences.  The results are in terms of key system parameters, and allow for arbitrary distributions of the power of each of the data streams and the eigenvalues of the channel correlation matrix. In the case of the ALS receiver, we allow a diagonal loading constant and an arbitrary data windowing function.  For \iid training sequences and no diagonal loading, we give a fundamental relationship between the transient/steady-state SINR of the ALS and the MMSE receivers. We demonstrate that for a particular ratio of receive to transmit dimensions and window shape, all channels which have the same MMSE SINR have an identical transient ALS SINR response. We demonstrate several applications of the results, including an optimization of information throughput with respect to training sequence length in coded block transmission.
\end{abstract}

\begin{keywords}
Large System, MMSE, Recursive Least Squares, MIMO, CDMA.
\end{keywords}

\setcounter{page}{1}

\section{Introduction}

Large-system analysis of linear receivers for random matrix channels has attracted significant attention in recent years, and has proven to be a powerful tool in their understanding and design (e.g., see \cite{Tulino04Verdu} and references therein).  In particular, large-system analysis of the matched filter, decorrelator, and minimum-mean-squared-error (MMSE) receivers, which have knowledge of the channel state information, has been exhaustively studied for the downlink, using results such as the Silverstein-Bai theorem \cite{silverstein95bai}, Girko's law \cite{Girko90}, and free probability \cite{Hiai00Petz}.  In this paper we take a different approach to the problem, which allows us to consider random \iid and orthogonal channels (or matrix of signatures) in the same treatment, in contrast to the existing separate analyses of \iid \cite{Li04Tulino} and orthogonal \cite{Debbah03Hachem,Chaufray03Hachem} channels.  The new approach also allows us to consider a more general class of signal models, and a receiver which does not have the benefit of channel state information, namely the adaptive least-squares (ALS) receiver.

In this paper, we consider both the MMSE and ALS receivers. The ALS receiver approximates the MMSE receiver and requires training symbols \cite{Haykin96}. In particular, the autocorrelation matrix of the received vector  (an ensemble average), which is used in the MMSE receiver, is replaced in the ALS receiver by a sample autocorrelation matrix (a time average). Given sufficient training symbols, the performance of the ALS receiver approaches that of the MMSE receiver. In this paper, two types of adaptive training modes are considered, based on those presented in \cite{Poor97Wang}, where either the training sequence is known at the receiver, or a semi-blind method is employed. In a time-varying environment, weighting can be applied to the errors to create data windowing, which allows for tracking. When implemented online as a series of rank-1 updates with exponential windowing, this receiver is often referred to as the recursive least-squares (RLS) receiver \cite{Haykin96}.   To prevent ill-conditioning of the sample autocorrelation matrix with RLS filtering, diagonal loading can be employed, which refers to initializing this matrix with a small positive constant times the identity matrix.

Prior relevant work on ALS techniques (in particular, as applied to channel equalization and estimation) includes \cite{Ling84Proakis}, where an approximate expression is derived for the transient excess mean-squared-error (MSE) of the ALS receiver with respect to the MMSE receiver, often referred to as self-noise, for a general channel model and windowing function. Also, an approximate expression is given for the convergence time constant of the receiver when tracking non-stationary signals.  Steady-state and transient analysis of the RLS receiver was also considered in \cite{Eleftheriou86Falconer}. A comprehensive treatment of ALS techniques and variations is contained in \cite{Haykin96}.

The application of ALS to CDMA was considered in \cite{Honig95}, where the convergence of a blind multiuser detector based on a stochastic gradient-descent adaptation rule is also analyzed. In \cite{Poor97Wang}, approximate expressions are given for the relationship between the MMSE signal-to-interference-plus-noise-ratio (SINR) and the steady-state ALS SINR with exponential windowing for DS-CDMA in flat-fading. This was extended in \cite{Caire00} to a steady-state analysis of two-stage algorithms based on RLS using decision-directed adaptation. A review of adaptive interference mitigation techniques is given in \cite{Honig98Poor}.

Large-system analysis for ALS receivers was first considered in \cite{Xiao00Honig,Honig02Xiao}, which considered DS-CDMA in flat fading with \iid training sequences, and an ALS receiver with diagonal loading and rectangular or exponential weighting. The transient SINR (i.e., after a given number of training symbols) and steady-state SINR (i.e., after an unlimited number of training symbols) of the ALS receiver was derived in the limit where the number of transmit dimensions, the number of multiplexed data streams, and the number of training symbols all tend to infinity with fixed ratios. With rectangular windowing and no diagonal loading, a relationship between the MMSE SINR and the transient ALS SINR was given. In other work, large-system analysis has been applied to so-called subspace-based blind ALS receivers in \cite{Zhang02Wang,Xu04Wang,Host-Madsen04Wang}.

In this paper, we consider a more general complex-valued AWGN matrix-vector channel model of the form $\vr = \mH\mS\mA\vb + \vn$, where $\mH$ is an arbitrary matrix, $\mA$ is an arbitrary diagonal matrix, and $\mS$ is either an \iid or orthogonal matrix. For example, the model applies explicitly to downlink synchronous direct sequence (DS) or multi-carrier (MC) CDMA in frequency-selective fading, as well as multi-input multi-output (MIMO) channels. In fact, if we additionally require that the data vector $\vb$ is unitarily invariant, then the MMSE and ALS results we obtain \emph{apply to all AWGN matrix-vector systems} of the form $\vr = \mC\vb + \vn$, provided that the eigenvalues of $\mC\mC^\dag$ converge to a well-behaved deterministic distribution.  As such, the results can be applied to systems not previously considered in large-system analysis, such as equalization of single-user finite-impulse-response (FIR) channels.

%In fact, if we additionally require that the data vector $\vb$ is unitarily invariant, considering a general AWGN matrix-vector system of the form $\vr = \mC\vb + \vn$, the MMSE and ALS results we obtain can be formulated in terms of the \aed of $\mC\mC^\dag$, and therefore apply to any $\mC$ for which the \aed can be determined.

For this channel model, we derive the SINR of the MMSE receiver, with either \iid or orthogonal $\mS$, in the large-system limit where the number of transmit and receive dimensions, and the number of multiplexed data streams all tend to infinity with fixed ratios. The expression for the SINR is a function of these ratios and the received SNR, and allows for arbitrary asymptotic eigenvalue distributions (\aed's) of $\mA\mA^\dag$ and $\mH\mH^\dag$. This result can also be derived under the same set of assumptions using the $\ST$-transform from free probability. However, unlike the free probability technique, we show that the technique used to derive this result also applies to the ALS receiver.

For the ALS receiver, we extend the work of \cite{Honig02Xiao} to the general channel model described. We consider an arbitrary data windowing function and both \iid and orthogonal training sequences. That is, we determine both the transient and steady-state ALS SINR in the limit described for the MMSE receiver, and also as the number of training symbols tend to infinity.

Also, we present an expression which relates the SINR of the MMSE receiver to the transient and steady-state SINRs of the ALS receiver for the case of \iid training sequences and no diagonal loading.  We demonstrate that in this situation, for a particular ratio of receive to transmit dimensions and window shape, \emph{all channels, which have the same MMSE SINR, have an identical transient ALS SINR response}. Since our results hold for all well-behaved matrix-vector systems for which the data vector is unitarily invariant (as previously discussed), the MMSE-ALS relationship is in fact seen to be a \emph{fundamental property} of adaptive least-squares estimation.

It is interesting to compare our results to an approximate expression for the steady-state ALS SINR given in \cite{Poor97Wang} for the special case of DS-CDMA in a flat-fading channel, with \iid spreading and exponential weighting. A comparison of the expressions reveals the approximation in \cite{Poor97Wang} to be excellent, particularly for large window sizes.  Also, we note that for a general channel model, the study in \cite{Eleftheriou86Falconer} previously came to the conclusion that the ALS convergence rate is independent of the channel; however, their conclusion was based on making several approximations, and only exponential windowing (RLS) was considered. This conclusion was stated for DS-CDMA in \cite{Poor97Wang}; however, an explicit relationship between the MMSE SINR and the transient ALS SINR, such as derived in this paper, is not given in either of those papers.

Unfortunately, we have not determined a simple relationship between the MMSE SINR and ALS SINR with orthogonal training sequences and/or diagonal loading. This remains an open problem.  Inspection reveals that with orthogonal training sequences and no diagonal loading, the transient relationship will depend on the channel.

During the course of the analysis we solve for the \Stieltjes (or Cauchy) transforms of the \aed's of both the autocorrelation matrix and the weighted sample autocorrelation matrix of the received signal.  These expressions are found using matrix manipulations, which do not require the use of free probability. The first transform can also be derived using free probability; however, such techniques cannot be applied to derive the second transform, since the associated constituent matrices are not free.  As such, the results are of independent mathematical interest.

% XXX removed
%Our results are conditioned on the convergence of a number of quantities (as is the case when free probability is used). Here we do not attempt to specify technical conditions that guarantee convergence, although we suspect that convergence can be proved under mild conditions using standard techniques. Numerical studies support this position. Throughout the paper, we state the conditions, which are necessary for our analysis.

%We are confident that the convergence can be proved under some fairly mild conditions using standard techniques. Numerical studies support this position. Throughout the paper, we outline certain conditions that are necessary for our analysis, but a rigorous description of the existence and nature of the convergence is yet to be established.

Through numerical studies, we demonstrate the applicability of the large-system results to finite systems, and the benefits of orthogonal precoding and training is examined. We demonstrate an application of the results in the optimization of information throughput with respect to training sequence length in coded block transmission.

The paper is arranged as follows.  Section \ref{sec:sysmodel} outlines the general transmission model, the receivers considered, and defines the large-system asymptotic limit.  Section \ref{sec:preamble} discusses the general approach we take for the analysis.  Section \ref{sec:MMSE_soln} reviews analytical approaches to the MMSE SINR problem, and presents the general solution based on the unified analytical approach. An alternate expression for the MMSE SINR, different from those presented in \cite{Tulino04Verdu}, is also presented which allows the relationship between the MMSE SINR and ALS SINR to be derived in certain cases in later sections.  Section \ref{sec:als_results} presents the general result for the ALS SINR. A simple relationship between the ALS SINR and MMSE SINR with \iid training and without diagonal loading is then presented in Section \ref{sec:relationship}. Finally, numerical studies are presented in Section \ref{sec:simulations}.

\section{System Model} %
\label{sec:sysmodel} %

This paper considers a general matrix-vector transmission model.  It applies to a wide range of practical data communication systems, including frequency-flat fading MIMO channels, and both frequency-flat and frequency-selective fading downlink DS- or MC-CDMA channels.

\subsection{General Transmission Model}
In matrix notation\footnote{{\bf Notation:} All vectors are defined as column vectors and designated with bold lower case; all matrices are given in bold upper case; $(\cdot)^T$ denotes transpose; $(\cdot)^\dag$ denotes Hermitian (i.e.\ complex conjugate) transpose; $(\cdot)^\ddag$ denotes the operation $\mX^\ddag = \mX\mX^\dag$; $\tr{\cdot}$ denotes the matrix trace; $\abs{\cdot}$ and $\norm{\cdot}$ denote the Euclidian and induced spectral norms, respectively; $\mI_N$ denotes the $N\times N$ identity matrix; $\RR^* = \{x\in\RR \: : \: x\geq 0\}$; and, expectation is denoted $\Exp{\cdot}$.}, the received signal in the $m\thh$ symbol period is
\begin{equation}
\vr_m = \mH \mS \mA \vb_m + \vn_m \label{eq:rcv_sig}
\end{equation}
where
\begin{itemize}
\item $\mH$ is an $M\times N$ complex-valued channel matrix.
\item $\mS = \left[\vs_1 \cdots \vs_K \right]$ is an $N\times K$ complex-valued matrix which contains either
    \begin{itemize}
        \item random orthonormal columns, i.e., we assume that $\mS$
        is obtained by extracting $K\leq N$ columns from an $N\times N$ Haar-distributed\footnote{
        A unitary random matrix $\mOmega$ is Haar distributed if its probability distribution is
        invariant to left or right multiplication by any constant unitary matrix. If $\mX$ is a square random
        matrix with \iid complex Gaussian centered unit variance entries, then the unitary matrix
        $\mX\(\mX^\dag\mX\)^{-1/2}$ is Haar distributed.} unitary random matrix, or, %
        \item \iid complex elements\footnote{For technical reasons, we also
        require the elements have finite positive moments. Also, if $\mH\neq\mI_N$, we additionally
        require that $\mS$ is unitarily invariant, although we believe the results apply more generally.}
        with mean zero and variance $\frac{1}{N}$.
        For example, \iid real \& imaginary parts which are either $\pm 1/\sqrt{2N}$ with equal probability, or
        \iid Gaussian with zero mean and variance $1/(2N)$.
    \end{itemize}
    We shall call the first case `isometric $\mS$', and the second case `\iid $\mS$', as
    is done in \cite{Debbah03Hachem}.
\item $\mA$ is a $K\times K$, diagonal, complex-valued matrix of transmit coefficients, \ie $\mA = \diag(A_1 \lldots A_K)$. In fact, the results which follow depend only the values of $P_k = |A_k|^2$, and so to simplify notation, without lack of generality, we may assume $A_k$, $k=1\lldots K$, is non-negative and real valued. %
\item The complex vector $\vb_m$ is $K\times 1$ and either contains transmit data, or training symbols (for an ALS receiver).  Elements of $\vb_m$ can be either \iid with zero mean and unit variance\footnotemark[\value{footnote}] (for data and \iid training cases), or they can be drawn from a set of orthogonal sequences (for orthogonal training, as explained further in Section \ref{sec:defn_als_rcv}).
\item $\vn_m$ contains \iid, zero mean, circularly symmetric, complex Gaussian entries with variance per dimension $\sn/2$. %
\item $\mH$, $\mS$, $\mA$, $\vb_m$, and $\vn_m$ are mutually independent. %
\end{itemize}

\subsection{Discussion}

The general transmission model in (\ref{eq:rcv_sig}) is widely applicable, and in particular
includes the following systems.
\begin{itemize}

\item \emph{Downlink MC- or DS-CDMA:} In this case, $\mS$ represents the matrix of $K$ signatures with spreading gain $N$.  Typically, the number of output dimensions equals the number of input dimensions and hence $\mH$ is square.
\begin{itemize}
\item For flat-fading DS-CDMA, $\mH=\mI_N$ and $\mA$ represents the combined effect of each users' transmit power and channel coefficient. %
\item For MC-CDMA in frequency-selective fading, $\mH$ is the diagonal matrix of the channel frequency response in each subcarrier, and $\mA$ represents the transmit amplitude of each signature. %
\item For DS-CDMA in frequency-selective fading, $\mH$ is a circulant or Toeplitz matrix constructed from the channel impulse response, and $\mA$ represents the transmit amplitude of each signature.
\end{itemize}

\item \emph{`Rich' MIMO:} The standard point-to-point flat fading MIMO channel model is given by (\ref{eq:rcv_sig}), where $K$ and $N$ correspond to the number of transmit and receive antennas, respectively. The standard MIMO channel matrix with \iid circularly symmetric complex Gaussian coefficients between each pair of transmit and receive antennas corresponds here to setting $\mH=\mI_N$ and $\mS$ \iid Gaussian. The matrix $\mA$ defines the transmit amplitudes on each antenna. %

\item \emph{MIMO with richness parameter:}  The MIMO channel model introduced in \cite{Muller02} can also be described by (\ref{eq:rcv_sig}), with $\mS$ \iid and $\mH = \mPhi\mC$, where $\mPhi$ is  \iid, and $\mC$ is diagonal.  In this case, $\mS$ models the propagation from the transmitter to a `scattering array', modeled by $\mC$, and $\mPhi$ models the propagation from the scattering array to the receiver. The rank of the scattering array matrix determines the richness of the MIMO channel.

\item If we additionally require that the data vector $\vb$ is unitarily invariant, then the MMSE and ALS asymptotic SINR results we obtain \emph{apply to all AWGN matrix-vector systems} of the form $\vr = \mC\vb + \vn$,  under certain conditions on the channel matrix $\mC$, and the data and noise vectors, $\vb$ and $\vn$. A full explanation is given in Section \ref{sec:limit}.

\end{itemize}

\subsection{MMSE Receiver}

The output of the MMSE receiver with full channel state information (CSI) and knowledge of $\mS$
for stream $k$ at symbol interval $m$ is given by
\begin{equation}
\hat{\vb}_m(k) = \vc_k^\dag \vr_m
\end{equation}
where
\begin{align}
\vc_k &= \mR^{-1} \mH \vs_k \label{eq:mmse_filter} \\
\mR &= (\mH\mS\mA)^\ddag + \sn\mI_M \label{eq:defn_mR}
\end{align}
Now, identifying the signal and interference components of the received signal in (\ref{eq:rcv_sig}), \ie, $\vr_m = A_k\mH\vs_k\vb_m(k) + \vr_m^I$, the corresponding output SINR is defined as
\begin{align}
\SINR^{\text{MMSE}}_{k,N} &= \frac{\Exp{|\vc_k^\dag(\vr_m - \vr_m^I)|^2}}{\Exp{|\vc_k^\dag\vr_m^I|^2}} \label{eq:def_sinr}
\end{align}
where the expectation in (\ref{eq:def_sinr}) is with respect to $\vb_m$ and $\vn_m$. The subscript $N$ indicates that this is a non-asymptotic quantity.

\subsection{ALS Receiver}
\label{sec:defn_als_rcv} %

The output of the adaptive least-squares (ALS) receiver with $i$ training symbols for stream $k$ at
symbol interval $m>i$ is given by
\begin{align}
\hat{\vb}_m(k) &= \hat{\vc}_k^\dag \vr_m
\end{align}
where
\begin{align}
\hat{\vc}_k &= \mhR^{-1} \hat{\vs}_k \label{eq:ALS_filter} \\
\mhR &= \frac{1}{i}\sR \mW \sR^\dag + \frac{\mu}{\eta}\mI_M \label{eq:def_mhR} \\
%&= \frac{1}{i} \sum_{m=1}^i w_m \vr_m \vr_m^\dag + \frac{\mu}{\eta}\mI_M  \\
\sR &= \mH\mS\mA\mB^\dag + \mN \\
\hat{\vs}_k &= \begin{cases} %
\frac{1}{i}\sR \mW \vub_k & \text{, with training} \\
\mH\vs_k & \text{, semi-blind} %
\end{cases} \label{eq:ALS_shat}
\end{align}
and where
\begin{itemize}
\item $\mB$ is an $i\times K$ matrix of training data, where the $m\thh$ row of $\mB$ is $\vb_m^\dag$.
The $k\thh$ column of $\mB$ will be denoted  as $\vub_k$.  We consider both \iid and
orthogonal training sequences, i.e.,
\begin{itemize}
    \item $\mB$ contains \iid elements with zero mean, unit variance, and finite
    positive moments, or,
    \item $\mB$ contains either random orthogonal rows or columns. If $K<i$
    then $\mB^\dag\mB=i\mI_K$, and we assume that $\frac{1}{\sqrt{i}}\mB$ is obtained by extracting $K$
    columns from an $i\times i$ Haar-distributed unitary random matrix. If $K>i$,
    then $\mB\mB^\dag=K\mI_i$, and we assume that $\frac{1}{\sqrt{K}}\mB^\dag$ is obtained by extracting $i$
    columns from a $K\times K$ Haar-distributed unitary random matrix.
\end{itemize}
\item $\mN$ is an $M\times i$ matrix of noise, where the $m\thh$ column is $\vn_m$. %
\item $\mu$ is a real-valued non-negative diagonal loading constant, \ie, $\mu \in \RR^*$, and $\eta=i/N$. %
\item $\mW$ is an $i\times i$ diagonal real-valued data windowing matrix, \ie $\mW = \diag(w_1 \lldots w_i)$ where $w_m\in\RR^*$ for $m=1\lldots i$.  For example, with exponential weighting $w_m = \epsilon^{i-m}$, where $\epsilon\in(0,1]$, or, without data windowing $\mW=\mI_i$. %
\end{itemize}

Note that, although strictly speaking the model in (\ref{eq:rcv_sig}) applies only to time-invariant systems, we include windowing to allow for practical situations such as slowly time varying channels, or users entering/leaving the system.  The term `semi-blind' in (\ref{eq:ALS_shat}) refers to the case when $\mH$ and $\vs_k$ are known, and there is no training data, whereas `with training' refers to when just $\mB$ is known. For more details on the practical issues, see \cite{Poor97Wang}, where this ALS formulation is considered for DS-CDMA in flat-fading.

The SINR for the $k\thh$ stream at the output of the ALS receiver, $\SINR^{\text{ALS}}_{k,N}$, is defined by the right-hand side of (\ref{eq:def_sinr}), however, with $\vc_k$ replaced by $\hat{\vc}_k$.

\subsection{Large System Limit} %
\label{sec:limit}

We define $\alpha=K/N$ and $\beta = M/N$, and for the ALS receiver, $\eta = i/N$.

Throughout this paper we consider the asymptotic limit $(M,N,K,i)\to\infty$ with $K/N\to\alpha>0$, $i/N\to\eta>0$, and $M/N\to\beta>0$ constant.

With data windowing it is necessary to consider how $\mW$ is defined for each $i$ so that the empirical distribution function (\edf) of its diagonal values converges to something appropriate. Any finite window length becomes negligible in the large system limit as $i\to\infty$, therefore it is necessary to scale the window shape with the system size.  For example, as in \cite{Honig02Xiao}, with exponential windowing we define $L=\frac{1}{1-\epsilon}$ as the `average' window length, and take $L\to\infty$ with $L/N \to \Leff>0$ constant.

To facilitate the large system analysis, we also require that $\mH^\ddag$, $\mA^2$, and $\mW$ each have a uniformly bounded spectral norm,\footnote{In particular, this condition is required for the derivations in the appendices, which frequently rely on Lemma \ref{lem:lem3.1} in Appendix \ref{ap:precursor} along with other key lemmas as a precursor to the asymptotic analysis contained in the remaining appendices.} that is, a bound which is independent of the system dimension $N$.  Also, we require the empirical distribution functions of the eigenvalues of $\mA^2$, $\mH^\ddag$, and $\mW$ to converge in distribution almost surely to non-random distributions on the non-negative real axis, which will have compact support due to the previous assumption. We also assume that the limiting distributions of $\mA^2$, $\mH^\ddag$, and $\mW$ are non-trivial, \ie, do not have all mass at zero.

\begin{proposition}
\label{pr:allAWGNbinvar}
The large-system MMSE and ALS SINRs corresponding to the transmission model in (\ref{eq:rcv_sig}) are the same as the large-system MMSE and ALS SINRs, respectively, computed for any matrix-vector system of the form $\vr=\mC\vb+\vn$ in which
\begin{itemize}
\item $\mC$, $\vb$, and $\vn$ are mutually independent, %
\item the $M\times 1$ noise vector $\vn$ satisfies the same conditions as $\vn_m$ above, %
%\item the $M\times 1$ noise vector $\vn$ is proper complex Gaussian with variance per dimension $\sn/2$, %
\item the $K\times 1$ data/training vector $\vb$ satisfies the same conditions as $\vb_m$ above, and \emph{additionally is unitarily invariant},\footnote{That is, the elements of $\mU\vb$ have the same joint distribution as $\vb$ for any $K\times K$ unitary matrix $\mU$.  This is an extra restriction on \iid data vectors. For training vectors from Haar-distributed matrices, this condition is automatically satisfied, and is easily verified for $\alpha\geq\eta$ (that is, when $\vb$ is a column from a Haar-distributed matrix). For $\alpha<\eta$, note that $\mU\vb$ corresponds to the Hermitian transpose of a row of $\mB\mU^\dag$, and $\mB\mU^\dag$ can be written as $\mTheta\mE_K\mU^\dag$, where $\mTheta$ is $i\times i$ Haar, and $\mE_K = [\mI_K, \mZero_{K,i-K}]^\dag$. Alternately, we have $\mB\mU^\dag = \mTheta\tilde{\mU}\mE_K$, where $\tilde{\mU}$ is the unitary matrix created by replacing the upper left $K\times K$ sub-block of $\mI_i$ with $\mU^\dag$. Since $\mTheta$ is Haar, $\mTheta\tilde{\mU}$ is also Haar, and hence $\mB\mU^\dag$ has the same distribution as $\mB$.}
% and is either an \iid or isometric data vector with $\alpha \geq \eta$ satisfying the conditions above,
\item the $M\times K$ channel matrix $\mC$ is such that the \edf of $\mC^\ddag$ satisfies the conditions on $\mH^\ddag$ mentioned in the previous paragraph,
\end{itemize}
and corresponds to taking $\mH=\mC$, $\mS$ to be $K\times K$ isometric, and $\mA=\mI_K$, respectively.
\end{proposition}
\begin{proof}
Since $\vb$ is unitarily invariant, all data streams have identically distributed SINRs. Since $\vb$ has the same distribution as $\mU\vb$, where $\mU$ is a $K\times K$ Haar-distributed random unitary matrix, we see that the distribution of the MMSE and ALS SINR associated with $\vr' = \mC\mU\vb+\vn$ are, respectively, the same as that for $\vr$, and will share a common large-system limit, if it exists.
\end{proof}

%Suppose we performed a random $M\times M$ Haar-distributed unitary transformation $\mT$ on $\vr$ before MMSE filtering, \ie, $\vr' = \mT\vr$. The MMSE or ALS SINR is clearly unchanged. However, we may write $\vr' = (\mT\mU)\mD(\mV^\dag\vb) + \mT\vn$, where $\mU\mD\mV^\dag$ is the singular value decomposition of $\mC$. Since $\mT$ is Haar and $\mU$ is unitary, by definition $\mS' = \mT\mU$ is Haar-distributed. Since the all-zero columns of $\mD$ (of which there are at least $M-K$) contribute nothing to $\vr$, we may take $\mD'$ as the first $K$ columns of $\mD$. Now, since the noise is Gaussian, $\vn' = \mT\vn$ has the same distribution as $\vn$.  Finally, $\vb' = \mV^\dag\vb$ has the same distribution as $\vb$ under the assumption of unitarily invariant input vectors. Therefore we may identify $\vr'$, $\mI_M$, $\mS'$, $\mD'$, $\vb'$, and $\vn'$ with $\vr$, $\mH$, $\mS$, $\mA$, $\vb$ and $\vn$ in (\ref{eq:rcv_sig}) respectively, where we take $\mS$ as $M\times K$ isometric.

Proposition \ref{pr:allAWGNbinvar} implies that our model encompasses the classic equalization model. Namely, $\vr$  represents $N=M$ samples at the output of a single-input/single-output (SISO) FIR channel $\vh$ of length $L_c$, i.e., $\vr(n)=\sum_{\ell=0}^{L_c-1}\vh(\ell)\vb(n-\ell) + \vn(n)$.  If a cyclic prefix of appropriate length is used, we set $\mC$ defined in Proposition \ref{pr:allAWGNbinvar} equal to the circulant channel matrix.  Therefore, from Proposition \ref{pr:allAWGNbinvar}, the corresponding model (\ref{eq:rcv_sig}) takes $\mA=\mI_N$, $\mS$ as $N\times N$ isometric, and $\mH$ as $\mC$ or equivalently as an $N\times N$ diagonal matrix with the $N$-point DFT of $\vh$ on the diagonal.

\section{Unified Large System Analysis}
\label{sec:preamble}

In Sections \ref{sec:MMSE_soln} and \ref{sec:als_results}, we derive the asymptotic SINR for the model (\ref{eq:rcv_sig}) with both MMSE and ALS receivers. The SINR in both cases is directly related to the \Stieltjes transform\footnote{The \Stieltjes (or Cauchy) transform of the distribution of a real-valued random variable $X$ is the expected value of $1/(X - z)$, where $z\in\CC^+$ is the transform variable, and $\CC^+ = \{x \:|\:x\in\CC,\:\imag(x)>0 \}$ (\eg, see \cite{silverstein95bai}).} of the \aed of the received signal correlation matrix $\mR$ for the MMSE receiver, and $\mhR$ for the ALS receiver. For the MMSE case, there are a number of existing methods for finding such transforms directly (see e.g., \cite{Tulino04Verdu}).  However, those methods do not extend to the ALS problem.  We now discuss a general approach, which applies to both \iid and isometric $\mS$ for both the MMSE and the ALS receiver.

The aim is to derive a set of equations for each ``constituent'' dimension in $\mR$ (or $\mhR$), which can be solved for the \Stieltjes transform.  For example, $\mR$ has three constituent dimensions ($K$, $N$, and $M$), while $\mhR$ has four, since it also includes $i$.

Each equation is based on expanding the simple identity $\mR^{-1}\mR=\mI_M$ (or $\mhR^{-1}\mhR=\mI_M$) in each constituent dimension.  That is, since $\mR$ (or $\mhR$) is Hermitian, this term in the identity can be written as a sum of vector outer products, where the sum index runs up to the value of the dimension. Taking the normalized trace of both sides of the resulting equation can be simplified, and involves terms which have equivalent asymptotic forms, which can be evaluated using an asymptotic extension to the matrix inversion lemma.  These equivalent forms are in terms of scalar variables, some of which are mixed matrix moments.  Each of these moments can be expressed in terms of the other variables.

The result is a set of equations, which can be solved for the \Stieltjes transform, and other unknowns (e.g., various matrix moments).  Interestingly, in all cases we consider, the equations can be written in a form such that solving for the \Stieltjes transform numerically amounts to zero-finding in at most two dimensions.

\section{Analysis of MMSE Receiver}
\label{sec:MMSE_soln} %

It has been shown that, for both \iid and isometric $\mS$, the asymptotic SINR for the $k\thh$ stream at the output of the full-CSI MMSE receiver in (\ref{eq:mmse_filter}) satisfies \cite{verdu88MultiuserDetection,Chaufray03Hachem}
\begin{align}
\max_{k\leq K}\abs{ \SINR^{\text{MMSE}}_{k,N} - P_k \rhomNm } &\asto 0 \label{eq:mmse_sinr_N}
\end{align}
under the limit considered, where $\asto$ denotes almost-sure convergence,
\begin{align}
\rhomNm &=  \begin{cases} %
\frac{1}{N} \tr{\mH^\dag\mR^{-1}\mH} & \iidS \\
\frac{1}{N-K}\tr{\mPi\mH^\dag\mR^{-1}\mH} & \isoS,
\end{cases} \label{eq:sinr_mmse}
\end{align}
and $\mPi = \mI_N - \mS^\ddag$.

We now discuss some existing methods for computing the limit of the moment $\rhomNm$ in (\ref{eq:sinr_mmse}) with the MMSE full-CSI receiver, and note that the methods do not extend to the ALS receiver.  We then present the main result of this section, namely, a general SINR expression, which applies to both \iid and isometric $\mS$, derived using the approach discussed in Section \ref{sec:preamble}.

For \iid $\mS$, and square invertible $\mH$, the SINR can be obtained in terms of the limiting distribution of $\mH^\ddag$ and $\mA^2$, using the result of Silverstein and Bai \cite{silverstein95bai} after writing
\begin{align}
\rhomNm &= \frac{1}{N} \tr{(\mS\mA^2\mS^\dag + \sn(\mH^\dag\mH)^{-1})^{-1}}
\end{align}
as was done in \cite{PeacockCollingsMCCDMAJournal}, and for more general channel distributions in \cite[Theorem 2]{Debbah03Hachem}. More generally, a solution for arbitrary (non-square) channel models can be obtained for \iid $\mS$ via Girko's law (see e.g., \cite[Lemma 1]{Li04Tulino}, or \cite[Theorem IV.2]{Chuah02Tse} for just the \Stieltjes transform of $\mR$), again in terms of the limiting distributions of $\mH^\ddag$ and $\mA^2$. We note that neither of these techniques can be used to compute the output SINR for the MMSE
receiver with isometric $\mS$, or the ALS receiver.

For isometric $\mS$ and square $\mH$, the asymptotic SINR was first presented in \cite{Debbah03Hachem} with $\mA=\mI_K$, and was extended in \cite{Chaufray03Hachem} to include general $\mA$.  Although this approach could also be used to consider non-square $\mH$, it does not extend to the ALS receiver since it relies on the particular structure of $\mR$, which is not shared by $\mhR$.

\subsection{Asymptotic SINR for MMSE Receiver}

%The proof of the theorem requires replacing $\sn$ by the complex transform variable $-z$, where $z\in\CC^+$.

The following theorem allows us to evaluate the limit of $\rhomNm$ (and hence the asymptotic MMSE SINR) for general channels and for either \iid or isometric $\mS$. The theorem is in terms of the \Stieltjes transform of the \edf of the eigenvalues of $(\mH\mS\mA)^\ddag$. That is, we generalize the definition of $\mR$ from (\ref{eq:defn_mR}) by replacing $\sn$ by a complex variable $z\in\CC^+$ (\ie, $\mR = (\mH\mS\mA)^\ddag - z\mI_M$), such that the \Stieltjes transform of the \edf of the eigenvalues of $(\mH\mS\mA)^\ddag$ is given by $\Gmse^N(z) = \frac{1}{M}\tr{\mR^{-1}}$. The theorem is given in terms of the two additional random variables $\rhomNm\in\CC^+$, as defined in (\ref{eq:sinr_mmse}) using the redefinition of $\mR$, and $\taumNm\in\CC^+$.  The variable $\taumNm$ is defined in terms of matrix equations, and is given in Appendix \ref{sec:mmse_var_defns} since the definition is lengthy and is not needed to state the following result.
\begin{theorem}
\label{th:unified_mmse} %
Under the assumptions in Section \ref{sec:limit}, as $(M,N,K)\to\infty$ with $M/N\to\beta>0$ and $K/N\to\alpha>0$ fixed, the \Stieltjes transform of the \edf of the eigenvalues of $(\mH\mS\mA)^\ddag$, $\Gmse^N(z)$, $z\in\CC^+$, along with $\rhomNm$ and $\taumNm$ satisfy
\begin{align}
\abs{\Gmse^N(z) - \gmm} \asto 0, \\
\abs{\rhomNm - \rhomm} \asto 0, \\
\abs{\taumNm - \taumm} \asto 0,
\end{align}
where $\gmm,\rhomm,\taumm\in\CC^+$ are solutions to
\begin{align}
\gmm &=  -\frac{1}{z}\(1-\frac{\alpha}{\beta}\rhomm\sE_{1,1}\)
\label{eq:Gz_mmse_unified} \\
\rhomm &= \begin{cases} \label{eq:rm1_mmse_unified}
\displaystyle -z^{-1}\beta^*\sHm_{1,1} & \iidS \\
\displaystyle  \frac{-z^{-1}\beta^*\sHm_{1,1}}{1-\beta(1+z\gmm)} & \isoS,
\end{cases} \\
\taumm &= \alpha\EP - \begin{cases}
\displaystyle  \alpha\sE_{1,1} & \iidS \\
\displaystyle  \frac{\alpha\sE_{1,1}}{1-\beta(1+z\gmm)} & \isoS,
\end{cases} \label{eq:tm1_mmse_unified}
%\rhomm &= \begin{cases} \label{eq:rm1_mmse_unified}
%\displaystyle -z^{-1}\beta^*\sHm_{1,1} & \iidS \\
%\displaystyle  \frac{-z^{-1}\beta^*\sHm_{1,1}}{1-\beta(1+z\gmm)} & \isoS
%\end{cases} \\
%\taumm &= \begin{cases}
%\displaystyle  \alpha\EP - \alpha\sE_{1,1} & \iidS \\
%\displaystyle  \alpha\EP - \frac{\alpha\sE_{1,1}}{1-\beta(1+z\gmm)} & \isoS
%\end{cases} \label{eq:tm1_mmse_unified}
\end{align}
where $\beta^* = \min(\beta,1)$, and
\begin{align}
\sE_{m,1} &= \EXp{\frac{P^m}{1 + P\rhomm}} \label{eq:def_Emn_mmse} \\
\sHm_{m,1} &= \EXp{ \frac{H^m}{1+Hz^{-1}(\taumm-\alpha\EP)} } \label{eq:def_Hmn_mmse}
\end{align}
for $m\in\ZZ^*$. The expectations in (\ref{eq:def_Emn_mmse}) and (\ref{eq:def_Hmn_mmse}) are with respect to the scalar random variables $P$ and $H$, respectively, and the distributions of $P$ and $H$ are the asymptotic eigenvalue distributions of $\mA^2$ and the first $\beta^*N = \min(M,N)$ non-zero eigenvalues of $\mH^\ddag$, respectively, and $\EP=\Exp{P}$.
\end{theorem}
\begin{proof}
See Appendix \ref{ap:mmse_proofs}.
\end{proof}
Remarks:
\begin{itemize}
\item If (\ref{eq:Gz_mmse_unified})--(\ref{eq:tm1_mmse_unified}) has a unique solution\footnote{Certainly, for \iid $\mS$, the equations of Theorem \ref{th:unified_mmse} have a unique solution, since the same result can be obtained via Girko's law, for which the solution is known to be unique. Although it is not proved, we believe that this is also true for isometric $\mS$. Numerical studies support this.} $\gmm,\rhomm,\taumm\in\CC^+$ for any given $z\in\CC^+$, then Theorem \ref{th:unified_mmse} additionally implies that the \edf of the eigenvalues of $(\mH\mS\mA)^\ddag$ almost surely converges in distribution to a deterministic distribution, whose \Stieltjes transform is $\Gmse(z)$.

Moreover, we have that $\rhomNm$ converges almost surely to the deterministic value $\rhomm$ in the limit considered, and so, letting $z = -\sn + \epsilon j$ and taking $\epsilon\to 0$, as suggested by (\ref{eq:mmse_sinr_N}), the asymptotic SINR of the $k\thh$ data stream almost surely converges to $P_k\rhomm$. %
\item For \iid $\mS$, Theorem \ref{th:unified_mmse} can be obtained via Girko's law (see e.g., \cite[Theorem IV.2]{Chuah02Tse}). For isometric $\mS$, this result appears to be new. However, in both cases Theorem \ref{th:unified_mmse} can be derived (under the same set of assumptions) using the $\ST$-transform from free probability, or the method of \cite{Vasilchuk01}. We give a different proof, relying only on elementary matrix manipulations. The primary reason for presenting this result is to lead into the ALS analysis, which will follow the same general procedure outlined in the proof of Theorem \ref{th:unified_mmse} in Appendix \ref{ap:mmse_proofs}.
\item The following steps describe how to find $\Gmse(z)$ numerically via Theorem \ref{th:unified_mmse}
for specific distributions of $P$ and $H$, and given values of $z$, $\alpha$, and $\beta$.
    \begin{itemize}
    \item Consider (\ref{eq:rm1_mmse_unified}) as a scalar function of $\rhomm$, i.e., $X(\rhomm)=0$. %
    \item Numerically find the unique positive root of $X(\rhomm)$ using standard techniques (e.g.,
    using a routine such as \emph{fzero} in Matlab), where, for a given value of $\rhomm$, the
    corresponding values of $\taumm$ and $\gmm$ are directly evaluated using
    (\ref{eq:tm1_mmse_unified}) for $\taumm$, and (\ref{eq:Gz_mmse_unified}) for $\gmm$. %
    \end{itemize} %
\item In fact, (\ref{eq:Gz_mmse_unified}) is just one of many possible expressions which can be derived from the identity $\frac{1}{M}\tr{\mR\mR^{-1}} = 1$.  Other expressions involving $\gmm$ derived in this manner include
\begin{align}
\beta(1 + z\gmm) &=  \alpha(1 - \sE_{0,1}) \:=\: \alpha\rhomm\sE_{1,1} \label{eq:RRinv_mmse_dimK} \\
&= \beta^*(1-\sHm_{0,1}) \:=\: \beta^*z^{-1}(\taumm-\alpha\EP)\sH_{1,1} \label{eq:RRinv_mmse_dimN}
\end{align} These expressions, derived in Appendix \ref{ap:mmse_proofs}, are used in the proof of Theorem \ref{th:unified_mmse}.%
\item If $\mH$ is exponentially distributed with mean one (i.e., MC- or DS-CDMA in frequency-selective Rayleigh fading),  $\sHm_{1,1} = (1 - \ffunca(x))/x$, where $x = z^{-1}(\taumm-\alpha\EP)$, and $\ffunca(x) = x^{-1}\exp\(x^{-1}\)\Ei\(x^{-1}\)$ where $\Ei(x) = \int_1^\infty  e^{-xt}t^{-1}\text{d}t$ is the first-order Exponential Integral. %
\end{itemize}

%The \Stieltjes transform of the \aed of an $M\times M$ random matrix $\mX$, is equivalently the
%limit of the moment $\frac{1}{M}\tr{(\mX-z\mI_M)^{-1}}$, where $z$ is the transform variable.
%Therefore, if we define $\gmm^N = \frac{1}{M}\tr{\mR^{-1}}$ and $\gmm = \lim
%\gmm^N$, then by evaluating the \Stieltjes transform at $z=-\sn$ we get $\gmm$ and
%$\rhomm$ in the general channel and \iid/isometric case, and therefore the MMSE SINR follows
%directly from (\ref{eq:sinr_mmse_defn}).

\subsection{Alternate Representation of MMSE SINR}
\label{sec:alternate_mmse_sinr} %

%$\taumtNm$, and $\taumtNbm$,

We now present an alternate expression for the asymptotic value of $\SINR^{\text{MMSE}}_{k,N}$, which will allow us to determine the relationship between the asymptotic MMSE SINR and the asymptotic ALS SINR considered later in Sections \ref{sec:als_results}--\ref{sec:relationship}. This expression depends on the additional random variables $\rhok{j}^N\in\CC^+$, $j=2,3,4$, which, along with the auxiliary random variables $\taun{j}^N\in\CC^+$, $j=2,3$, are defined in terms of matrix equations in Appendix \ref{sec:mmse_var_defns} (The definitions of these variables are lengthy, and are not needed to state the following result; so to facilitate the flow of results they are not stated here.)

It is shown in Appendix \ref{ap:alternate_mmse_sinr} that, under the assumptions in Section \ref{sec:limit},
\begin{align}
\max_{k\leq K} \abs{ \SINR^{\text{MMSE}}_{k,N} \;-\; \frac{P_k \abs{\rhomNm}^2}{\rhomtNcm + \sn\rhomtNm}} \asto 0 \label{eq:alternate_mmse_sinr}
\end{align}
as $(M,N,K)\to\infty$ with $M/N\to\beta>0$ and $K/N\to\alpha>0$. Moreover, $\abs{\rhok{j}^N - \rhok{j}}\asto 0$, $j=2,3,4$, and $\abs{\taun{j}^N - \taun{j}}\asto 0$, $j=2,3$, in the limit considered, where $\rhok{j}\in\CC^+$, $j=2,3,4$, and $\taun{j}\in\CC^+$, $j=2,3$, are solutions to the following set of equations. For \iid $\mS$,
\begin{align}
\rhok{j} &= \begin{cases}
\beta^*\abs{z}^{-2}(\sH_{1,2} + \taumtm\sH_{2,2}) &,\quad j=2, \\
\beta^*\abs{z}^{-2}(1 + \taumtbm)\sH_{2,2} &,\quad j=3, \\
\alpha \rhomtbm \sE_{1,2} &,\quad j=4.
\end{cases} \label{eq:rhomtm_iid} \\
\taun{j} &= \alpha \rhok{j} \sE_{2,2} \quad,\quad j=2,3. \label{eq:taumtbm_iid}
%\taun{j} &= \begin{cases}
%\alpha \rhomtm \sE_{2,2} &,\quad j=2, \\
%\alpha \rhomtbm \sE_{2,2} &,\quad j=3.
%\end{cases}
\end{align}
and for isometric $\mS$,
\begin{align}
\rhok{j} &= \frac{\left.\begin{cases}
\beta^*\abs{z}^{-2}(\sH_{1,2} + \taumtm\sH_{2,2}) &,\quad j=2, \\
\beta^*\abs{z}^{-2}(1 + \taumtbm)\sH_{2,2} &,\quad j=3, \\
\alpha(\rhomtbm \sE_{1,2} - \abs{\rhomm}^2\sE_{1,2}) &,\quad j=4. \\
\end{cases}\right\}}{\alpha(\sE_{0,2}-1)+1} \label{eq:rhomtm_iso} \\
%\rhomtm &= \frac{\beta^*\abs{z}^{-2}(\sH_{1,2} + \taumtm\sH_{2,2})}{\alpha(\sE_{0,2}-1)+1} \label{eq:rhomtm_iso} \\
%\rhomtbm &= \frac{\beta^*\abs{z}^{-2}(1 + \taumtbm)\sH_{2,2}}{\alpha(\sE_{0,2}-1)+1} \\
%\rhomtcm &= \frac{\alpha(\rhomtbm \sE_{1,2} - \abs{\rhomm}^2\sE_{1,2})}{\alpha(\sE_{0,2}-1)+1} \\
\taun{j} &= \frac{\left.\begin{cases}
\alpha\rhomtm\sE_{2,2}-\beta^*\abs{z}^{-2}\abs{\alpha\EP-\taumm}^2\sH_{1,2} &,\quad j=2, \\
\alpha\rhomtbm\sE_{2,2}-\beta^* \abs{z}^{-2}\abs{\alpha\EP-\taumm}^2\sH_{2,2} &,\quad j=3.
\end{cases}\right\}}{\beta^*(\sHm_{0,2}-1) + 1} \label{eq:taumtbm_iso}
%\taumtm &=\frac{\alpha\rhomtm\sE_{2,2}-\beta^*\abs{z}^{-2}\abs{\alpha\EP-\taumm}^2\sH_{1,2}}{\beta^*(\sHm_{0,2}-1) + 1}\\
%\taumtbm &= \frac{\alpha\rhomtbm\sE_{2,2}-\beta^* \abs{z}^{-2}\abs{\alpha\EP-\taumm}^2\sH_{2,2}}{\beta^*(\sH_{0,2}-1) + 1} \label{eq:taumtbm_iso}
\end{align}
Also,
\begin{align}
\sE_{m,2} &= \EXp{\frac{P^m}{\abs{1 + P\rhomm}^2}} \\
\sHm_{m,2} &= \EXp{ \frac{H^m}{\abs{1+Hz^{-1}(\taumm-\alpha\EP)}^2} }
\end{align}
for $m\in\ZZ$, and $\rhomm$, $\taumm$, $\EP$, $\beta^*$, $P$ and $H$ are determined from Theorem \ref{th:unified_mmse}.

Again, assuming that (\ref{eq:Gz_mmse_unified})--(\ref{eq:tm1_mmse_unified}) and also (\ref{eq:rhomtm_iid})--(\ref{eq:taumtbm_iso}) have unique solutions, then we also have that the SINR of the $k\thh$ data stream converges almost surely to the deterministic quantity $P_k \abs{\rhomm}^2/(\rhomtcm + \sn\rhomtm)$.

\section{Analysis of ALS Receiver}
\label{sec:als_results}

In this section we derive the asymptotic SINR for the adaptive receiver, using the general approach discussed in Section \ref{sec:preamble}.

Firstly, we derive the asymptotic transient ALS SINR after a specified number of training intervals (either with training, or semi-blind). Our aim is to characterize the typical transient response of the receiver as a function of $\eta$, \ie, as the number of training symbols increases. The resulting expression is in terms of several large matrix variables involving the sample autocorrelation matrix. We present a theorem which gives the \Stieltjes transform of the \aed of the sample autocorrelation matrix, and fixed-point expressions for each variable required to compute the asymptotic SINR.

Secondly, from the transient SINR solution we determine the steady-state asymptotic ALS SINR, that is, the SINR as the number of training intervals (either with a training sequence, or semi-blind) goes to infinity (i.e., $\eta\to\infty$). Without data windowing, we verify that the solution for the ALS SINR converges to the MMSE SINR. Then we determine the steady-state SINR when an arbitrary windowing function is used.

\subsection{Transient ALS SINR}
\label{sec:transient_als_sinr}

The following result relates the transient SINR of the ALS receiver to six auxiliary random variables, $\rhomn{j}^N$ and $\psimn{j}^N\in\CC^+$, $j=1,2,4$.  The definitions of these variables are in terms of matrix traces and quadratic forms, and are quite lengthy. So, for clarity of presentation, and also since asymptotically equivalent values of these variables can be calculated from subsequent results, the definitions are given in Appendix \ref{sec:als_var_defns}.

\begin{theorem}
\label{th:sinr_als} %
In the limit as $(M,N,K,i)\to\infty$ with $M/N\to\beta>0$, $K/N\to\alpha>0$, and $i/N\to\eta>0$ fixed,
\begin{align}
\max_{k\leq K} \abs{ \SINR^{\text{ALS}}_{k,N} \;-\; \frac{P_k \abs{a_{k,1}}^2 \abs{\rhomNi}^2}{\abs{a_{k,1}}^2(\rhomtNci + \sn\rhomtNi) + \abs{a_{k,2}}^2(\psimtNci + \sn\psimtNi)} } \asto 0 \label{eq:asympSINR}
\end{align}
where
\begin{align}
(a_{k,1},\; a_{k,2}) &= \begin{cases}
(1,\;  -A_k\rhomNi) & \text{, semi-blind LS} \\
(A_k(\EW-\psimNi),\; 1) & \text{, LS with training} ,
\end{cases} \label{eq:def_a1a2}
\end{align}
$\EW$ is the mean of the \aed of $\mW$, and the definitions of $\rhomn{j}^N$ and $\psimn{j}^N$, $j=1,2,4$ are given in Appendix \ref{sec:als_var_defns}.
\end{theorem}
\begin{proof}
See Appendix \ref{ap:derivationSINR}. Note that the definitions and derivations of Appendix \ref{ap:proof_unified_als} necessarily precede Appendix \ref{ap:derivationSINR}.
\end{proof}

Remarks:
\begin{itemize}
\item Expressions, which can be used to compute asymptotically equivalent values of $\rhomn{j}^N$ and $\psimn{j}^N$, $j=1,2,4$, are presented in Theorem \ref{th:unified_als} and Lemma \ref{lem:extra_moments}. %
\item The preceding ALS SINR expression resembles the alternate MMSE SINR expression (\ref{eq:alternate_mmse_sinr}) derived in Section \ref{sec:alternate_mmse_sinr}. However, a simplified expression for the ALS SINR, such as that presented for the MMSE SINR in Theorem \ref{th:unified_mmse}, is not possible. This is due to the fact that a simplification of the interference power, as discussed in the proof of (\ref{eq:alternate_mmse_sinr}) in Appendix \ref{ap:alternate_mmse_sinr}, is not possible for the ALS receiver.
\end{itemize}

The following Theorem and Lemma give a sufficient number of relations to calculate the asymptotic moments required for the asymptotic SINR in (\ref{eq:asympSINR}) of Theorem \ref{th:sinr_als}.

In a similar manner to Section \ref{sec:MMSE_soln}, firstly we determine expressions for the \Stieltjes transform of the \edf of the eigenvalues of the \emph{sample} autocorrelation matrix.  That is, we generalize the definition of $\mhR$ as follows, $\mhR = \frac{1}{i}\sR\mW\sR^\dag - z\mI_M$, where $z\in\CC^+$, such that the \Stieltjes transform of the \edf of the eigenvalues of $\frac{1}{i}\sR\mW\sR^\dag$ is given by $\Gals^N(z) = \frac{1}{M}\tr{\mhR^{-1}}$. The result is necessarily stated in terms of the additional random variables $\rhomNi$, $\taumNi$, $\psimNi$, $\omegmNi$, $\nmNi$, and $\rmNi\in\CC^+$. As in Theorem \ref{th:sinr_als}, the definitions of these variables are in terms of matrix traces and quadratic forms, and are lengthy. To facilitate the presentation of results, the definitions of these variables are given in Appendix \ref{sec:als_var_defns}.

%Firstly, the following Theorem gives the \Stieltjes transform of the \aed of the sample autocorrelation matrix, which is used to determine $\rhomi$ and $\psimi$ required for Theorem \ref{th:sinr_als}. In the theorem, we replace $-\frac{\mu}{\eta}$ by the transform variable $z\in\CC^+$.

\begin{theorem}
\label{th:unified_als} %
Under the assumptions in Section \ref{sec:limit}, as $(M,N,K,i)\to\infty$, with $M/N\to\beta>0$, $K/N\to\alpha>0$, and  $i/N\to\eta>0$ fixed, the \Stieltjes transform of the \edf of the eigenvalues of $\frac{1}{i}\sR\mW\sR^\dag$, $\Gals^N(z)$, $z\in\CC^+$, along with $\rhomNi$, $\taumNi$, $\psimNi$, $\omegmNi$, $\nmNi$, and $\rmNi\in\CC^+$ satisfy
\begin{align}
\abs{\Gals^N(z) - \gmi} \asto 0,
\end{align}
\begin{align}
\abs{\rhomNi - \rhomi} &\asto 0, &
\abs{\taumNi - \taumi} &\asto 0, \\
\abs{\psimNi - \psimi} &\asto 0, &
\abs{\omegmNi - \omegmi} &\asto 0, \\
\abs{\nmNi - \nmi} &\asto 0, &
\abs{\rmNi - \rmi} &\asto 0.
\end{align}
%$\abs{\rhomNi - \rhomi} \asto 0$, $\abs{\taumNi - \taumi} \asto 0$, $\abs{\psimNi - \psimi} \asto 0$, $\abs{\omegmNi - \omegmi} \asto 0$, $\abs{\nmNi - \nmi} \asto 0$, and $\abs{\rmNi - \rmi} \asto 0$,
where $\gmi$, $\rhomi$, $\taumi$, $\psimi$, $\omegmi$, $\nmi$, and $\rmi\in\CC^+$ are solutions to
\begin{align}
\gmi &= \nmi\(1 + \frac{\alpha}{\beta}\rhomi(\psimi-\EW)\sEa_{1,1}\) \label{eq:gmi_gen_als} \\
\rhomi &= \begin{cases} \label{eq:rhomi_gen}
\displaystyle  \nmi \beta^* \sHa_{1,1} & \iidS \\
\displaystyle \frac{\nmi\beta^*\sHa_{1,1}}{1-\beta(1-\gmi/\nmi)} & \isoS,
\end{cases} \\
\taumi &= \alpha\EP\EW +  \begin{cases} \label{eq:taumi_gen}
\displaystyle  \alpha (\psimi-\EW) \sEa_{1,1} & \iidS \\
\displaystyle  \frac{\alpha(\psimi-\EW)\sEa_{1,1}}{1-\beta(1-\gmi/\nmi)} & \isoS,
\end{cases}
\end{align}
$\beta^* = \min(\beta,1)$, and
\begin{align}
\psimi &= \EW - \begin{cases} \label{eq:psimi_gen}
\displaystyle  \sW_{1,1} & \iidB \\
\displaystyle  \frac{\sW_{1,1}}{1 - \frac{\beta}{\eta^*}(1 - \gmi/\nmi)} & \isoB
\end{cases} \\
\omegmi &= \begin{cases} \label{eq:omegmi_gen}
\displaystyle  \frac{\alpha}{\eta} \rhomi \sEa_{1,1} & \iidB \\
\displaystyle  \frac{\frac{\alpha}{\eta}\rhomi\sEa_{1,1}}{1-\frac{\beta}{\eta^*}(1 - \gmi/\nmi)} & \isoB
\end{cases} \\
\nmi &= \frac{1}{\sn\sW_{1,1} - z} \\
\rmi &= \omegmi + \frac{\beta}{\eta}\sn\gmi \label{eq:nmi_gen}
\end{align}
$\eta^* = \max(\eta,\alpha)$, and
\begin{align}
\sEa_{m,1} &= \EXp{\frac{P^m}{1-P\rhomi(\psimi-\EW)}} \label{eq:def_Emn} \\
\sHa_{m,1} &= \EXp{ \frac{H^m} {1 - H \nmi(\taumi-\alpha\EP\EW)} } \label{eq:def_Hmn} \\
\sW_{m,1} &= \EXp{\frac{W^m}{1 + W \rmi}} \label{eq:def_Wmn}
\end{align}
for $m\in\ZZ^*$. The expectations in (\ref{eq:def_Emn}), (\ref{eq:def_Hmn}), and (\ref{eq:def_Wmn}) are with respect to the scalar random variables $P$, $H$, and $W$, respectively, where the distributions of $P$, $H$, and $W$ are the \aed{s} of $\mA^2$, the first $\beta^*N$ eigenvalues of $\mH^\ddag$, and $\mW$, respectively. Also, $\EP=\Exp{P}$ and $\EW = \Exp{W}$.
\end{theorem}
\begin{proof}
See Appendix \ref{ap:proof_unified_als}.
\end{proof}
Remarks:
\begin{itemize}
\item If (\ref{eq:gmi_gen_als})--(\ref{eq:nmi_gen}) has a unique solution $\gmi$, $\rhomi$, $\taumi$, $\psimi$, $\omegmi$, $\nmi$, $\rmi\in\CC^+$ for any given $z\in\CC^+$, then Theorem \ref{th:unified_als} additionally gives that the \edf of the eigenvalues of $\frac{1}{i}\sR\mW\sR^\dag$ almost surely converges in distribution to a deterministic distribution, whose \Stieltjes transform is $\Gals(z)$.
\item Note that for \iid $\mB$ and $\mu\to 0$ (\ie, \iid training sequences and no diagonal loading), $\sW_{1,1}\gmi$, $\sW_{1,1}\rhomi$, and $\sW_{1,1}(\taumi - \alpha\EP\EW)$ satisfy the same equations as $\gmm$, $\rhomm$, and $(\taumm - \alpha\EP)$ from Theorem \ref{th:unified_mmse}. Moreover, due to (\ref{eq:RRinv_dimi_a}), we have that $\sW_{1,1}$ is a function of only $\beta$, $\eta$, and the window shape. This observation, along with the alternate MMSE SINR expression of Section \ref{sec:alternate_mmse_sinr}, are the key elements in determining the relationship between the ALS and MMSE SINRs, outlined later in Section \ref{sec:relationship}.
\item For exponential weighting with $\Leff < \infty$ (where $\Leff$ is the large-system window size defined in Section \ref{sec:limit}), in Appendix \ref{ap:dist_W} the \edf of $\mW$ is shown to converge in distribution to the fixed distribution
\begin{align}
F_W(w) &= 1 + \frac{\Leff}{\eta}\ln w\;,\hspace{5mm} e^{-\eta/\Leff}\leq w\leq1 \label{eq:dist_W}
\end{align}
which is the relevant distribution of $W$ required in (\ref{eq:def_Wmn}). Also, for $z\to\real(z)$, values of $\sW_{m,1}$ which are required can be evaluated using (\ref{eq:dist_W}), and are given by
\begin{align}
\sW_{0,1} &=  \frac{\Leff}{\eta}\log\(\frac{1+e^{-\eta/\Leff}\rmi}{1+\rmi}\) + 1 \label{eq:def_W01} \\
\sW_{1,1} &= \frac{1}{\rmi}(1 - \sW_{0,1}) \label{eq:def_W11}
%%\frac{\Leff}{\eta \rmi}\log\(\frac{1+\rmi}{1+\rmi e^{-\eta/\Leff}} \)\label{eq:def_W11} \\
%\sW_{1,2} &= \frac{\EW}{(1+\rmi)(1+e^{-\eta/\Leff} \rmi)} \label{eq:def_W12} \\
%%\sW_{2,1} &= \frac{\frac{\Leff}{\eta}(1-e^{-\eta/\Leff})-\sW_{1,1}}{\rmi} \\
%\sW_{2,2} &=  \frac{1}{\rmi}(\sW_{1,1} - \sW_{1,2}) \label{eq:def_W22}
\end{align}

%\item We could also do the Ralf channel this way. % ZZZ
%Note: we can obtain the moments for the MIMO channel with richness parameter (that is, \iid $\mH$
%in place of isometric $\mH$) by replacing $\nmi = \gmi$, $\nmti = \gmti$ and $\nmtbi =
%\rhomti/\beta$, and some other things. Maybe I'll put this separately.

\item In fact, (\ref{eq:gmi_gen_als}) is one of many possible expressions, which can be derived
from the identity $\frac{1}{M}\tr{\mhR\mhR^{-1}} = 1$.  Other expressions
involving $\gmi$ derived in this manner include
\begin{align}
\beta(1+z\gmi) &= \eta(1-\sW_{0,1}) \label{eq:RRinv_dimi_a} \\
\beta(1 - \gmi/\nmi) &= \eta \omegmi\sW_{1,1} \label{eq:RRinv_dimi} \\
&= \alpha(1 - \sEa_{0,1}) \:=\: -\alpha\rhomi(\psimi - \EW)\sEa_{1,1} \label{eq:RRinv_dimK} \\
&= \beta^*(1-\sHa_{0,1}) \:=\: -\nmi(\taumi - \alpha\EP\EW)\beta^*\sHa_{1,1} \label{eq:RRinv_dimN}
\end{align}
Note the similarity to the expressions (\ref{eq:RRinv_mmse_dimK})--(\ref{eq:RRinv_mmse_dimN}),
derived in a similar manner for the MMSE receiver.  These expressions, derived in Appendix \ref{ap:proof_unified_als}, are used in the proof of Theorem \ref{th:unified_als}.

\item Note that $\sEa_{m,1}$, $\sHa_{m,1}$, and $\sW_{m,1}$ are all of the form $\sX_{m,1} = \EXp{X^m/(1+xX)}$,
for which the following simple and useful identity holds.
\begin{align}
\Exp{X^m} &= \EXp{X^m \frac{1 + x X}{1 + xX}} \:=\: \sX_{m,1} + x \sX_{m+1,1} \label{eq:id_WEHa}
\end{align}
for $m\in\ZZ^*$. Observe also that the last equality in each of (\ref{eq:RRinv_mmse_dimK}), (\ref{eq:RRinv_mmse_dimN}), (\ref{eq:RRinv_dimK}) and (\ref{eq:RRinv_dimN}) follows from the identity (\ref{eq:id_WEHa}), as does (\ref{eq:def_W11}). This identity also relates (\ref{eq:RRinv_dimi_a}) to (\ref{eq:RRinv_dimi}), although in a less obvious way. %

\item The set of equations in Theorem \ref{th:unified_als} can be solved numerically in a similar manner to that discussed for the MMSE case after Theorem \ref{th:unified_mmse}.  Here, however, it is advantageous to consider (\ref{eq:rhomi_gen}) and (\ref{eq:omegmi_gen}) as a two-dimensional equation in the variables\footnote{That is, unless $z\to 0$ is being considered (\ie, no diagonal loading), in which case it is necessary to instead consider $\rhomi$ and $\omegmi$ as the search variables, since $\rmi$ depends only on $\beta$, $\eta$, and $\sW_{0,1}$ due to (\ref{eq:RRinv_dimi_a}).} $\rhomi$ and $\rmi$. During zero finding, given these values, the remaining variables $\gmi$, $\omegmi$, $\taumi$, $\psimi$, $\nmi$, $\sW_{1,1}$, $\sEa_{1,1}$, and $\sHa_{1,1}$ can be directly calculated.

\item Similar to \cite[Section 3]{silverstein95bai}, it is possible that through a process of truncation and centralization, the condition that the moments of $\mS$ and $\mB$ are bounded when either matrix is \iid may be removed.

%\item A further valuable extension to the above result would be to consider the transient response in time-varying channels. For example, a quasi-static channel model where $\sR = [\sR_1, \sR_2]$ such that $\sR_\ell=\mH_\ell\mS\mA\mB_\ell + \mN_\ell$, $\ell=1,2$. Problem: Need $\mH_\ell^\ddag$ jointly diagonalizable. XXX
\end{itemize}

Using Theorem \ref{th:unified_als}, we may now calculate $\rhomi$ and $\psimi$, which are asymptotically equivalent to $\rhomNi$ and $\psimNi$, two of the quantities required to compute the SINR in Theorem \ref{th:sinr_als}.

The following Lemma gives expressions, which may be solved for quantities asymptotically equivalent to $\rhomn{j}^N$ and $\psimn{j}^N\in\CC^+$, $j=2,4$, and which occur in (\ref{eq:asympSINR}) of Theorem \ref{th:sinr_als}. This Lemma introduces more auxiliary random variables in $\CC^+$, namely, $\gmn{j}^N$, $\rhomn{j}^N$, $\psimn{j}^N$, and $\omegmn{j}^N$, $\rmn{j}^N$, $j=2,3,4$, plus $\taun{j}^N$ and $\nmn{j}^N$, $j=2,3$, which are defined in terms of matrix traces and quadratic forms in Appendix \ref{sec:als_var_defns}.
\begin{lemma}
\label{lem:extra_moments} %
In addition to the assumptions and definitions of Theorem \ref{th:unified_als}, under the limit specified,  $\abs{\gmn{j}^N - \gmn{j}}\asto 0$, $\abs{\rhomn{j}^N-\rhomn{j}}\asto 0$, $\abs{\psimn{j}^N-\psimn{j}}\asto 0$,  $\abs{\omegmn{j}^N-\omegmn{j}}\asto 0$, and $\abs{\rmn{j}^N-\rmn{j}}\asto 0$, $j=2,3,4$, and $\abs{\taun{j}^N-\taun{j}}\asto 0$ and $\abs{\nmn{j}^N-\nmn{j}}\asto 0$, $j=2,3$, where $\gmn{j}$, $\rhomn{j}$, $\psimn{j}$, $\omegmn{j}$, and $\rmn{j}$, $j=2,3,4$, and $\taun{j}$ and $\nmn{j}$, $j=2,3$, are solutions to the following equations.  For \iid $\mS$,
\begin{align}
\rhomn{j} &= \begin{cases}
\beta^*(\nmti \sHa_{1,2} + \abs{\nmi}^2\taumti \sHa_{2,2})  &,\quad j=2, \\
\beta^*(\nmtbi\sHa_{1,2} + \abs{\nmi}^2(1 + \taumtbi)\sHa_{2,2}) &,\quad j=3, \\
\alpha ( \rhomtbi \sEa_{1,2} + \psimtbi \abs{\rhomi}^2 \sEa_{2,2} )  &,\quad j=4,
\end{cases} \label{eq:rhomnj_iid} \\
\taumn{j} &= \alpha (\psimn{j} \sEa_{1,2} + \abs{\EW-\psimi}^2 \rhomn{j} \sEa_{2,2})  \quad,\quad j=2,3, \label{eq:taumnj_iid}
\end{align}
and for isometric $\mS$,
\begin{align}
\rhomn{j} &= \frac{\left.\begin{cases}
\displaystyle \beta^*(\nmti \sHa_{1,2} + \taumti\abs{\nmi}^2\sHa_{2,2}) -\alpha\psimti\abs{\rhomi}^2\sEa_{1,2} &,\quad j=2, \\
\displaystyle \beta^*(\nmtbi\sHa_{1,2} + (1 + \taumtbi)\abs{\nmi}^2\sHa_{2,2}) - \alpha\psimtbi \abs{\rhomi}^2\sEa_{1,2}  &,\quad j=3, \\
\displaystyle  \alpha(\rhomtbi\sEa_{1,2} + \psimtbi\abs{\rhomi}^2\sEa_{2,2} - (1 + \psimtci)\abs{\rhomi}^2\sEa_{1,2}) &,\quad j=4,
\end{cases}\right\}}{\alpha(\sEa_{0,2}-1) + 1} \label{eq:rhomnj_iso} \\
\taumn{j} &= \frac{\left.\begin{cases}
\displaystyle \alpha(\psimti \sEa_{1,2} + \abs{\EW-\psimi}^2 \rhomti\sEa_{2,2}) -\beta^*\nmti\abs{\alpha\EP\EW-\taumi}^2\sHa_{1,2} &,\quad j=2, \\
\displaystyle \alpha (\psimtbi \sEa_{1,2} + \abs{\EW-\psimi}^2 \rhomtbi\sEa_{2,2}) -\beta^*\abs{\alpha\EP\EW-\taumi}^2(\nmtbi\sHa_{1,2} + \abs{\nmi}^2\sHa_{2,2}) &,\quad j=3,
\end{cases}\right\}}{\beta^*(\sHa_{0,2}-1) + 1} \label{eq:taumnj_iso}
\end{align}
where for \iid $\mB$,
\begin{align}
\psimn{j} &= \rmn{j}\sW_{2,2} \label{eq:psimnj_iid} \\
\omegmn{j} &= \begin{cases}
\frac{\alpha}{\eta} (\rhomn{j} \sEa_{1,2} + \abs{\rhomi} \psimn{j}  \sEa_{2,2}) &,\quad j=2,3, \\
\frac{\alpha}{\eta} (\rhomtci \sEa_{1,2} + \abs{\rhomi}^2 (1+\psimtci)\sEa_{2,2}) &,\quad j=4,
\end{cases} \label{eq:omegmnj_iid}
\end{align}
while for orthogonal $\mB$
\begin{align}
\psimn{j} &= \frac{\left.\begin{cases}
\displaystyle \rmn{j}\sW_{2,2} - \frac{\alpha}{\eta^*}|\EW-\psimi|^2\rhomn{j}\sEa_{1,2} &,\quad j=2,3, \\
\displaystyle  \rmn{4}\sW_{2,2} - \frac{\alpha}{\eta^*}|\EW-\psimi|^2(\rhomtci\sEa_{1,2}+\abs{\rhomi}^2\sEa_{2,2})&,\quad j=4,
\end{cases}\right\}}{\frac{\alpha}{\eta^*}(\sEa_{0,2}-1)+1} \label{eq:psimnj_iso} \\
\omegmn{j} &= \frac{\left.\begin{cases}
\displaystyle \frac{\alpha}{\eta} (\rhomn{j} \sEa_{1,2}  +  \abs{\rhomi}^2 \psimn{j} \sEa_{2,2}) - \sn\frac{\beta}{\eta^*}\gmn{j}\abs{\omegmi}^2\sW_{2,2} &,\quad j=2,3, \\
\displaystyle \frac{\alpha}{\eta} (\rhomtci \sEa_{1,2}  +  \abs{\rhomi}^2 (1+\psimtci)\sEa_{2,2}) -
\sn\frac{\beta}{\eta^*}\gmtci\abs{\omegmi}^2\sW_{2,2} &,\quad j=4,
\end{cases}\right\}}{\frac{\eta}{\eta^*}(\abs{\omegmi}^2\sW_{2,2}-2\real(\omegmi\sW_{1,1}))+1} \label{eq:omegmnj_iso}
\end{align}
Also,
\begin{align}
\rmn{j} &= \omegmn{j} + \sn\frac{\beta}{\eta}\gmn{j} \quad,\quad j=1\lldots 4, \label{eq:rmnj_eq} \\
\gmn{j} &= \begin{cases}
\displaystyle z^{-*}(\frac{\eta}{\beta}\rmti\sW_{1,2} - \gmi)  &,\quad j=2, \\
\displaystyle \frac{\beta^*}{\beta}(\nmti\sHa_{1,2}+\abs{\nmi}^2\taumti\sHa_{2,2})) &,\quad j=3, \\
\displaystyle \frac{\alpha}{\beta} (\rhomti \sEa_{1,2}  +  \psimti \abs{\rhomi}^2 \sEa_{2,2}) &,\quad j=4,
\end{cases} \label{eq:gmnj_eq} \\
\nmn{j} &= \begin{cases}
\abs{\nmi}^2( 1 + \sn\rmti\sW_{2,2}) &,\quad j=2, \\
\sn \abs{\nmi}^2 \rmtbi \sW_{2,2} &,\quad j=3,
\end{cases} \label{eq:nmnj}
\end{align}
and
\begin{align}
\sEa_{m,2} &= \EXp{\frac{P^m}{\abs{1-P\rhomi(\psimi-\EW)}^2}}, \\
\sHa_{m,2} &= \EXp{ \frac{H^m} {\abs{1 - H \nmi(\taumi-\alpha\EP\EW)}^2} }, \\
\sW_{m,2} &= \EXp{\frac{W^m}{\abs{1 + W \rmi}^2}},
\end{align}
where $\gmi$, $\rhomi$, $\taumi$, $\psimi$, $\omegmi$, $\nmi$, $\rmi$, $\sEa_{m,n}$, $\sHa_{m,n}$, and $\sW_{m,n}$ are determined by Theorem \ref{th:unified_als}, and again $z\to -\mu/\eta$.

\end{lemma}
\begin{proof}
The proof of Lemma \ref{lem:extra_moments} follows the same approach as the derivation of Theorem \ref{th:unified_als}. That is, expressions for $\rhomn{j}$, $\taumn{j}$, $\psimn{j}$, $\omegmn{j}$ and $\nmn{j}$ for $j=2,3,4$ are derived in the same manner as the expressions for $\rhomn{1}$, $\taumn{1}$, $\psimn{1}$, $\omegmn{1}$ and $\nmn{1}$ in Theorem \ref{th:unified_als}, respectively.  Also, as the expression for $\gmi$ in Theorem \ref{th:unified_als} is derived from the identity $\mhR\mhR^{-1}=\mI_M$, so the expression for $\gmti$ in (\ref{eq:gmnj_eq}) is derived from the identity $\mhR^\dag\mhR^{-\dag}\mhR^{-1}=\mhR^{-1}$.  A full derivation can be found in \cite{Peacock05}. %
%See Appendix \ref{ap:secondorder}.
\end{proof}

Remarks:
\begin{itemize}

\item For exponential weighting with $\Leff < \infty$, as $z\to\real(z)$, the required expressions for $\sW_{m,2}$ can be evaluated using (\ref{eq:dist_W}), and are given by
\begin{align}
\sW_{1,2} &= \frac{\EW}{(1+\rmi)(1+e^{-\eta/\Leff} \rmi)} \label{eq:def_W12} \\
\sW_{2,2} &=  \frac{1}{\rmi}(\sW_{1,1} - \sW_{1,2}) \label{eq:def_W22}
\end{align}
where $\EW = \frac{\Leff}{\eta}(1 - e^{-\eta/\Leff})$.

\item Note that $\sEa_{m,2}$, $\sHa_{m,2}$, and $\sW_{m,2}$ are all of the form $\sX_{m,2} = \EXp{X^m/\abs{1+xX}^2}$,
for which the following simple and useful identity holds.
\begin{align}
\sX_{m,1} &= \EXp{\frac{(1 + x^* X)X^m}{\abs{1 + xX}^2}} \:=\: \sX_{m,2} + x^* \sX_{m+1,2} \label{eq:id_WEHb}
\end{align}
for $m\in\ZZ^*$. This identity can be used to simplify the calculation of certain terms in Lemma \ref{lem:extra_moments}, and also gives (\ref{eq:def_W22}). %

\item These equations can be solved numerically using three zero-finding routines, two of which are for two variables, while the third is for one variable.  Specifically,
\begin{enumerate}
    \item First solve the subset of equations given by
    (\ref{eq:rmnj_eq}), (\ref{eq:gmnj_eq}), (\ref{eq:nmnj}),
    and depending on the type of $\mS$,
    (\ref{eq:rhomnj_iid}) and (\ref{eq:taumnj_iid}), or
    (\ref{eq:rhomnj_iso}) and (\ref{eq:taumnj_iso}),
    and depending on the type of $\mB$,
    (\ref{eq:psimnj_iid}) and (\ref{eq:omegmnj_iid}), or
    (\ref{eq:psimnj_iso}) and (\ref{eq:omegmnj_iso}),
    to find $\rmti$, $\gmti$, $\nmti$, $\rhomti$, $\taumti$, $\psimti$, and $\omegmti$.
    This can be done numerically using a zero-finding routine for the two variables $\rmti$ and $\rhomti$.

    \item Solve the subset of equations given by
    (\ref{eq:rmnj_eq}), (\ref{eq:gmnj_eq}), (\ref{eq:nmnj}),
    and depending on the type of $\mS$,
    (\ref{eq:rhomnj_iid}) and (\ref{eq:taumnj_iid}), or
    (\ref{eq:rhomnj_iso}) and (\ref{eq:taumnj_iso}),
    and depending on the type of $\mB$,
    (\ref{eq:psimnj_iid}) and (\ref{eq:omegmnj_iid}), or
    (\ref{eq:psimnj_iso}) and (\ref{eq:omegmnj_iso}),
    to find $\rmtbi$, $\gmtbi$, $\nmtbi$, $\rhomtbi$, $\taumtbi$, $\psimtbi$, and $\omegmtbi$.
    This can be done numerically using a zero-finding routine for the two variables $\omegmtbi$ and $\rhomtbi$.

    \item Now solve the subset of equations given by, (\ref{eq:rmnj_eq}), (\ref{eq:gmnj_eq}),
    and depending on the type of $\mS$,
    (\ref{eq:rhomnj_iid}) or (\ref{eq:rhomnj_iso}),
    and depending of the type of $\mB$,
    (\ref{eq:psimnj_iid}) and (\ref{eq:omegmnj_iid}), or
    (\ref{eq:psimnj_iso}) and (\ref{eq:omegmnj_iso}),
    to find $\rmtci$, $\gmtci$, $\rhomtci$, $\psimtci$, and $\omegmtci$. This can be done numerically with a zero-finding routine for one variable, namely $\psimtci$.
\end{enumerate}

\item The solution (\ref{eq:gmnj_eq}) is one of many possible expressions, which can be derived from the identity $\frac{1}{M}\tr{\mhR\mhR^{-2}} = \frac{1}{M}\tr{\mhR^{-1}}$.  All possible expressions involving $\gmti$ derived in this manner include
\begin{align}
%\gmi + z\gmti &= (\frac{\eta}{\beta}\omegmti + \sn\gmti) \sW_{1,2} \\
%\gmti \nmi &=  \gmi\nmti + \frac{\alpha}{\beta}( (\psimi-\EW)\rhomti + \rhomi\psimti )\nmi^2\sEa_{1,2} \\
%&= \gmi\nmti + \frac{\beta^*}{\beta}( (\taumi-\alpha\EP\EW)\nmti + \nmi\taumti)\nmi^2\sHa_{1,2} \\
\gmi + z^*\gmti &= \frac{\eta}{\beta}\rmti \sW_{1,2}  \\
\gmti\nmi - \gmi\nmti &= \frac{\alpha}{\beta}( (\psimi-\EW)^*\rhomti + \rhomi\psimti )\abs{\nmi}^2\sEa_{1,2} \\
&= \frac{\beta^*}{\beta}( (\taumi-\alpha\EP\EW)^*\nmti + \nmi\taumti)\abs{\nmi}^2\sHa_{1,2}
\end{align}
When solving the set of equations in Lemma \ref{lem:extra_moments}, both of these expressions are more useful than (\ref{eq:gmnj_eq}) when considering $z \to 0$ (i.e.\ no diagonal loading).
\end{itemize}

\subsection{Steady-State ALS SINR}
\label{sec:unlimited_training} %

We now determine the steady-state ALS SINR, that is, the SINR as the number of training intervals $\eta \to \infty$
(either with a training sequence, or semi-blind), from the transient ALS SINR expressions in Section \ref{sec:transient_als_sinr}. Of course, if there is no windowing (i.e., $\mW=\mI_i$), and diagonal loading $\mu/\eta$ for any $\mu>0$, then the output SINR converges to that of an MMSE receiver with full CSI.  We first verify this result, and then turn to the more interesting case of data windowing and (optionally) diagonal loading.  We will see that the steady-state response is the same for both \iid and orthogonal training sequences, which matches intuition, since \iid training sequences become orthogonal as $\eta \to \infty$.

An approximate analysis of the steady-state performance of the ALS receiver with exponential windowing was presented in \cite{Poor97Wang} for DS-CDMA with flat fading. The large-system steady-state ALS performance is considered in \cite[Corollary 2]{Honig02Xiao}. In \cite{Honig02Xiao}, results from asymptotic analysis of reduced rank filters are used, which rely on arguments related to non-crossing partitions. Here we give a more direct derivation of the large-system steady-state performance of the ALS receiver for the general transmission model (\ref{eq:rcv_sig}).

Strictly speaking, Theorems \ref{th:unified_mmse} and \ref{th:unified_als} require $z\in\CC^+$, however for the following discussion we shall implicitly consider $z \to -\sn$ and $z \to -\mu/\eta$, respectively.

\subsubsection{No Windowing}

We first consider the limit of the equations in Theorem \ref{th:unified_als} as $\eta\to\infty$, and show that
without windowing (i.e.,\ $\mW = \mI_i$) the ALS SINR converges to the MMSE SINR.

First note that $\omegmi\to 0$ for both \iid and orthogonal $\mB$, which means $\sW_{m,n}\to 1$, and therefore $\psimi\to 0$ and $\nmi\to 1/\sn$.  Moreover, $\sEa_{m,n}\to\sE_{m,n}$ and $\sHa_{m,n}\to\sHm_{m,n}$ from Theorem \ref{th:unified_mmse}.  We see that the expressions for the ALS moments $\gmi$, $\rhomi$ and $\taumi$ from Theorem \ref{th:unified_als} converge to the MMSE moments $\gmm$, $\rhomm$, and $\taumm$, respectively, in Theorem \ref{th:unified_mmse} at $z\to-\sn$ as $\eta\to\infty$.

Now consider the limit of the equations in Lemma \ref{lem:extra_moments} with no windowing as $\eta\to\infty$. Clearly, $\omegmti$, $\omegmtbi$, and $\omegmtci$ all $\to 0$, and hence also $\psimti$, $\psimtbi$, and $\psimtci$ also $\to 0$ for either \iid or orthogonal $\mB$. It follows that $\nmti\to(1/\sn)^2$ and $\nmtbi\to 0$.

Substituting the preceding limits into (\ref{eq:rhomnj_iid})--(\ref{eq:taumnj_iso}), we see immediately that as $\eta\to\infty$, the variables $\rhomti$, $\rhomtbi$, $\rhomtci$, $\taumti$, and $\taumtbi$ satisfy the same set of equations as $\rhomtm$, $\rhomtbm$, $\rhomtcm$, $\taumtm$, and $\taumtbm$ for the MMSE receiver, which appear
in the SINR expression in Section \ref{sec:alternate_mmse_sinr}, given by (\ref{eq:rhomtm_iid})--(\ref{eq:taumtbm_iso}). Therefore, $\rhomtci + \sn\rhomti \to \rhomm^*$ as $\eta\to\infty$, and therefore the ALS SINR converges to the MMSE SINR as $\eta\to\infty$. The diagonal loading constant $\mu/\eta$
disappears in this limit.

\subsubsection{Fixed-Length Data Windowing}

The following results apply to fixed-length windowing functions. That is, the large-system window size does scale with $\eta$.\footnote{Of course, the actual window size should increase with $N$ in order to define a meaningful large system limit, as explained in Section \ref{sec:limit}. Here we are referring to the large-system window size \emph{after} the large-system limit has been determined. With exponential windowing this is the difference between $L$ and $\Leff$.} For example, in the case of exponential windowing this corresponds to fixed $\Leff$.  More precisely, we define fixed-length windowing as
\begin{align}
%\Exp{W_\eta} &\neq 0 \quad,\; \eta<\infty \\
& \lim_{\eta\to\infty} \eta\Exp{W_\eta}  \:>\: 0 \\
& \lim_{\eta\to\infty} F_{W_\eta}(w) = \begin{cases}
0 &,\; w<0 \\
1 &,\; w\geq 0
\end{cases}
\end{align}
where, for given $\eta$, $W_\eta$ denotes a scalar r.v.\ with distribution $F_{W_\eta}(w)$, given by the (compactly supported) \aed of $\mW$. In other words, $F_{W_\eta}$ converges in distribution to a delta-distribution at zero as $\eta\to\infty$, with the mean of $W_\eta$ of order $\eta^{-1}$. For example, for exponential windowing, the mean window size is $\EW = \frac{\Leff}{\eta}(1 - e^{-\eta/\Leff})$.

%\begin{eqnarray}
%\lim_{\eta\to\infty}\sW_{m,n} &= 0  \\
%\lim_{\eta\to\infty}\eta\sW_{m,n} &\;\text{ is finite.}
%\end{eqnarray}
%for any finite $\rmi\in\CC^+$ and for $(m,n)\in\{(1,0),(0,1),(1,1),(1,2),(2,2)\}$, where $\sW_{m,n}$ is defined in (\ref{eq:def_Wmn}). These conditions are sufficient for the proof of the subsequent corollary.

We have the following corollary to Theorem \ref{th:sinr_als}, Theorem \ref{th:unified_als}, and Lemma \ref{lem:extra_moments}, which specifies the steady-state ALS SINR with fixed-length windowing.
\begin{corollary} %
\label{cor:etainf_sinr} %
Under the limit specified in Theorem \ref{th:sinr_als}, and also as $\eta\to\infty$, provided all asymptotic moments exist, the asymptotic steady-state SINR for the ALS receiver for stream $k$ with fixed-length windowing is given by the asymptotic ALS SINR specified in Theorems \ref{th:sinr_als}, \ref{th:unified_als}, and Lemma \ref{lem:extra_moments}, where
\begin{itemize}
\item the \iid $\mB$ relations are used for both \iid and isometric $\mB$, %
\item $z$ is replaced by $-\mu$, %
\item $\eta$ is replaced by one, and,
\item the variables $\EW$ and $\sW_{m,n}$ are replaced by $\tEW$ and $\tsW_{m,n}$, where
\begin{align}
\tEW &= \lim_{\eta\to\infty} \eta\EW \\
\tsW_{m,n} &= \lim_{\eta\to\infty} \eta\sW_{m,n} \label{eq:def_tsW}
\end{align}
\end{itemize}
\end{corollary}
\begin{proof}
See Appendix \ref{ap:unlimited_training}.
\end{proof}

Remark:
\begin{itemize}

\item With exponential windowing, $\tEW = \Leff$, and from (\ref{eq:def_tsW}) we have
\begin{align}
\tsW_{1,1} &= \frac{\beta(1-\mu\tgmi)}{\trmi} \label{eq:W11_etainf_expwindow} \\
\tsW_{1,2} &= \frac{\Leff}{1+\trmi} \label{eq:def_tsW12} \\
\tsW_{2,2} &= \frac{1}{\trmi}\(\tsW_{1,1} - \tsW_{1,2}\) \label{eq:def_tsW22}
\end{align}
where $\trmi = \exp(\frac{\beta}{\Leff}(1-\mu\tgmi))-1$, and we have used (\ref{eq:RRinv_dimi_a}) to obtain (\ref{eq:W11_etainf_expwindow}). %

\end{itemize}

\section{Relationship Between MMSE and ALS Receivers}
\label{sec:relationship}

In this section we present a simple relationship between the SINRs of the MMSE and ALS receivers given in the previous sections.  We note that this relationship has recently been studied in the special case $\mH=\mI_N$, \iid $\mS$, \iid training, and no diagonal loading. An approximate relation was given for rectangular windowing in \cite{Zhang02Wang}, with a corresponding exact large system expression given in \cite[Corollary 1]{Honig02Xiao}.  Also, \cite{Poor97Wang} obtained approximate expressions for the steady-state SINR relationship ($\eta\to\infty$) with exponential windowing.

\subsection{Transient Response}
The following theorem applies to any $\mH$, both \iid and isometric $\mS$, and any windowing shape.  The only restrictions are that there is \iid training and no diagonal loading (i.e., $\mu=0$). The theorem relates the expressions in Theorem \ref{th:unified_als} and Lemma \ref{lem:extra_moments} to the alternate MMSE SINR expression of Section \ref{sec:alternate_mmse_sinr}.

\begin{theorem}
\label{th:relationship_mmse_als} %
For the $k\thh$ data stream, the asymptotic SINR of the full-CSI MMSE receiver $\SINR^{\text{MMSE}}_k$ is related to the asymptotic SINR of the ALS receiver with \iid training sequences, data windowing, and no diagonal loading, $\SINR^{\text{ALS}}_k$, according to
\begin{align}
\SINR^{\text{ALS}}_k &= \frac{\SINR^{\text{MMSE}}_k}{\zeta + \frac{\zeta-1}{\SINR^{\text{MMSE}}_k}} \label{eq:sinr_als_from_mmse_train}
\end{align}
with training sequences, and
\begin{align}
\SINR^{\text{ALS}}_k &= \frac{\SINR^{\text{MMSE}}_k}{\zeta + (\zeta-1) \SINR^{\text{MMSE}}_k}
\label{eq:sinr_als_from_mmse_blind}
\end{align}
for semi-blind training,
with either \iid or isometric $\mS$, where
\begin{align}
\zeta &= \frac{\sW_{1,1}}{\sW_{1,2}} \label{eq:def_zeta}
\end{align}
which depends \emph{only} on $\eta$, $\beta$, and the window shape.
\end{theorem}
\begin{proof}
See Appendix \ref{ap:relationship_mmse_als}.
\end{proof}

Remarks:
\begin{itemize}
\item To calculate $\zeta$, note that from (\ref{eq:RRinv_dimi_a}) we have $\sW_{0,1}=1-\frac{\beta}{\eta}$ for any window shape.  Since $\sW_{0,1}$ is a fixed known function of $\rmi$, we can invert this equation to find $\rmi$.  For example, with exponential windowing, we obtain from (\ref{eq:def_W01}) that
\begin{align}
\rmi &= \frac{e^{\beta/\Leff}-1}{1-e^{(\beta-\eta)/\Leff}}
\end{align}
and with rectangular windowing, $\rmi = \beta/(\eta-\beta)$. Given $\rmi$, we can directly calculate $\sW_{m,n}$ from the definition given in (\ref{eq:def_Wmn}), and $\zeta$ from (\ref{eq:def_zeta}). The point here is that $\rmi$, $\sW_{m,n}$, and $\zeta$ are essentially constants, depending only on $\beta$, $\eta$, and the window shape.

%In addition, from (\ref{eq:RRinv_dimi_a}) and the identity (\ref{eq:id_WEH}) we determine that $\sW_{1,1} = \frac{\beta}{\eta\rmi}$.

\item Remarkably, Theorem \ref{th:relationship_mmse_als}
implies that $\SINR^{\text{ALS}}_k$ only depends on  $\beta$, $\eta$, the window shape, and $\SINR^{\text{MMSE}}_k$. That is, \emph{the convergence rate of the ALS SINR to the steady-state value is independent of the channel} (of course, the steady-state value itself depends on the channel). Stating this another way, for a particular $\beta$ and window shape, all channels, which have the same MMSE SINR, have an identical transient ALS SINR response. `Channel' here refers to the product $\mH\mS\mA$. This has been observed in \cite{Eleftheriou86Falconer,Poor97Wang}, although a transient SINR relationship, such as that given in Theorem \ref{th:relationship_mmse_als}, has not previously been determined. %
\item Recall that due to Proposition \ref{pr:allAWGNbinvar}, Theorem \ref{th:relationship_mmse_als} also holds for the general AWGN model $\vr=\mC\vb+\vn$ for which $\vb$ is unitarily invariant and the eigenvalues of $\mC^\ddag$ are well behaved. In that sense Theorem \ref{th:relationship_mmse_als} is a fundamental property of linear estimation.

\item In fact, our assumption that the additive noise $\vn_m$ is \iid complex Gaussian distributed is an unnecessary restriction, as all results presented hold for any distribution such that $\vn_m$ is unitarily invariant, and the elements of $\vn_m$ are \iid with zero mean and variance $\sn$.

\item With exponential windowing we have
\begin{align}
\zeta &= \frac{\beta(1 - e^{-\eta/\Leff})}{\Leff(1 -  e^{(\beta-\eta)/\Leff})(1-e^{-\beta/\Leff})} \label{eq:zeta_exponential_transient}
\end{align}
and for rectangular windowing, we have $\zeta = 1 + \beta/(\eta-\beta)$.  With rectangular windowing and $\beta=1$ (i.e., square $\mH$) this matches the expression derived in \cite{Honig02Xiao} for DS-CDMA with \iid signatures in flat fading.
\end{itemize}

Unfortunately, we do not have a compact expression, analogous
to Theorem \ref{th:relationship_mmse_als},
which relates the ALS and MMSE SINRs
with orthogonal training sequences and/or diagonal loading,
although it seems likely that such a relationship exists.
What we can say is that with orthogonal training sequences and no diagonal loading the moments of Theorem \ref{th:unified_mmse} and Theorem \ref{th:unified_als} are related via
\begin{align}
\rhomi &= \frac{D_\mB}{\sW_{1,1}} \rhomm \\
\gmi &= \frac{1}{\sW_{1,1}} \gmm \label{eq:gmi_gmm_transient} \\
\taumi &= \frac{\sW_{1,1}}{D_\mB}(\taumm - \alpha\EP) + \alpha\EW\EP
\end{align}
where $D_\mB = 1 - \frac{\alpha}{\eta^*}\rhomm\sE_{1,1}$. Interestingly, these relationships depend on the channel through $D_{\mB}$ (which was not the case with \iid training).  Finding a corresponding relation for the SINRs
with orthogonal training remains an open problem.

\subsection{Steady-State Response}

For the steady-state response ($\eta\to\infty$) with fixed-length data windowing, Theorem \ref{th:relationship_mmse_als} holds with
\begin{align}
\zeta &= \frac{\tsW_{1,1}}{\tsW_{1,2}} \label{eq:def_zeta_etainf}
\end{align}
where $\tsW_{m,n} = \lim_{\eta\to\infty}\eta\sW_{m,n}$.  This result is proved simply by letting $\eta\to\infty$ in Theorem \ref{th:relationship_mmse_als}. Of course, this steady-state relationship also holds for orthogonal training sequences.

% XXX Say something about window shape optimization?  Can't really think of a good scenario, because rectangular is always going to be the best for a static channel model!

With exponential weighting, we have from (\ref{eq:zeta_exponential_transient}) that $\zeta = \frac{\beta}{\Leff(1 - e^{-\beta/\Leff})}$.   Note that as $\Leff\to\infty$ (i.e., as we increase the window size) $\zeta\to 1$ (using
L'H\^{o}pital's rule), and $\SINR^{\text{ALS}}_k \to \SINR^{\text{MMSE}}_k$, as expected. %

\begin{figure}
\centering %
\includegraphics[width=\figw]{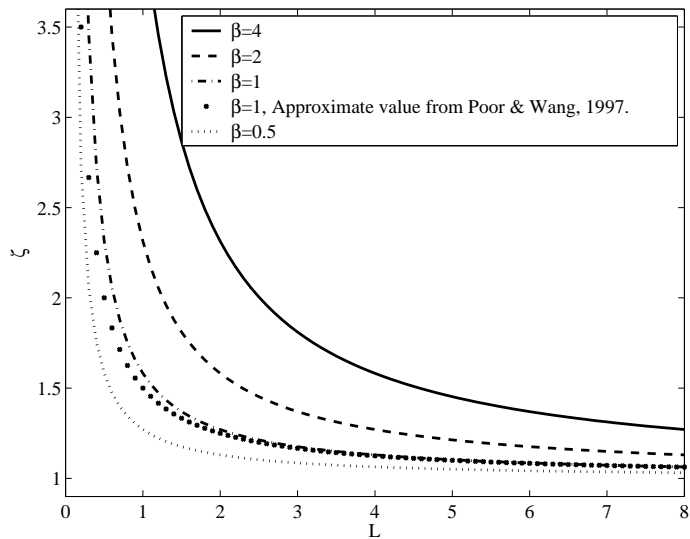} %
\caption{$\zeta$ vs.\ $\Leff$ for $\eta=\infty$ and a range of $\beta$, from (\ref{eq:def_zeta}).} %
\label{fig:zeta_vs_Lbar}
\end{figure}

In \cite{Poor97Wang}, similar approximate relationships were derived for the steady state performance of the ALS receiver with exponential windowing for DS-CDMA in flat fading.  The equivalent value of $\zeta$ there is $\zeta' = \frac{1-\epsilon}{2\epsilon}(N-1)$, which converges to $\zeta' \to \frac{1}{2\Leff}+1$ as $N\to\infty$ after substituting $\epsilon = 1 - \frac{1}{N\Leff}$. \Fig \ref{fig:zeta_vs_Lbar} shows a plot of this approximation, which is quite accurate when compared to the exact large-system value at $\beta=1$, particularly for large $\Leff$.

\subsection{Capacity Relationship}

Consider the difference in capacity per data stream\footnote{That is, we are assuming
each data stream is independently coded and decoded. Also, we are assuming that the residual
multi-access interference (MAI) is Gaussian, which one would expect to be valid in the asymptotic
limit considered due to the central limit theorem.} of the MMSE and the ALS receivers, defined as
$\vartriangle\hspace{-1mm}\sC_{\text{ALS}}^k = \log(1 + \SINR^{\text{MMSE}}_k) - \log(1 +
\SINR^{\text{ALS}}_k)$.  We have from (\ref{eq:sinr_als_from_mmse_train}) and
(\ref{eq:sinr_als_from_mmse_blind})
\begin{align}
\label{eq:caploss_blind} %
\vartriangle\hspace{-1mm}\sC_{\text{ALS}}^k &= \ln\(1 + \(1-\frac{1}{\zeta}\)\SINR^{\text{MMSE}}_k\)
\end{align}
with semi-blind training, and
\begin{align}
\label{eq:caploss_train} %
\vartriangle\hspace{-1mm}\sC_{\text{ALS}}^k &= \ln\(1 +
\(\frac{1-\frac{1}{\zeta}}{\frac{\SINR^{\text{MMSE}}_k-1}{\zeta}+1}\)\SINR^{\text{MMSE}}_k\)
\end{align}
with training sequences.

As $\zeta\to 1$, the capacity difference approaches zero, whereas as $\zeta\to\infty$, the difference approaches $\ln\(1 + \SINR^{\text{MMSE}}_k\)$.  Recall that $\zeta$ only depends on $\eta$ (the ratio of training symbols to transmit dimensions), $\beta$ (the ratio of receive to transmit dimensions), and the window shape, and does not depend on the SNR, the (normalized) number of data streams $\alpha$, or the channel distribution. Nor does this value depend on whether $\mS$ is \iid or isometric. \Fig \ref{fig:zeta_vs_Lbar} shows the steady-state value of $\zeta$ vs.\ $\Leff$ with exponential windowing and a range of $\beta$ values.

\section{Numerical Studies}
\label{sec:simulations}

%If the first $\beta^*N$ eigenvalues of $\mH\mH^\dag$ are exponentially distributed.
%\begin{align}
%\sHa_{0,1} &= \ffunca(x) \\
%\sHa_{1,1} &= \frac{1 - \ffunca(x)}{j} \:=\: \sHa_{0,2} \\
%\sHa_{1,2} &= \frac{\ffunca(x)(x+1)-1}{x^2} \\
%\sHa_{2,2} &= \frac{x + 1 - \ffunca(x)(2x+1)}{x^3}
%\end{align}

% XXX new
We now present various applications of the results presented in previous sections.  We shall focus on three example systems:
\begin{itemize}
\item The first is the standard model of a MIMO channel with rich scattering, for which we set $\mH=\mI_N$ and $\mS$ \iid, so that $K$ and $N$ represent the number of transmit and receive antennas, respectively.
\item The second example system is CDMA in frequency-selective Rayleigh fading, for which $\mS$ contains either \iid or isometric signatures, and $\mH$ is a square $N\times N$ matrix (hence $\beta=1$), for which the \aed of the channel correlation matrix $\mH^\ddag$ is exponential with mean one (\ie, $f_H(h) = \exp(-h)$ for $h>0$ is the density used to compute the $\sHa_{m,n}$ values).
\item The third example system is a SISO FIR channel with a cyclic prefix, as described after Proposition \ref{pr:allAWGNbinvar} in Section \ref{sec:limit}, where $\vh = \left[0.227,\: 0.46,\: 0.688,\: 0.46,\: 0.227\right]^\dag$ (\ie, where the ALS and MMSE receiver is used to equalize the so-called Proakis Channel-C \cite[pp.\ 616]{Proakis01}). That is, the empirical results will be obtained using $\mH$ given by the circulant matrix obtained from $\vh$, and $\mS=\mA=\mI_N$. As described in Proposition \ref{pr:allAWGNbinvar}, the analytic results are obtained from the isometric $\mS$ equations with $\alpha=1$, and $H$ distributed according to the spectra of $\vh$, \ie, $F_H(h) = \frac{1}{N}\sum_{n=1}^{N}u(h-\sQ(n))$, where $u(t)$ is the step function, and $\sQ(n) = \abs{\operatorname{DFT}_{N}^n(\vh)}^2$, where $\operatorname{DFT}_{N}^n(\vh)$ denotes the $n\thh$ element of the $N$-point discrete Fourier transform of $\vh$.
\end{itemize}

Unless otherwise stated, we shall assume equal transmit power per data stream (i.e., $\mA=\mI_K$), and SNR$=10$ dB, where SNR is defined as the energy transmitted per data stream in each symbol interval, divided by $\sn$.

In the following plots we determine empirical values from averages of a size $N=32$ system with QPSK modulation for comparison with the large system results. The asymptotic values for the MMSE curves have been determined from Theorem \ref{th:unified_mmse}, and the asymptotic values for the ALS curves have been determined from Theorem \ref{th:sinr_als}, Theorem \ref{th:unified_als}, and Lemma \ref{lem:extra_moments}.  The steady-state values of the ALS receiver have been determined from Corollary \ref{cor:etainf_sinr}.  Where possible, the ALS SINR has been determined from the MMSE SINR using Theorem \ref{th:relationship_mmse_als} (\ie, any situation with \iid training sequences and no diagonal loading).

% XXX defined somewhere?
%\begin{align}
%{\rm SNR} &= A_k^2 \Exp{\vb_m(k)^*\vs_k^\dag \mH^\dag\mH \vs_k \vb_m(k)}/\sn \\
%&\asymp P_k\frac{1}{N}\tr{\mP}/\sn
%\end{align}
%where $P_k = A_k^2$ and $\mP=\mH^\dag\mH$.

\subsection{Transient ALS SINR response and comparison with empirical values}
\subsubsection{MIMO example}
\begin{figure}
    \centering
    \mbox{
    \subfigure[ALS with training]
    {\label{fig:sinr_vs_eta_train_beta1_alpha050_1_2_SNR10dB_Linf_eps01_Peq_idHH}
    \includegraphics[width=\subfigw]{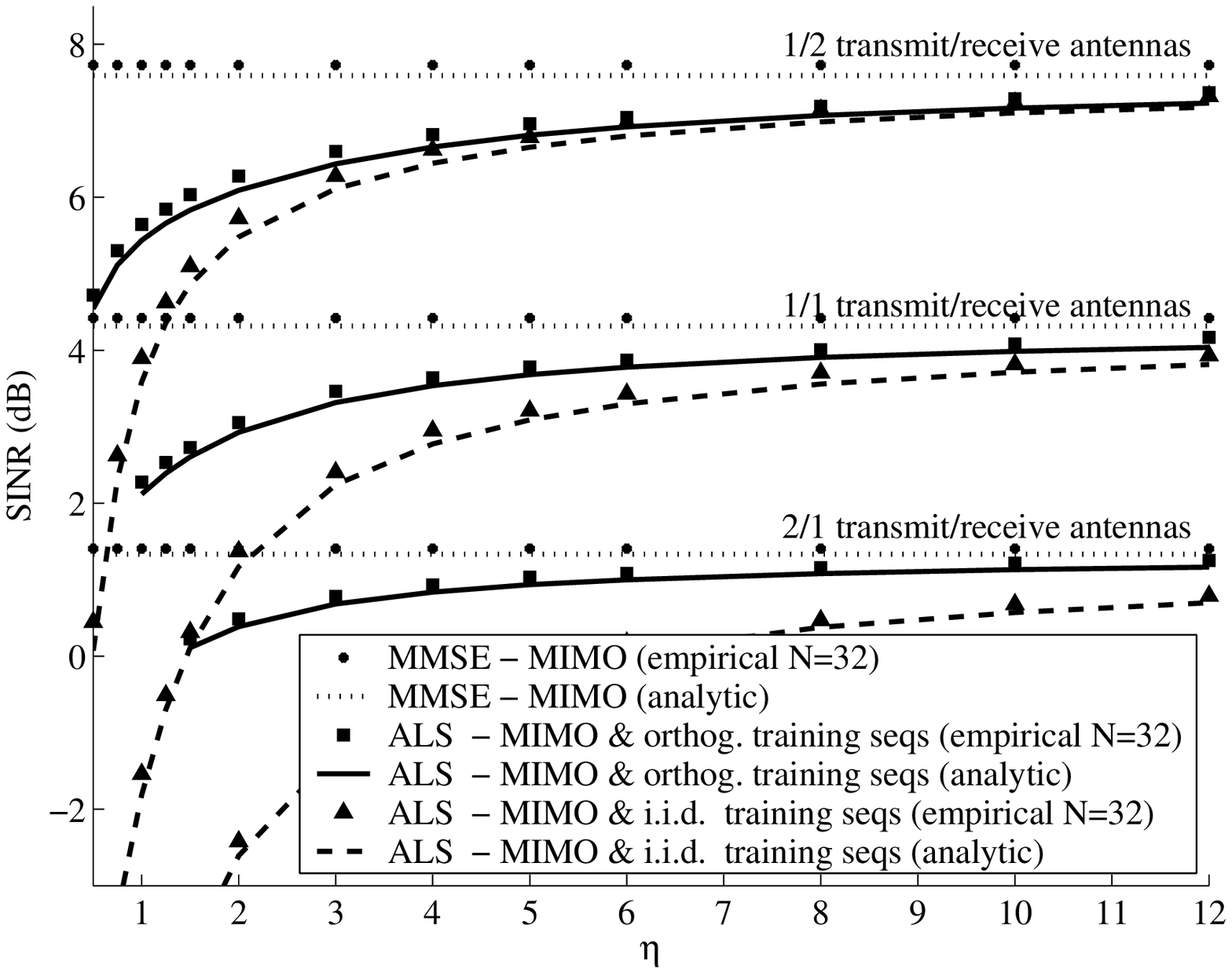}} \quad
    \subfigure[semi-blind ALS]
    {\label{fig:sinr_vs_eta_blind_beta1_alpha050_1_2_SNR10dB_Linf_eps01_Peq_idHH}
    \includegraphics[width=\subfigw]{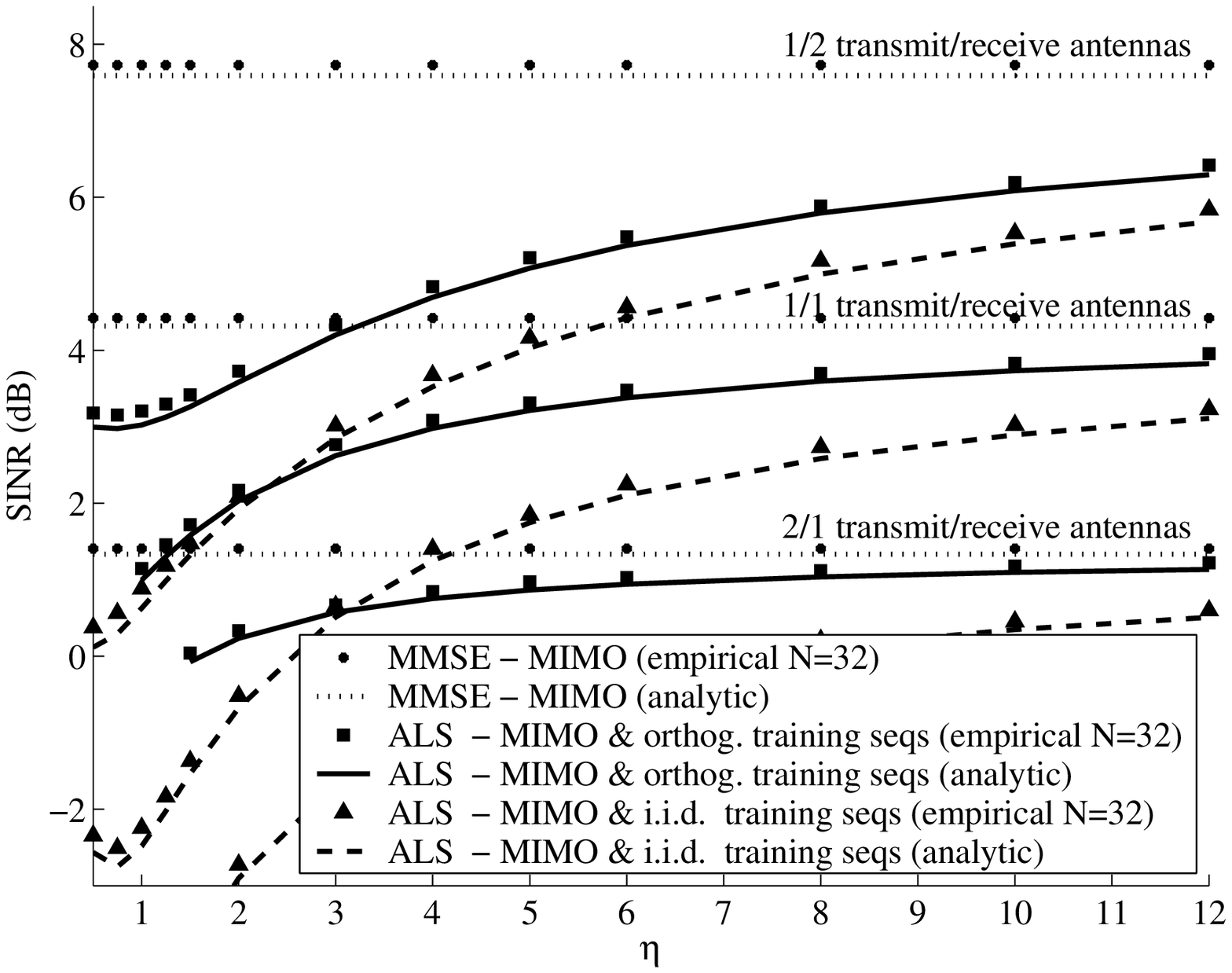}}
    }
    \caption{SINR vs.\ training length: (Rich MIMO) equal transmit power per antenna, SNR$=10$dB, $\mu=0.1$, rectangular windowing. Comparison with empirical values $N=32$, QPSK modulation.}
    \label{fig:sinr_vs_eta_beta1_alpha050_1_2_SNR10dB_Linf_eps01_Peq_idHH}
\end{figure}

Firstly, we demonstrate the relevance of the large-system limit to practical finite systems.  \Fig \ref{fig:sinr_vs_eta_beta1_alpha050_1_2_SNR10dB_Linf_eps01_Peq_idHH} shows both asymptotic and empirical values of MMSE and ALS SINR vs.\ training length for the example MIMO system with rich scattering. For the ALS receiver, the diagonal loading value is $\mu=0.1$, and rectangular windowing is used. Clearly, the empirical (finite) values match the analytic (asymptotic) values very closely.

Note that the orthogonal training sequences clearly outperform the \iid training sequences, particularly for `small' $\eta$.  This gap also widens as the number of receive dimensions decreases.  Also, the performance of the semi-blind ALS receiver is comparable to the performance of the ALS receiver with training for the 2 to 1 transmit to receive antennas ratio case, but is significantly worse in the 1 to 2 transmit to receive antennas ratio case.

\subsubsection{CDMA in frequency-selective fading}

\begin{figure}
    \centering
    \mbox{
    \subfigure[ALS with training]
    {\label{fig:sinr_vs_eta_train_beta1_alpha050_SNR10dB_Linf_eps01_Peq_expHH}
    \includegraphics[width=\subfigw]{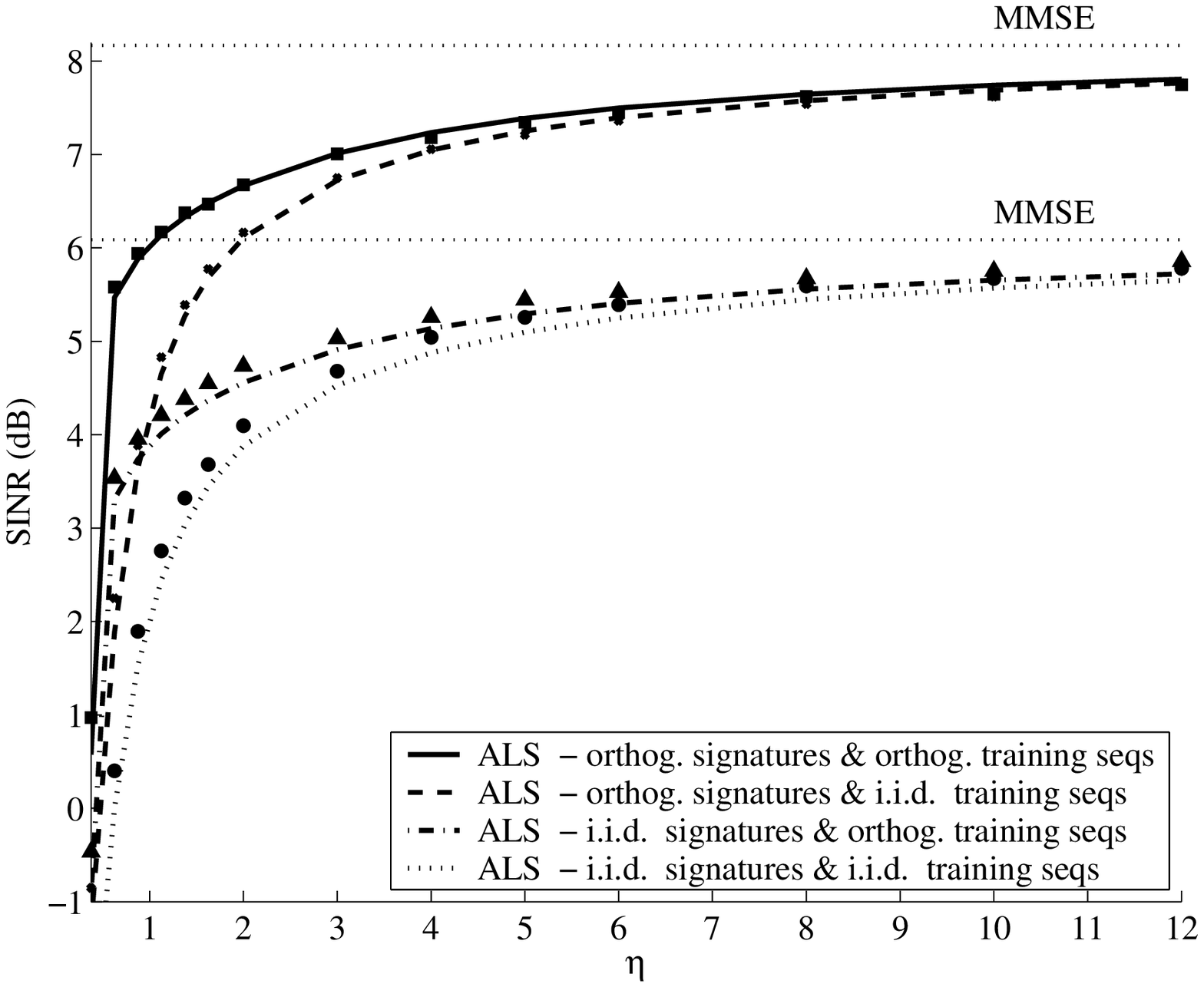}} \quad
    \subfigure[semi-blind ALS]
    {\label{fig:sinr_vs_eta_blind_beta1_alpha050_SNR10dB_Linf_eps01_Peq_expHH}
    \includegraphics[width=\subfigw]{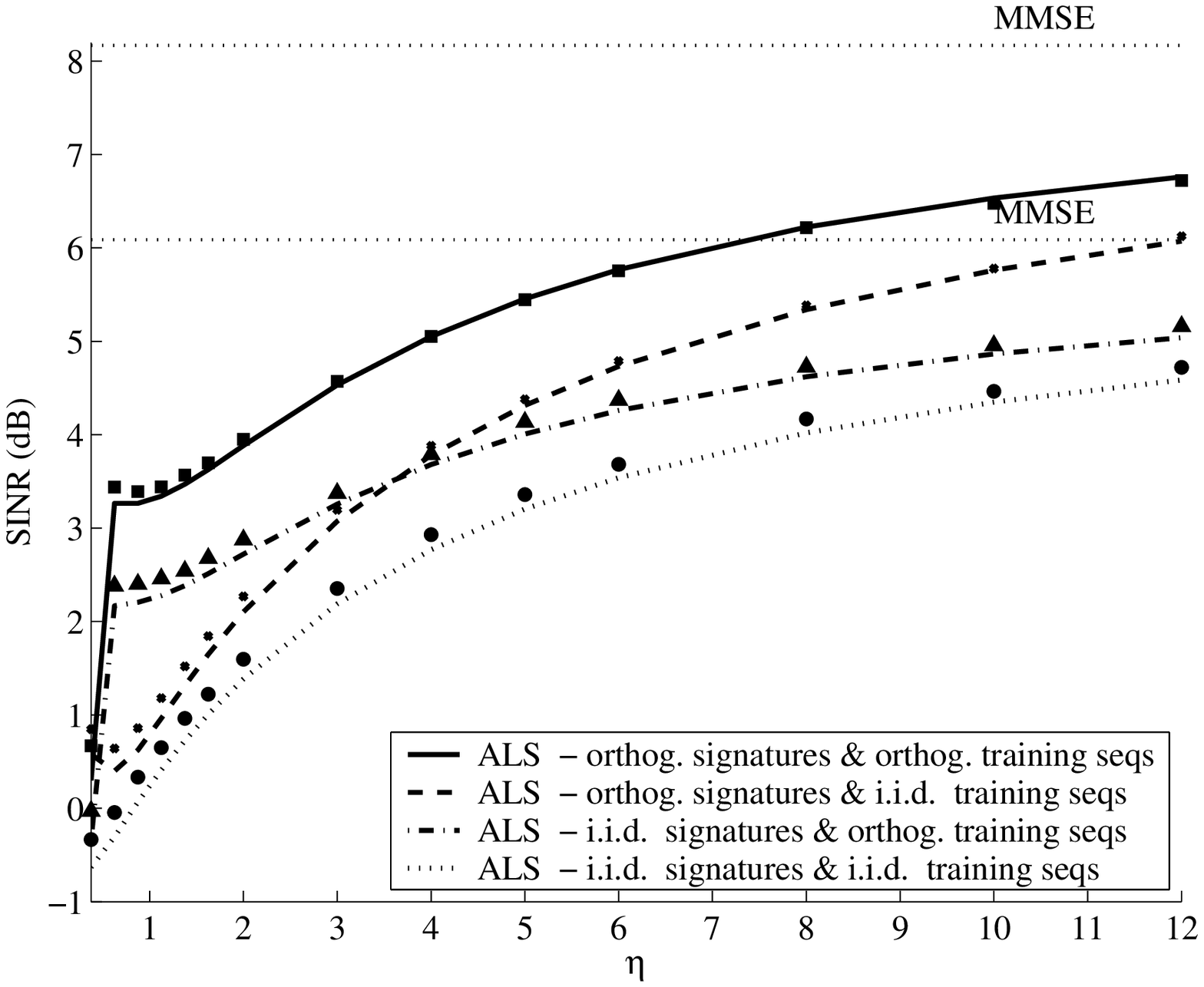}}
    }
    \caption{SINR vs.\ training length: CDMA in frequency-selective fading, SNR=10dB, $\alpha=0.50$, $\mu=0.1$, rectangular windowing, equal power per signature, exponential distribution for \aed of $\mH^\ddag$. Comparison with empirical values $N=32$, QPSK modulation.}
    \label{fig:sinr_vs_eta_beta1_alpha050_SNR10dB_Linf_eps01_Peq_expHH}
\end{figure}

\Fig \ref{fig:sinr_vs_eta_beta1_alpha050_SNR10dB_Linf_eps01_Peq_expHH} shows empirical and asymptotic values of MMSE and ALS SINR vs.\ $\eta$ for the example CDMA system in frequency-selective Rayleigh fading with $\alpha=0.50$. The ALS receiver uses rectangular windowing and a diagonal loading constant $\mu=0.1$.  Curves are shown for both \iid and isometric signatures, and \iid and orthogonal training sequences.  Again, the empirical (finite) values match the analytic (asymptotic) values.

Figure \ref{fig:sinr_vs_eta_blind_beta1_alpha050_SNR10dB_Linf_eps01_Peq_expHH} shows the intuitively pleasing result that for a small number of training symbols (i.e., small $\eta$), orthogonal training sequences improve the performance of the ALS receiver more than isometric signatures, and as $\eta$ increases, this situation is quickly reversed. This is due to the fact that the $K$ \iid training sequences of length $i$ become `more orthogonal' as $i$ increases, and also since isometric signatures consistently outperform \iid signatures.

In subsequent plots, we shall omit the empirical values, and concentrate on applications of the analytical results.

% Thesis.
%\begin{figure}
%    \centering
%    \mbox{
%    \subfigure[ALS with training]
%    {\label{fig:sinr_vs_eta_train_beta1_alpha075_SNR10dB_Linf_eps01_Peq_expHH}
%    \includegraphics[width=\subfigw]{sinr_vs_eta_train_beta1_alpha075_SNR10dB_Linf_eps01_Peq_expHH}} \quad
%    \subfigure[semi-blind ALS]
%    {\label{fig:sinr_vs_eta_blind_beta1_alpha075_SNR10dB_Linf_eps01_Peq_expHH}
%    \includegraphics[width=\subfigw]{sinr_vs_eta_blind_beta1_alpha075_SNR10dB_Linf_eps01_Peq_expHH}}
%    }
%    \caption{SINR vs.\ training length: same parameters as \Fig \ref{fig:sinr_vs_eta_beta1_alpha050_SNR10dB_Linf_eps01_Peq_expHH}, however with $\alpha=0.75$.}
%    \label{fig:sinr_vs_eta_beta1_alpha075_SNR10dB_Linf_eps01_Peq_expHH}
%\end{figure}

% Thesis.  This doubles up on the graph below.
%\begin{figure}
%    \centering
%    \mbox{
%    \subfigure[ALS with training]
%    {\label{fig:sinr_vs_Lbar_train_beta1_alpha075_SNR10dB_L1-20_eps0_P0_075_-3_025_expHH_iidS_iidB}
%\includegraphics[width=\subfigw]{sinr_vs_Lbar_train_beta1_alpha075_SNR10dB_L1-20_eps0_P0_075_-3_025_expHH_iidS_iidB}} \quad
%    \subfigure[semi-blind ALS]
%    {\label{fig:sinr_vs_Lbar_blind_beta1_alpha075_SNR10dB_L1-20_eps0_P0_075_-3_025_expHH_iidS_iidB}
%    \includegraphics[width=\subfigw]{sinr_vs_Lbar_blind_beta1_alpha075_SNR10dB_L1-20_eps0_P0_075_-3_025_expHH_iidS_iidB}}
%    }
%    \caption{}
%    \label{fig:sinr_vs_eta_beta1_alpha075_SNR10dB_Linf_eps01_Peq_expHH}
%\end{figure}

% XXX new
\subsubsection{Equalization}

\begin{figure}
    \centering
%    \mbox{
%    \subfigure[ALS with training]
%    {
%    \label{fig:sinr_vs_eta_SNR20dB_L5-10-inf_Kpsk2_eps01_FIR_train}
    \includegraphics[width=\subfigw]{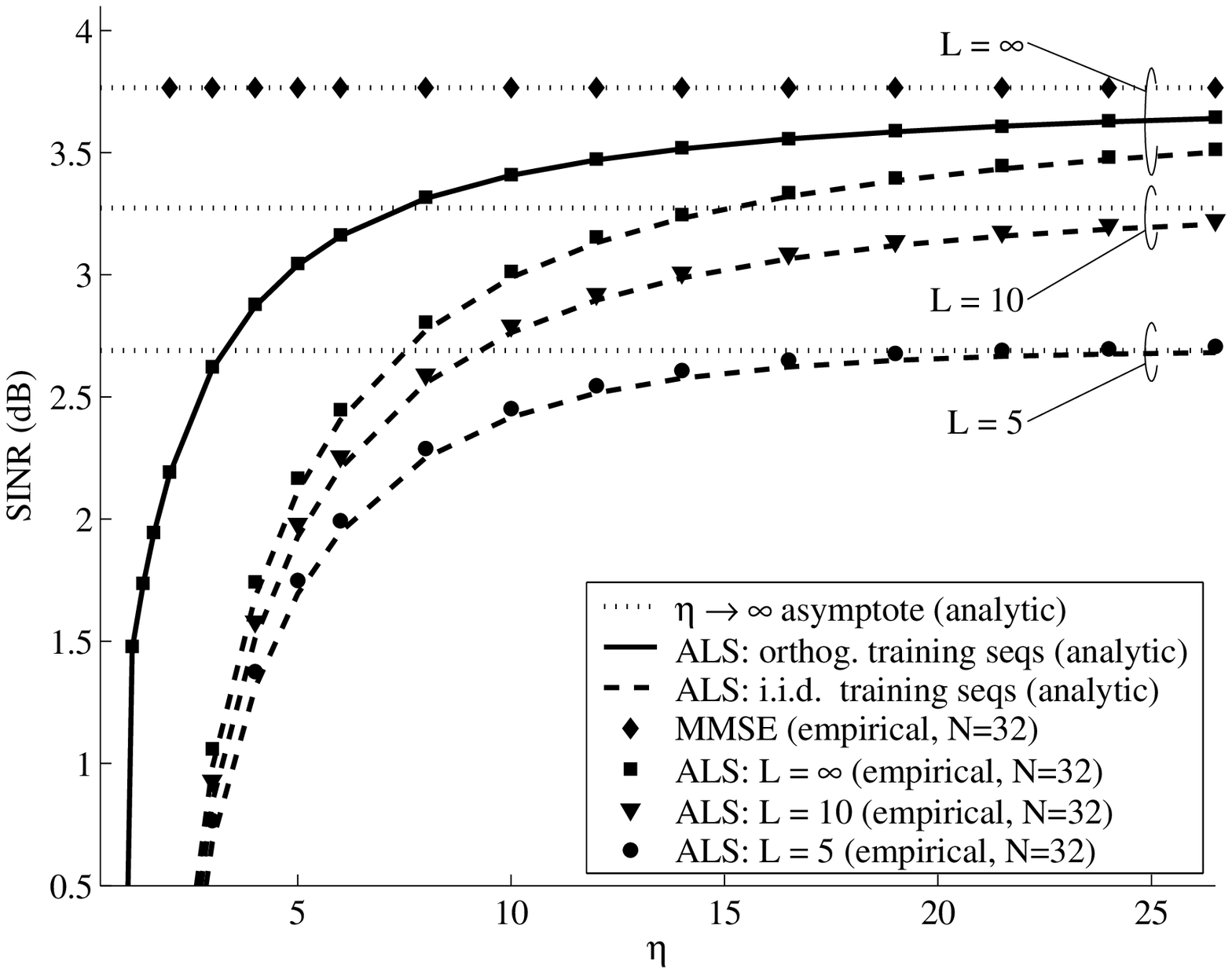}
%    } \quad
%    \subfigure[semi-blind ALS]
%    {\label{fig:sinr_vs_eta_SNR20dB_L5-10-inf_Kpsk2_eps01_FIR_blind}
%    \includegraphics[width=\subfigw]{sinr_vs_eta_SNR20dB_L5-10-inf_Kpsk2_eps01_FIR_blind}}
%    }
    \caption{SINR vs.\ training length: Equalization of Proakis C-channel (\ie, $\mH$ circulant matrix constructed from impulse response, and $\mS=\mA=\mI_N$) at SNR=20dB. The ALS receiver uses $\mu=0.1$, exponential windowing, with training using \iid and orthogonal training sequences. Comparison with empirical values $N=32$, QPSK modulation. Analytic values are obtained using isometric $\mS$ equations with $\alpha=\beta=1$, as specified by Proposition \ref{pr:allAWGNbinvar}.}
    \label{fig:sinr_vs_eta_SNR20dB_L5-10-inf_Kpsk2_eps01_FIR}
\end{figure}

\Fig \ref{fig:sinr_vs_eta_SNR20dB_L5-10-inf_Kpsk2_eps01_FIR} shows empirical and asymptotic values of MMSE and ALS SINR vs.\ $\eta$ for the example SISO FIR system at 20dB SNR. The ALS receiver uses exponential windowing and a diagonal loading constant $\mu=0.1$.  Curves are shown for both \iid and orthogonal training sequences.  Note that Proposition \ref{pr:allAWGNbinvar} requires that $\vb_m$ is unitarily invariant, whereas the empirical values in the figure are based on standard QPSK modulation (\ie, $\vb_m$ is not unitarily invariant). Clearly, at least in this case, the asymptotic results are a very good approximation for non-unitarily invariant data vectors.

\subsection{Capacity with exponential windowing}

\begin{figure}
    \centering
    \mbox{
    \subfigure[ALS with training]
    {\label{fig:Closs_vs_Lbar_train_beta1_alpha075_SNR10dB_L1-20_eps0_P0_075_-3_025_expHH_iidS_iidB}
\includegraphics[width=\subfigw]{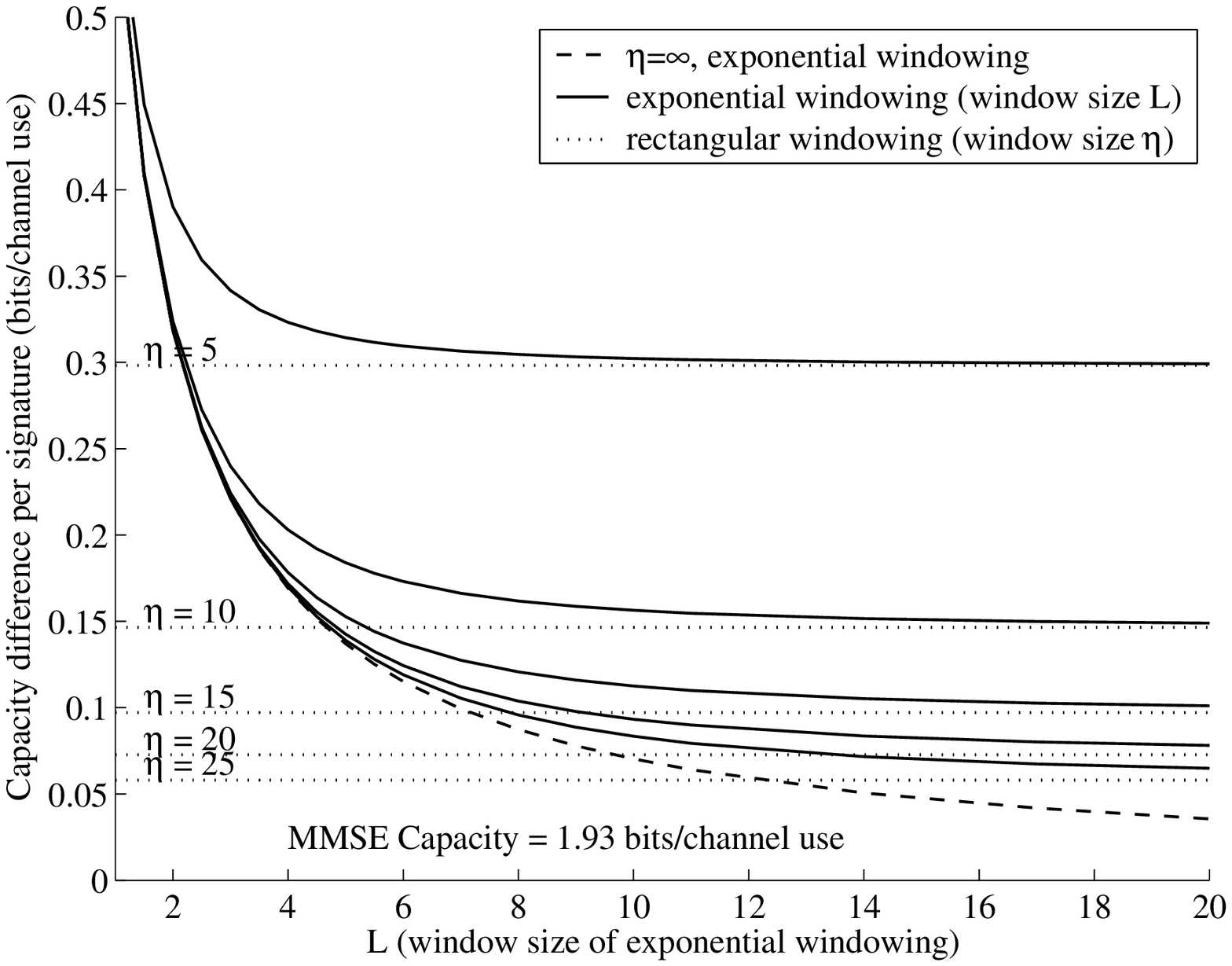}} \quad
    \subfigure[semi-blind ALS]
    {\label{fig:Closs_vs_Lbar_blind_beta1_alpha075_SNR10dB_L1-20_eps0_P0_075_-3_025_expHH_iidS_iidB}
    \includegraphics[width=\subfigw]{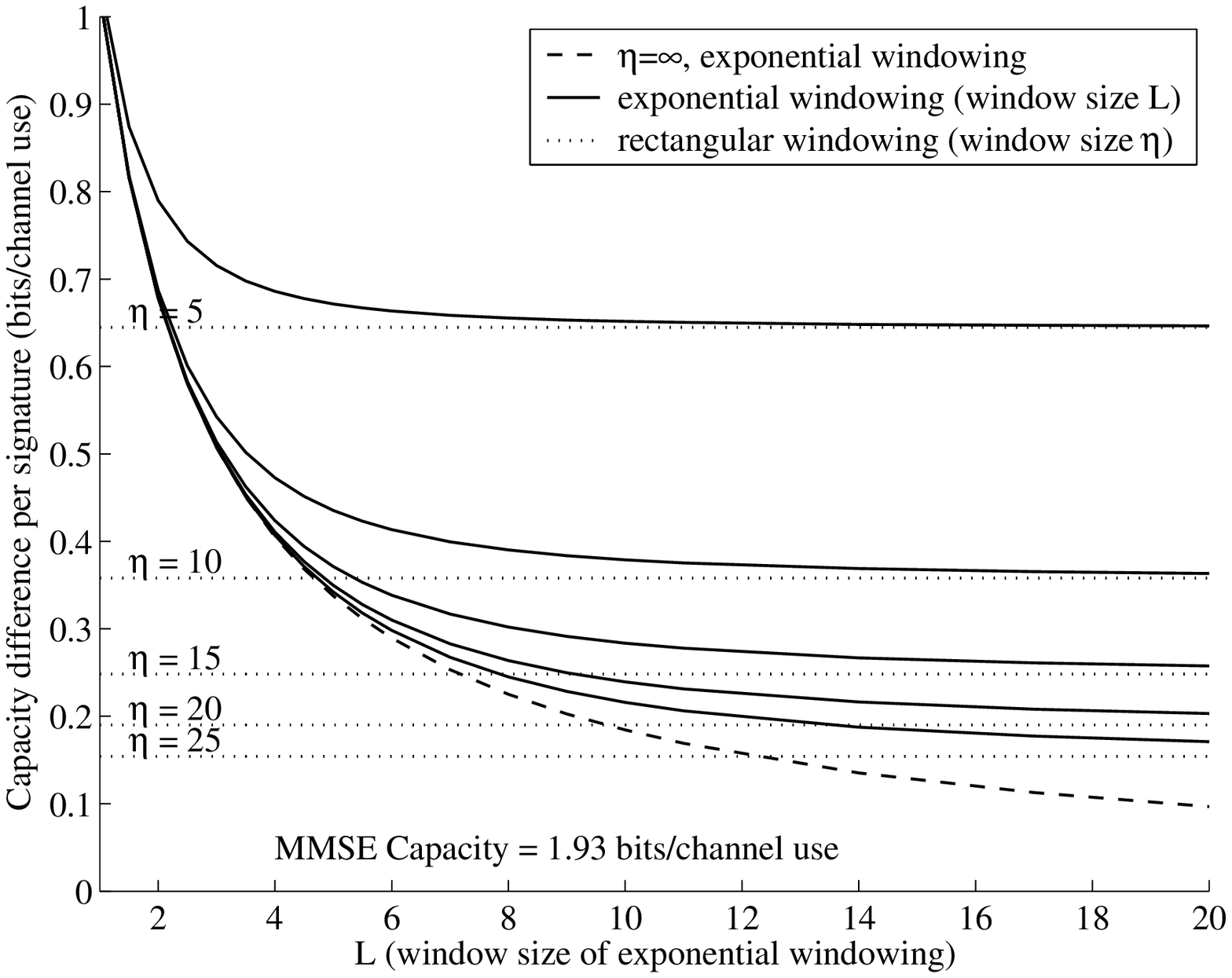}}
    }
    \caption{Capacity difference between ALS and MMSE receivers vs.\ window size of exponential window: CDMA in frequency-selective fading, \iid signatures, \iid training, SNR=10dB, $\alpha=0.75$, $\mu=0$, rectangular and exponential windowing. Note the window size for rectangular windowing is $\eta$.  Also, $f_{P}(p) = \frac{3}{4}\delta(p-1) + \frac{1}{4}\delta(p-\frac{1}{2})$. Curves shown correspond to the first $3/4$ of the signatures. Note that the scale of the vertical axis of Fig.\ \ref{fig:Closs_vs_Lbar_blind_beta1_alpha075_SNR10dB_L1-20_eps0_P0_075_-3_025_expHH_iidS_iidB} is twice that of Fig.\ \ref{fig:Closs_vs_Lbar_train_beta1_alpha075_SNR10dB_L1-20_eps0_P0_075_-3_025_expHH_iidS_iidB}. }
    \label{fig:Closs_vs_Lbar_beta1_alpha075_SNR10dB_L1-20_eps0_P0_075_-3_025_expHH_iidS_iidB}
\end{figure}

Now we examine the performance of the ALS receiver with both rectangular and exponential windowing, relative to the MMSE receiver.  \Fig \ref{fig:Closs_vs_Lbar_beta1_alpha075_SNR10dB_L1-20_eps0_P0_075_-3_025_expHH_iidS_iidB} shows the capacity difference per-signature as a function of the window size $\Leff$ (determined from (\ref{eq:caploss_blind}) and (\ref{eq:caploss_train})) for the example CDMA system in frequency-selective Rayleigh fading with \iid signatures, \iid training, and a system load of $\alpha=0.75$. Curves for the ALS receiver are shown with both rectangular and exponential windowing, and diagonal loading constant $\mu=0$. Also, $f_{P}(p) = \frac{3}{4}\delta(p-1) + \frac{1}{4}\delta(p-\frac{1}{2})$, that is, one quarter of the signatures are transmitted at half the power of the remaining signatures.

In this figure, we do not take into account the loss in rate due to the training.  This is considered in the following subsection.  Rather, for a single channel use at a certain SNR, we wish to see the relative capacity difference between the MMSE receiver (using full CSI), and that obtained by the ALS receiver as a function of the number of training symbols used to generate the filter. Also, for the ALS receiver, we wish to compare exponential windowing with rectangular windowing at a given value of $\eta$ as a function of the exponential windowing window size, $\Leff$.

Firstly, we see that for either type of windowing, increasing the number of training symbols is an exercise in diminishing returns.  Also, we see that as the window size increases, exponential windowing asymptotes to rectangular windowing, as would be expected for the time-invariant system model (\ref{eq:rcv_sig}).  Of course, exponential windowing is included to allow for time-varying channels. As such, the curves for exponential windowing are a valid approximation for a time-varying system in which the coherence time of the system\footnote{`Coherence time' here refers to the number of symbols over which $\mH\mS\mA$ and $\sn$ remain approximately constant.} is at least as large as the effective window size created by the exponential windowing. As such, the values of capacity or SINR obtained represent \emph{best possible} values, which are only attained if the system remains static for the duration of the ALS training period. Extending these results to time-varying systems is an open problem, and is likely to be difficult.

\subsection{Application: Throughput Optimization}

We now demonstrate how the results can be used to optimize the throughput with packet transmissions. More training symbols gives a higher ALS SINR, but leaves less room for data-carrying symbols in the packet. Clearly, there is an optimal ratio of training symbols to data-carrying symbols. Such an optimization has been considered for MIMO block fading channels and SISO FIR channels in \cite{Hassibi03Hochwald,VikaloHassibi} with an optimal (maximum-likelihood) receiver. In that work, the training symbols are used to estimate the channel directly. A lower bound on the capacity is derived, and is used to optimize the training length. Related work in \cite{Sun03} applies the large-system transient analysis in \cite{Honig02Xiao} for the MIMO \iid channel to optimize the training length with an ALS receiver (without exponential windowing or diagonal loading). It is shown there that for large packet lengths ($\ell$) the training length that maximizes capacity grows as $\Order(\sqrt{\ell})$. Optimization of power levels between the training and data symbols is also investigated.

Suppose we consider a packet containing $T>i$ symbols, of which the first $i$ are training symbols, and the remainder consists of data-carrying symbols. There are $K$ equal power data streams, which are coded independently with capacity-achieving\footnote{Here we assume that the residual interference at the receiver output is \iid circularly symmetric complex Gaussian.} codes with rate $R_c = \log_2(1+\SINR^{\text{ALS}})$.  We focus on the ALS receiver with known training symbols.  The number of information bits per block is therefore $K R_c (T-i)$, while the number of transmit dimensions per block is $N T$.  Therefore, the number of information bits per transmit dimension (hereafter referred to as `normalized capacity') is $C = \alpha \Reff$, where $\Reff = R_c(1-\eta/\ell)$ and $\ell = T/N$.  We shall consider the additional limit $T\to\infty$ with $T/N\to\ell>0$ in order to optimize $C$ with respect to the normalized training length $\eta$. We shall keep $E_b/\sn = \SNR/\Reff$ constant, and unless otherwise stated, in the numerical examples $E_b/\sn$ = 10 dB.

%We define $\SNR$ as the energy transmitted per data stream in each symbol interval, divided by $\sn$,

%% Thesis: full version of below paragraph.
%\Fig \ref{fig:capacity_vs_eta_and_alpha_train_beta1_EbN0_10dB_Linf_eps0_Peq_expHH_iidB} shows the normalized capacity of the example CDMA system in frequency-selective fading as a function of $\eta$ and $\alpha$ for a normalized block length of $\ell = T/N = 15$. The ALS receiver uses rectangular windowing and no diagonal loading. \Fig \ref{fig:capacity_gain_vs_eta_and_alpha_train_beta1_EbN0_10dB_Linf_eps0_Peq_expHH} shows the additional normalized capacity obtained, relative to the results for \iid training in \Fig \ref{fig:capacity_vs_eta_and_alpha_train_beta1_EbN0_10dB_Linf_eps0_Peq_expHH_iidB}, if orthogonal training sequences are used.

\begin{figure}
    \centering
    \mbox{
    \subfigure[\iid training signatures]
    {\label{fig:capacity_vs_eta_and_alpha_train_beta1_EbN0_10dB_Linf_eps0_Peq_expHH_iidS_iidB}
    \includegraphics[width=\subfigw]{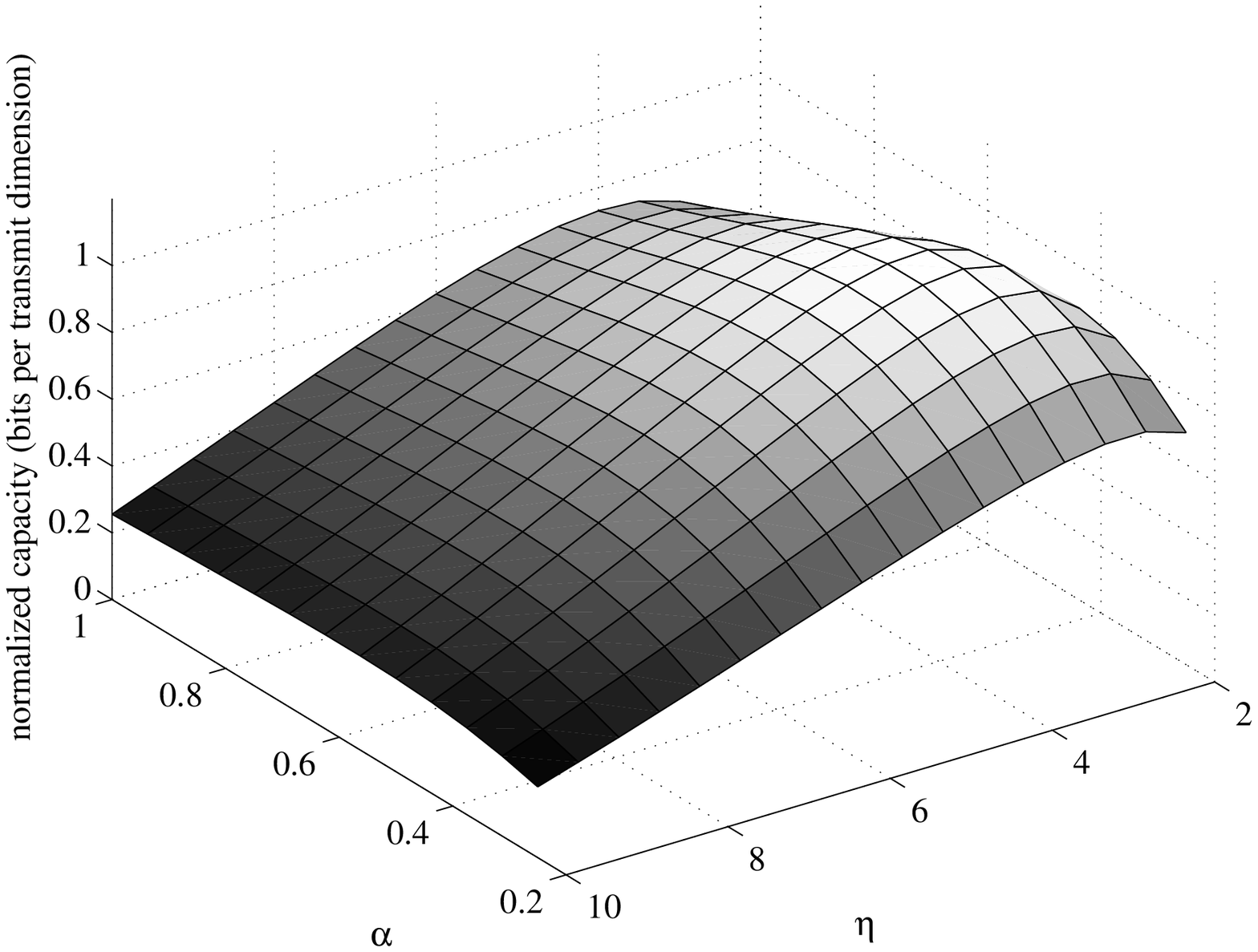}}
    \quad
    \subfigure[Additional normalized capacity obtained using orthogonal training signatures]
    {\label{fig:capacity_gain_vs_eta_and_alpha_train_beta1_EbN0_10dB_Linf_eps0_Peq_expHH_iidS}
    \includegraphics[width=\subfigw]{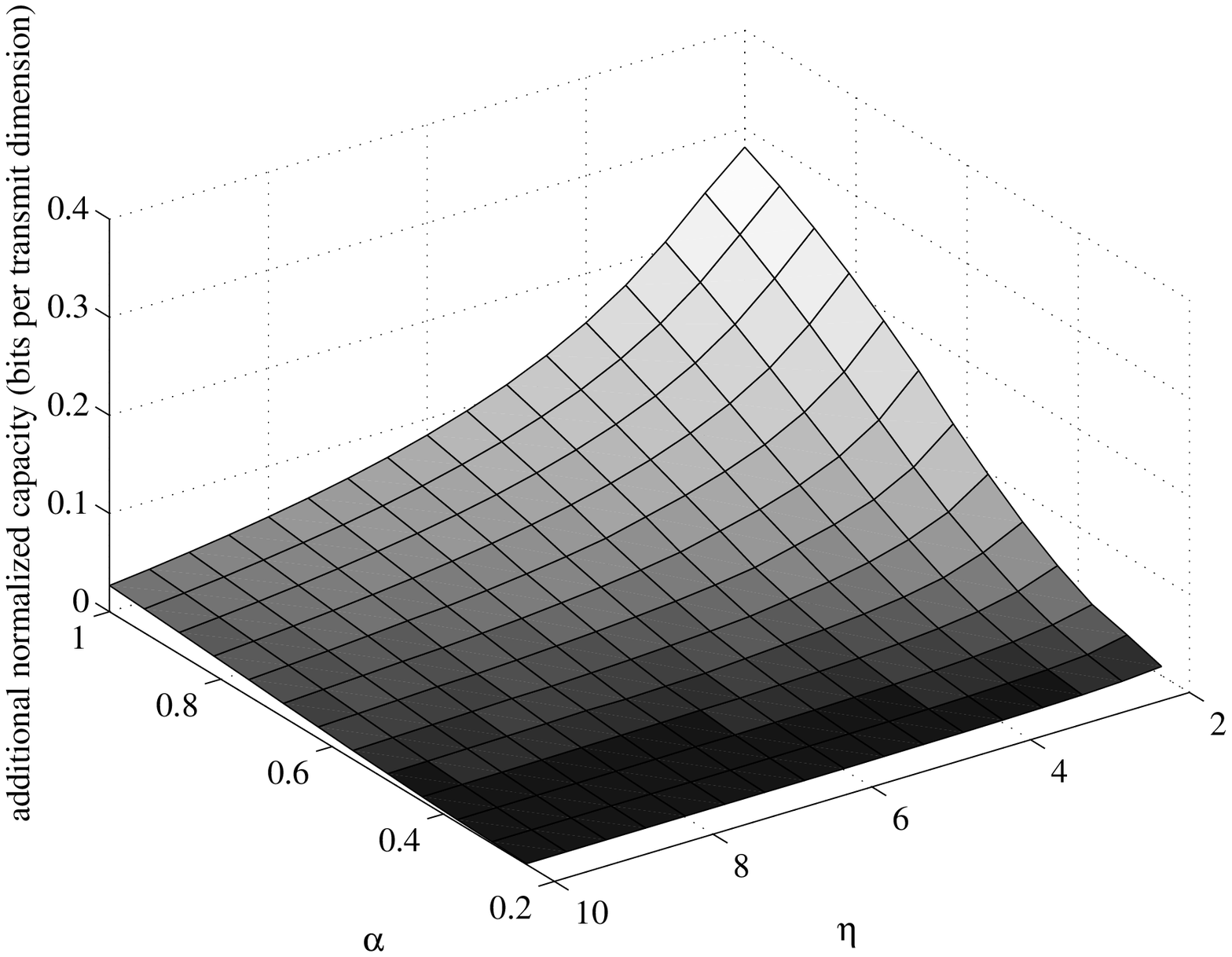}}
    }
    \caption{Throughput optimization: CDMA in frequency-selective fading, \iid $\mS$, normalized block length $T/N=15$, rectangular windowing, no diagonal loading, $\frac{E_b}{\sn}$=10 dB. \Fig \ref{fig:capacity_gain_vs_eta_and_alpha_train_beta1_EbN0_10dB_Linf_eps0_Peq_expHH_iidS} shows the additional normalized capacity (with respect to \Fig \ref{fig:capacity_vs_eta_and_alpha_train_beta1_EbN0_10dB_Linf_eps0_Peq_expHH_iidS_iidB}) obtained if orthogonal training sequences are used.}
    \label{fig:capacity_vs_eta_and_alpha_train_beta1_EbN0_10dB_Linf_eps0_Peq_expHH_iidS}
\end{figure}

\Fig \ref{fig:capacity_vs_eta_and_alpha_train_beta1_EbN0_10dB_Linf_eps0_Peq_expHH_iidS_iidB} shows the normalized capacity of the example CDMA system in frequency-selective fading as a function of $\eta$ and $\alpha$ for a normalized block length of $\ell = T/N = 15$. The ALS receiver uses rectangular windowing and no diagonal loading. \Fig \ref{fig:capacity_gain_vs_eta_and_alpha_train_beta1_EbN0_10dB_Linf_eps0_Peq_expHH_iidS} shows the additional normalized capacity obtained, relative to the results for \iid training in \Fig \ref{fig:capacity_vs_eta_and_alpha_train_beta1_EbN0_10dB_Linf_eps0_Peq_expHH_iidS_iidB}, if orthogonal training sequences are used. Although not shown, plots analogous to \Fig \ref{fig:capacity_vs_eta_and_alpha_train_beta1_EbN0_10dB_Linf_eps0_Peq_expHH_iidS} may also be obtained for isometric $\mS$. % XXX new.

\begin{figure}
    \centering
    \mbox{
    \subfigure[Optimal Throughput]
    {\label{fig:Copt_Ttrain15_expHH_beta1_EbN0_10_eqPw_Lbar_inf}
\includegraphics[width=\subfigw]{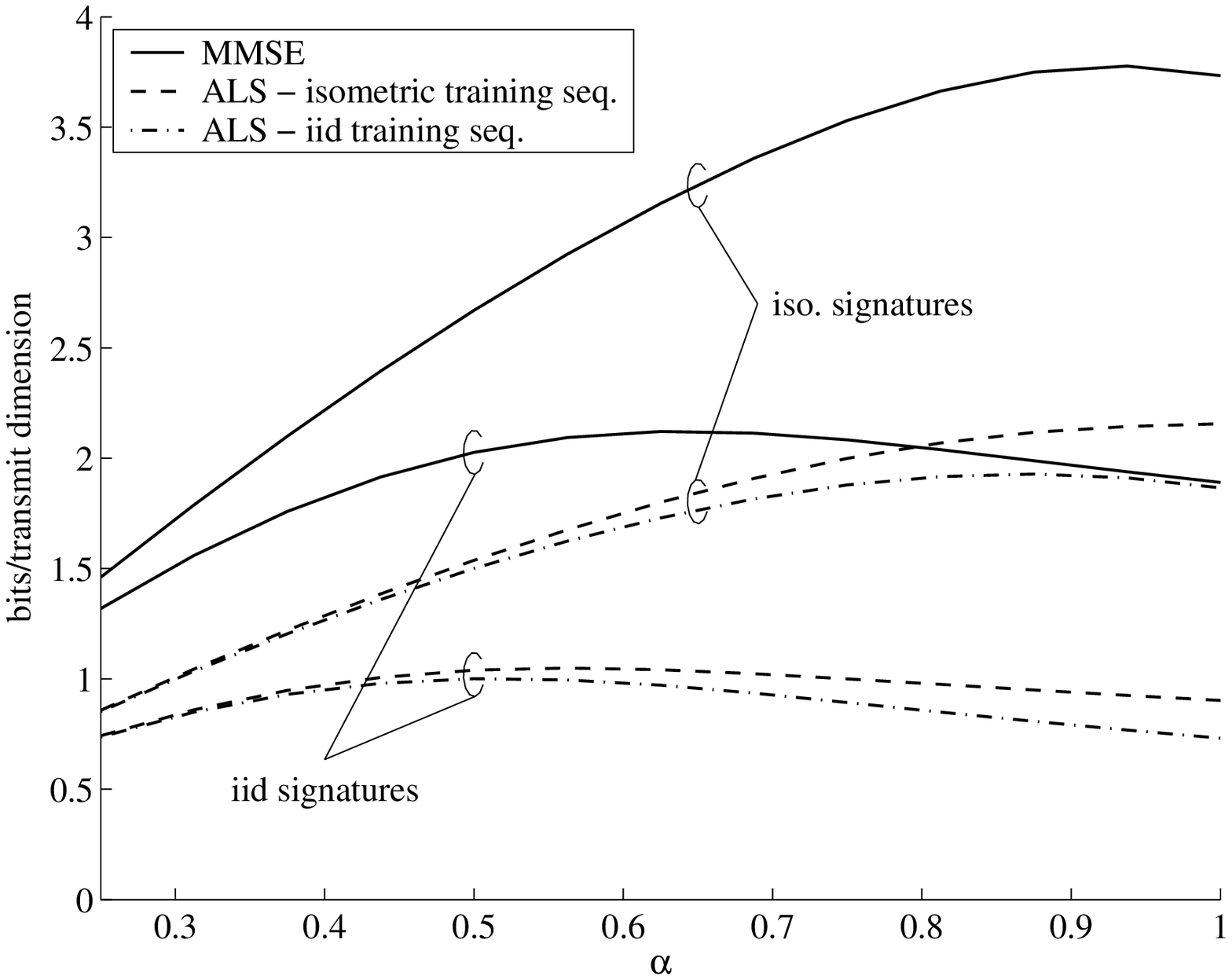}} \quad
    \subfigure[Optimal Training Length]
    {\label{fig:etaopt_Ttrain15_expHH_beta1_EbN0_10_eqPw_Lbar_inf}
    \includegraphics[width=\subfigw]{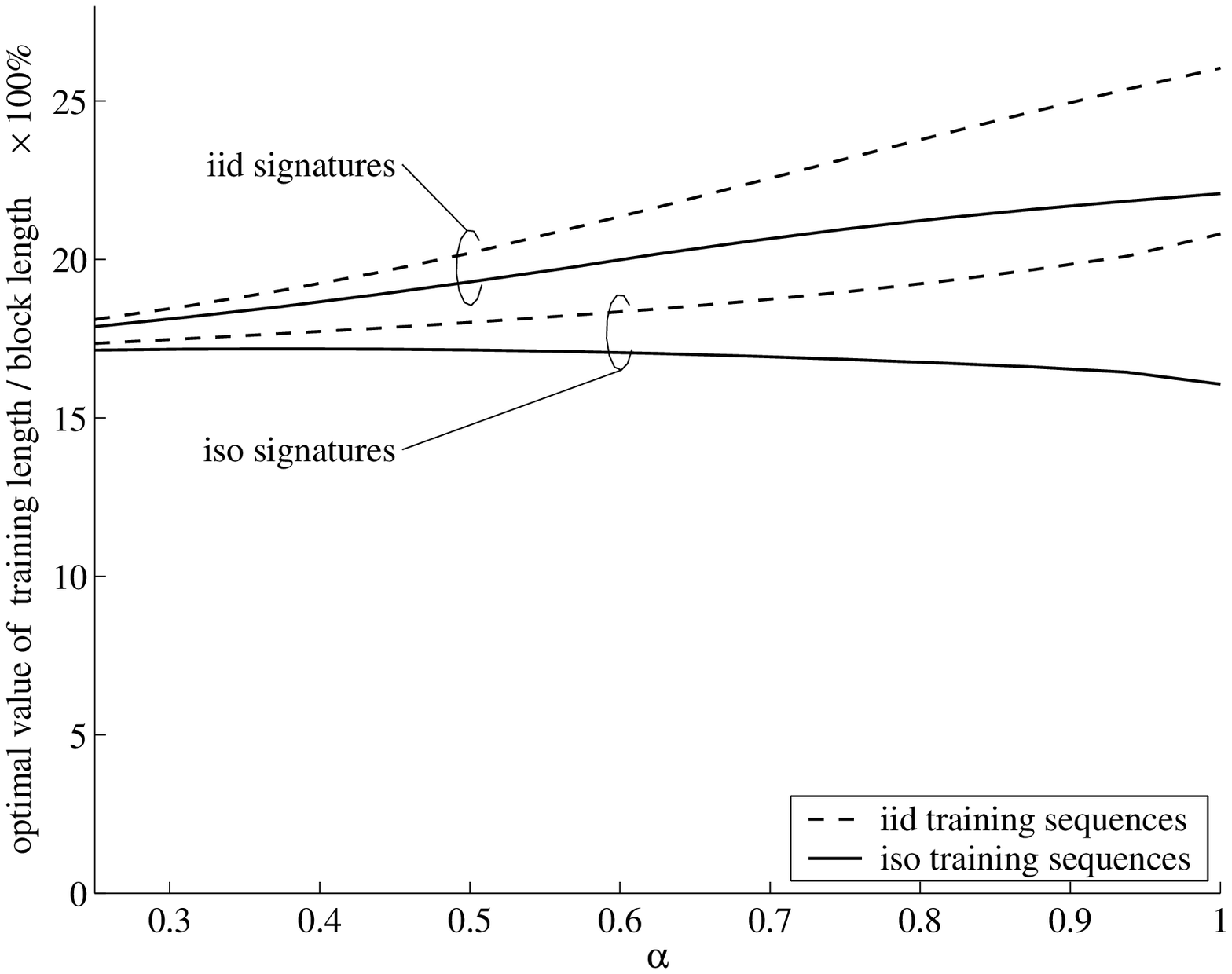}}
    }
    \caption{Throughput optimization: CDMA in frequency-selective fading, normalized block length $T/N=15$, rectangular windowing, no diagonal loading, $\frac{E_b}{\sn}$=10 dB, equal power per signature, exponential distribution for \aed of $\mH^\ddag$.}
    \label{fig:opt_Ttrain15_expHH_beta1_EbN0_10_eqPw_Lbar_inf}
\end{figure}

\Fig \ref{fig:capacity_vs_eta_and_alpha_train_beta1_EbN0_10dB_Linf_eps0_Peq_expHH_iidS_iidB} shows that there is an optimum value of $\eta/\ell$, i.e., the ratio of training length to block length, for each value of system load, $\alpha$. \Fig \ref{fig:Copt_Ttrain15_expHH_beta1_EbN0_10_eqPw_Lbar_inf} shows the value of normalized capacity at the optimum value of $\eta$, again for $\ell = 15$, as a function of the system load $\alpha$. \Fig \ref{fig:etaopt_Ttrain15_expHH_beta1_EbN0_10_eqPw_Lbar_inf} shows the corresponding value of $\eta$ (expressed as a percentage of $\ell$) which maximizes the normalized capacity.  Also shown in \Fig \ref{fig:Copt_Ttrain15_expHH_beta1_EbN0_10_eqPw_Lbar_inf} is the normalized capacity of the MMSE receiver at the same value of $E_b/\sn$, for both types of signatures.  Of course, the MMSE receiver assumes perfect CSI.

%Of course, this comparison is a little unfair as whatever method the CSI used by the MMSE receiver is determined, it would significantly lower the MMSE normalized capacity, not to mention the additional associated loss due to imperfect CSI.

\begin{figure}
    \centering
    \mbox{
    \subfigure[Throughput]
    {\label{fig:cap_vs_Ttrain_train_beta1_alpha4_2_1_05_025_EbN0_10dB_Linf_eps0_Peq_idHH_iidS}
\includegraphics[width=\subfigw]{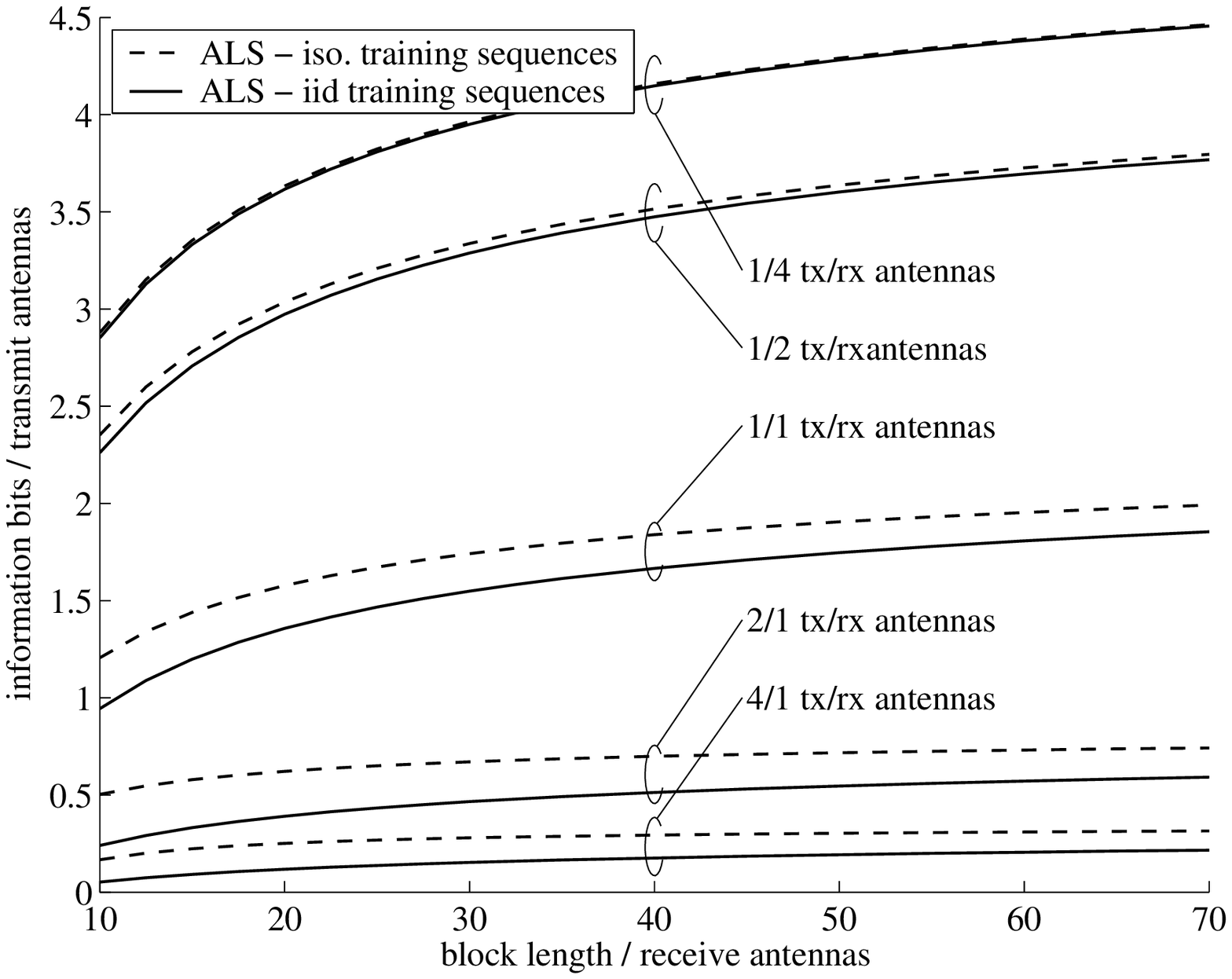}} \quad
    \subfigure[Optimal percentage of training in a block (\iid training)]
    {\label{fig:opteta_vs_Ttrain_train_beta1_alpha4_2_1_05_025_EbN0_10dB_Linf_eps0_Peq_idHH_iidS_iidB}
    \includegraphics[width=\subfigw]{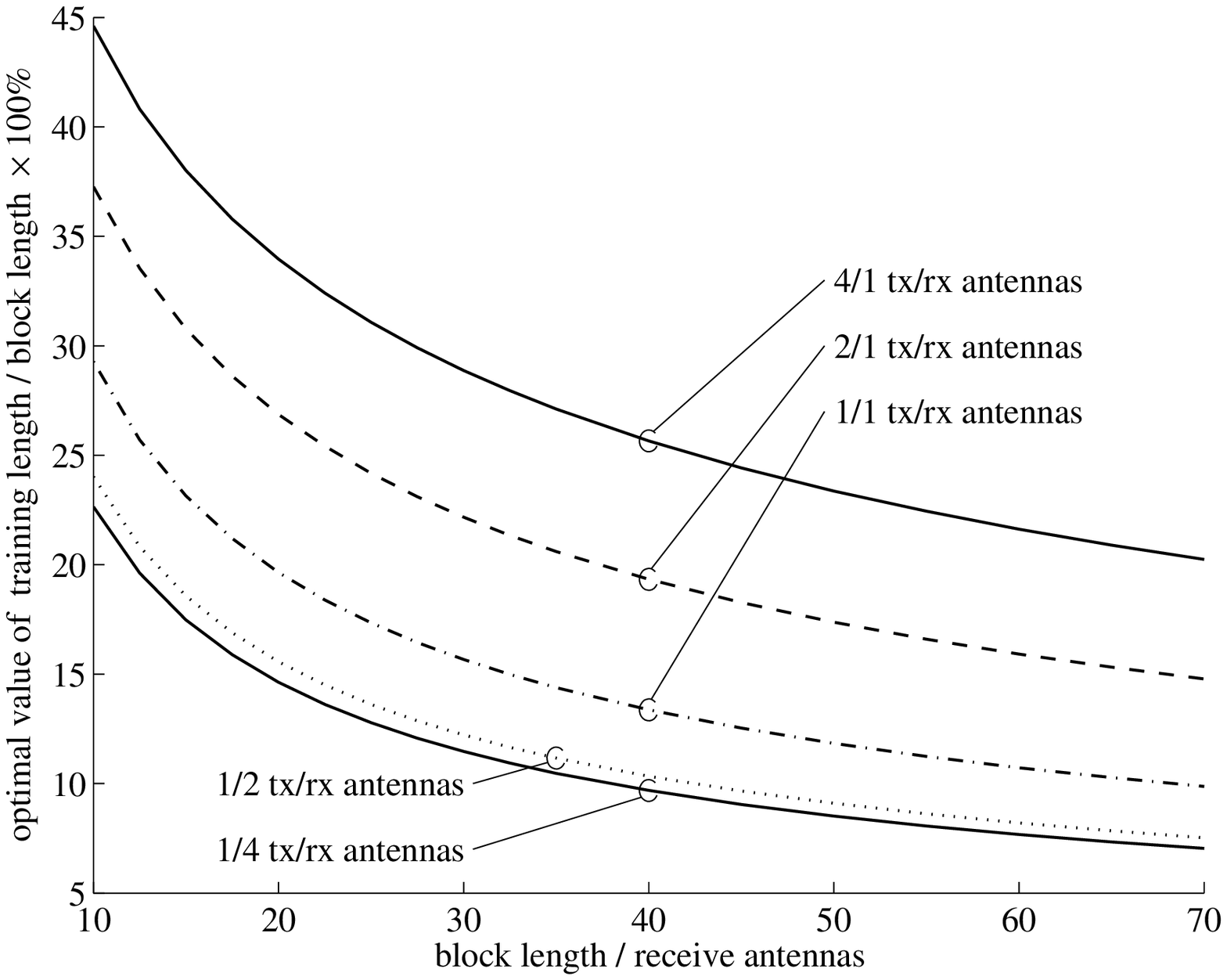}}
    }
    \caption{Throughput optimization: MIMO channel, rectangular windowing, no diagonal loading, $\frac{E_b}{\sn}$=10 dB, equal transmit power per antenna.}
    \label{fig:cap_opteta_vs_Ttrain_train_beta1_alpha4_2_1_05_025_EbN0_10dB_Linf_eps0_Peq_idHH_iidS}
\end{figure}

We now consider throughput optimization for the example MIMO system, and consider the growth in normalized capacity with respect to the normalized block length, $\ell$. In this example, since $K$ represents the number of transmit dimensions, the number of transmit dimensions per block is $K T$, and hence the number of information bits per transmit dimension is $C = \Reff$. Figure \ref{fig:cap_vs_Ttrain_train_beta1_alpha4_2_1_05_025_EbN0_10dB_Linf_eps0_Peq_idHH_iidS} shows the growth in normalized capacity, optimized with respect to $\eta/\ell$. These results show that the gain in using orthogonal training sequences appears to be more pronounced in situations where there is a high ratio of transmit antennas to receive antennas (i.e., $\alpha > 1$). Figure \ref{fig:opteta_vs_Ttrain_train_beta1_alpha4_2_1_05_025_EbN0_10dB_Linf_eps0_Peq_idHH_iidS_iidB} shows the associated optimal training length $\eta$, expressed as a percentage of the block length $\ell$, for \iid training sequences.

\begin{figure}
\centering
\includegraphics[width=\figw]{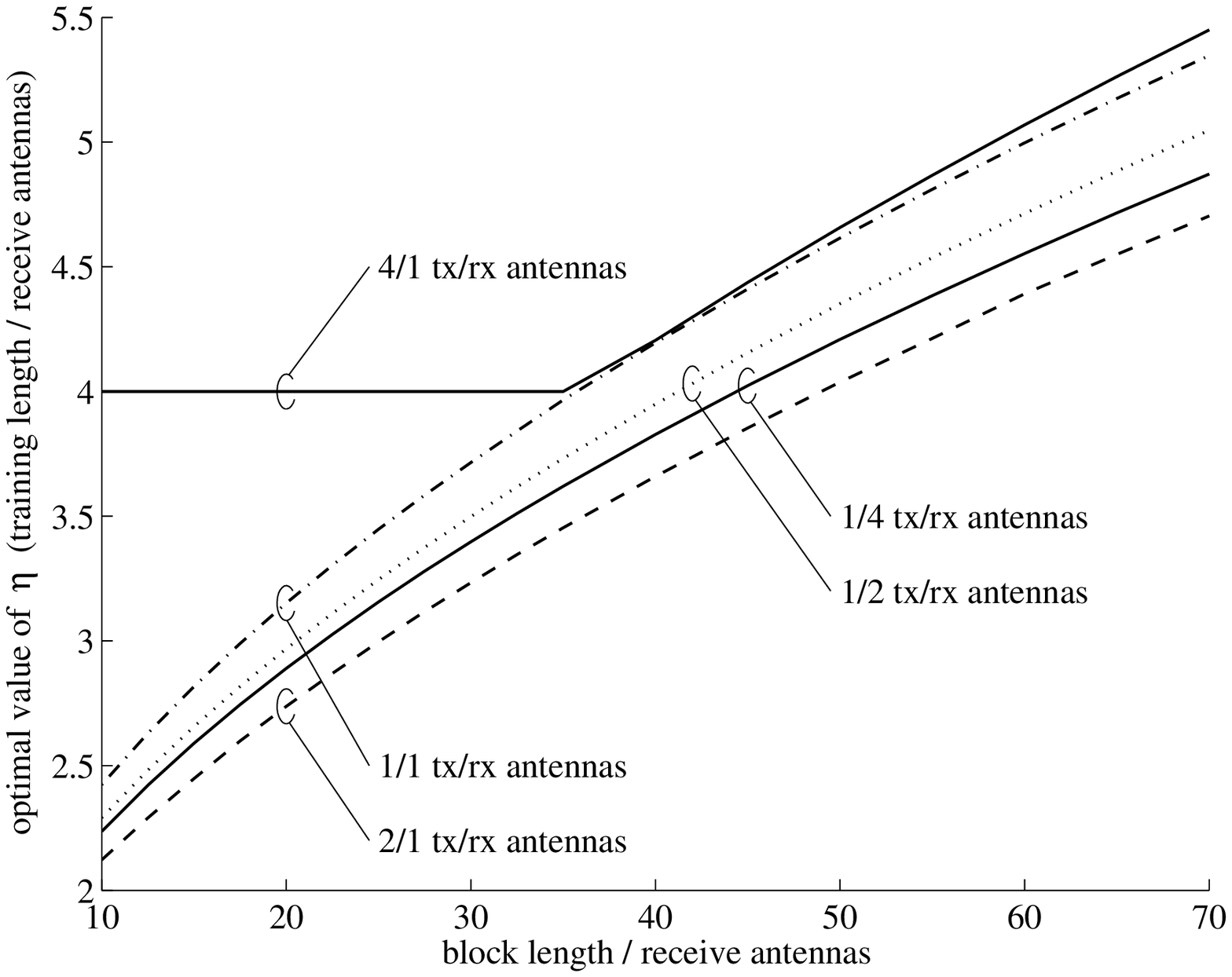}
\caption{Optimal value of normalized training length $\eta$ for orthogonal training sequences corresponding to Figure \ref{fig:cap_vs_Ttrain_train_beta1_alpha4_2_1_05_025_EbN0_10dB_Linf_eps0_Peq_idHH_iidS}.}
    \label{fig:opteta_vs_Ttrain_train_beta1_alpha4_2_1_05_025_EbN0_10dB_Linf_eps0_Peq_idHH_iidS_isoB}
\end{figure}

Figure \ref{fig:opteta_vs_Ttrain_train_beta1_alpha4_2_1_05_025_EbN0_10dB_Linf_eps0_Peq_idHH_iidS_isoB} shows the optimal value of $\eta$ for orthogonal training sequences corresponding to the curves in Figure \ref{fig:cap_vs_Ttrain_train_beta1_alpha4_2_1_05_025_EbN0_10dB_Linf_eps0_Peq_idHH_iidS}.

It is interesting to note the case $\alpha = 4$, where we see that $\eta$ is never chosen less than $4$.  Recall that for $\alpha > \eta$, we have orthogonal rows of $\mB$ and for $\alpha < \eta$, we have orthogonal columns of $\mB$. Clearly, orthogonal columns are preferable.  If the axis were extended, we would see the same behavior in the other curves.

%\begin{figure}
%    \centering
%    \mbox{
%    \subfigure[Throughput]
%    {\label{fig:cap_vs_Ttrain_train_beta1_alpha05_075_EbN0_10dB_Linf_eps0_Peq_idHH_iidS_iidB}
%\includegraphics[width=\subfigw]{cap_vs_Ttrain_train_beta1_alpha05_075_EbN0_10dB_Linf_eps0_Peq_idHH_iidS_iidB}} \quad
%    \subfigure[Optimal percentage of training in a block]
%    {\label{fig:opteta_vs_Ttrain_train_beta1_alpha05_075_EbN0_10dB_Linf_eps0_Peq_idHH_iidS_iidB}
%    \includegraphics[width=\subfigw]{opteta_vs_Ttrain_train_beta1_alpha05_075_EbN0_10dB_Linf_eps0_Peq_idHH_iidS_iidB}}
%    }
%    \caption{Throughput optimization: MIMO channel (i.e., \iid $\mS$, $\mH=\mI_N$, $\beta=1$), rectangular windowing, no diagonal loading, $\frac{E_b}{\sn}$=10 dB, \iid training sequences, equal transmit power per antenna.}
%    \label{fig:cap_opteta_vs_Ttrain_train_beta1_alpha05_075_EbN0_10dB_Linf_eps0_Peq_idHH_iidS_iidB}
%\end{figure}

%\subsection{Orthogonal Signatures or Orthogonal Training?}
%
%We now consider the slightly contrived situation in which we may have either isometric signatures or orthogonal training.  For $\beta\geq 1$, typically a system with isometric signatures significantly outperforms a comparable system with \iid signatures.  For $\mH=\mI$, clearly a matched filter with isometric signatures is best. Therefore, we consider the CDMA example with $\beta=0.5$, i.e., in which half the channel output are unavailable. Figure \ref{fig:example2}
%
%\begin{figure}
%\centering
%\epsfig{file=example2.eps,width=\figw}
%\caption{SINR vs.\ training length: (CDMA with non-square channel matrix) SNR = 10 dB, $\alpha=0.5$, $\beta=0.5$, $\mu=0.1$, rectangular windowing, equal transmit power per signature, exponential distribution for \aed of $\mH\mH^\dag$.}
%    \label{fig:example2}
%\end{figure}

\section{Conclusions}

Determining the transient behavior of ALS algorithms with random inputs is a classical problem, which is relevant to many communications applications, such as equalization and interference suppression. The large system results presented here are the first set of {\em exact} results, which characterize the transient performance of ALS algorithms for a wide variety of channel models of interest. Namely, our results apply to any linear input-output model (see Proposition \ref{pr:allAWGNbinvar}), where the input is unitarily invariant, and the channel matrix has a well-defined \aed with finite moments. As such, these results can be used to evaluate adaptive equalizer performance in the context of space-time channels. This represents a significant generalization of the previous large system results in \cite{Honig02Xiao}, which apply only to an i.i.d. channel matrix. Furthermore, the analytical approach relies only on elementary matrix manipulations, and is general enough to allow for orthonormal spreading and/or training sequences, in addition to i.i.d. sequences. Numerical results were presented, which show that orthogonal training sequences can perform significantly better than i.i.d. training sequences.

For the general ALS algorithm and model considered, the output SINR can be expressed as the solution to a set of nonlinear equations. These equations are complicated by the fact that they depend on a number of auxiliary variables, each of which is a particular large matrix moment involving the sample covariance matrix. Still, it is relatively straightforward to solve these equations numerically. Illustrative examples were presented showing the effect of
training length on the capacity of a block fading channel.

In the case of i.i.d. sequences without diagonal loading, the set of equations for output SINR yields a simple relationship between the SINRs for ALS and MMSE receivers, which accounts for an arbitrary data shaping window. This relation shows that ALS performance depends on the channel matrix {\em only through the MMSE}. In other words, ALS performance is {\em independent} of the channel shape given a target output MMSE. Whether or not an analogous relation holds with diagonal loading, orthogonal training and/or spreading sequences remains an open problem. Application of the analysis presented here to more general channel models (e.g., multi-user/multi-antenna) is also a topic for further study.

\begin{appendices}

\section{Precursor to Asymptotic Analysis}
\label{ap:precursor}%

\begin{defbox} \label{df:asymp}
Let $\{a_N\}_{N=1,\ldots}$ and $\{b_N\}_{N=1,\ldots}$ denote a pair of infinite sequences of complex-valued random variables indexed by $N$. These sequences are defined to be \emph{asymptotically equivalent}, denoted $a_N \asymp b_N$,  iff $\abs{a_N - b_N}\asto 0$ as $N\to\infty$, where $\asto$ denotes almost-sure convergence in the limit considered.
\end{defbox}
Clearly $\asymp$ is an equivalence relation, transitivity being obtained through the triangle inequality.  We shall additionally define asymptotic equivalence for sequences of $N\times 1$ vectors and $N\times N$ matrices in an identical manner as above, where the absolute value is replaced by the Euclidean vector norm and the associated induced spectral norm, respectively.

\begin{lemma} \label{lem:ax_by}
If $a_N \asymp b_N$ and $x_N \asymp y_N$, and if $\abs{a_N}$, $\abs{y_N}$ and/or $\abs{b_N}$, $\abs{x_N}$ are almost surely uniformly bounded above\footnote{A sequence $\{a_N\}_{N=1\ldots}$ of complex-valued $N\times 1$ vectors or scalars is uniformly bounded above over $N$ if $\sup_N \abs{a_N}<\infty$, or in the case of complex-valued $N\times N$ matrices, $\sup_N \norm{a_N}<\infty$.} over $N$, then $a_N x_N \asymp b_N y_N$.  Similarly, $a_N/ x_N \asymp b_N / y_N$ if $\abs{a_N}$ or $\abs{b_N}$ is uniformly bounded above over $N$, and at least one of $\inf_N \abs{x_N}$ and $\inf_N \abs{y_N}$ is positive almost surely.
\end{lemma}
\begin{proof}
The fact that $a_N x_N \asymp b_N y_N$ can be seen after writing $a_N x_N - b_N y_N = a_N x_N - b_N y_N + a_N y_N - a_N y_N$ and hence $\abs{a_N x_N - b_N y_N} \leq \abs{a_N}\abs{x_N -  y_N} + \abs{y_N}\abs{a_N  - b_N}$. Alternatively, we may add and subtract $b_N x_N$ from $a_N x_N - b_N y_N$ to obtain $\abs{a_N x_N - b_N y_N} \leq \abs{x_N}\abs{a_N - b_N} + \abs{b_N}\abs{x_N  - y_N}$. The division property, $a_N/ x_N \asymp b_N / y_N$, can be shown in the same way
\begin{align}
\abs{\frac{a_N}{x_N} - \frac{b_N}{y_N}}
%&\leq \frac{\abs{a_N}\abs{x_N - y_N}}{\abs{x_N}\abs{y_N}} + \frac{\abs{a_N - b_N}}{\abs{y_N}}
&\leq \begin{cases} \frac{\abs{a_N}\abs{x_N - y_N}}{\abs{x_N}\abs{y_N}} + \frac{\abs{a_N  - b_N}}{\abs{y_N}}  \\
\frac{\abs{b_N}\abs{x_N - y_N}}{\abs{x_N}\abs{y_N}} + \frac{\abs{a_N  - b_N}}{\abs{x_N}}
\end{cases}
\end{align}
Suppose $\delta = \inf_N \abs{y_N}$. Given a realization for which $\abs{x_N - y_N}\to 0$ and $\delta > 0$, we may take $N$ sufficiently large such that $\abs{x_N - y_N} \leq \delta/2$ and hence $\abs{x_N} \geq \delta/2$. Alternatively, for a realization for which $\abs{x_N - y_N}\to 0$  and $\delta' = \inf_N \abs{x_N}>0$, for $N$ sufficiently large we may show $\abs{y_N} \geq \delta'/2$. Using these facts, and the uniform upper bounds for $\abs{a_N}$ or $\abs{b_N}$ we obtain the result.
\end{proof}
Note that the multiplicative part of Lemma \ref{lem:ax_by} holds for any mixture of matrices, vectors or scalars for which the dimensions of $a_N$ and $x_N$ are such that $a_Nx_N$ makes sense, due to the submultiplicative property of the spectral norm. The following definition and related results, however, are concerned with scalar complex sequences.
\begin{defbox} \label{df:uasymp}
Let $\{\{a_{N,n}\}_{n=1\ldots N}\}_{N=1,\ldots}$ and $\{\{b_{N,n}\}_{n=1\ldots N}\}_{N=1,\ldots}$ denote a pair of infinite sequences, indexed by $N$. The $N\thh$ element is a  complex-valued sequences of length $N$, indexed by $n$.  These sequences are defined to be \emph{uniformly asymptotically equivalent}, denoted $a_{N,n} \uasymp{n} b_{N,n}$, iff $\max_{n\leq N}\abs{a_{N,n} - b_{N,n}} \asto 0$ as $N\to\infty$.
\end{defbox}
Also we define $a_N$ and $b_{N,n}$ (as defined in Definitions \ref{df:asymp} and \ref{df:uasymp} above) as being uniformly asymptotically equivalent (denoted $a_N \uasymp{n} b_{N,n}$), if $a_{N,n} \uasymp{n} b_{N,n}$ where $a_{N,n} = a_N$ for all $n=1\lldots N$.

Also, analogous to Lemma \ref{lem:ax_by}, we have
\begin{lemma} \label{lem:ax_by_ext}
If $a_{N,n} \uasymp{n} b_{N,n}$ and $x_{N,n} \uasymp{n} y_{N,n}$, and if $\abs{a_{N,n}}$, $\abs{y_{N,n}}$ and/or $\abs{b_{N,n}}$, $\abs{x_{N,n}}$ are almost surely uniformly bounded above over $N$ and $n$, then $a_{N,n} x_{N,n} \uasymp{n} b_{N,n} y_{N,n}$. Similarly, $a_{N,n}/ x_{N,n} \uasymp{n} b_{N,n} / y_{N,n}$ if $\abs{a_{N,n}}$ or $\abs{b_{N,n}}$ is almost surely uniformly bounded above over $N$ and $n$, and at least one of $\inf_{N,n} \abs{x_{N,n}}$ and $\inf_{N,n} \abs{y_{N,n}}$ is positive almost surely.
\end{lemma}

\begin{lemma} \label{lem:sums_converge}
If $a_{N,n} \uasymp{n} b_{N,n}$, then $\frac{1}{N}\sum_{n=1}^N a_{N,n} \asymp \frac{1}{N}\sum_{n=1}^N b_{N,n}$.
\end{lemma}
\begin{proof}
This follows immediately from
\begin{align}
\abs{\frac{1}{N}\sum_{n=1}^N ( a_{N,n} - b_{N,n} )} &\leq \max_{n\leq N}\abs{a_{N,n} - b_{N,n}}
\end{align}
\end{proof}
\begin{lemma} \label{lem:uN_in_Cplus}
For $N=1,\ldots$, let $\mX_N=\mM_N-z\mI_N$, where $\mM_N$ is an $N\times N$ Hermitian matrix and $z\in\CC^+$, and suppose $\vu_N \in \CC^N$. Denote $u_N = \vu_N^\dag\mX_N^{-1}\vu_N$. If
\begin{align}
b = \inf_N& \abs{\vu_N} > 0 \as \label{eq:uN_nz_lem} \\
B = \sup_N& \norm{\mX_N} < \infty \as \label{eq:normXNbound}
\end{align}
Then
\begin{align}
\imag(u_N) \geq \imag(z) \frac{b^2}{B^2} \as \label{eq:uNgeqimz}
\end{align}
and hence $u_N \in \CC^+$, almost surely.
\end{lemma}
\begin{proof}
First, we note the inequality (from the proof of \cite[Lemma 16.5]{Girko01})
\begin{align}
\frac{\vx^\dag\vx}{\vx^\dag\mY^{-1}\vx} & \leq \frac{\vx^\dag\mY\vx}{\vx^\dag\vx} \label{eq:girko_ineq}
\end{align}
for any $\vx\in\CC^N$, $\vx\neq\mZero$, and Hermitian positive definite $N\times N$ complex-valued matrix $\mY$.

Now,
\begin{align}
\imag(u_N) &= \frac{1}{2\mathbf{j}}\vu_N^\dag(\mX^{-1}-\mX^{-\dag})\vu_N \:=\: \imag(z)\vu_N^\dag(\mX^\dag\mX)^{-1}\vu_N \label{eq:imuN_as_qf}
\end{align}
and hence using (\ref{eq:girko_ineq}) and a realization for which (\ref{eq:uN_nz_lem}) and (\ref{eq:normXNbound}) holds we obtain
\begin{align}
\frac{1}{\imag(u_N)} &= \frac{1}{\imag(z)\vu_N^\dag(\mX^\dag\mX)^{-1}\vu_N} \;\leq\; \frac{\vu_N^\dag\mX^\dag\mX\vu_N}{\imag(z)(\vu_N^\dag\vu_N)^2} \;\leq\; \frac{\norm{\mX}^2}{\imag(z)\abs{\vu_N}^2}
\end{align}
which, with (\ref{eq:uN_nz_lem}), (\ref{eq:normXNbound}), and $z\in\CC^+$ gives the result.
\end{proof}

The following lemma is an asymptotic extension of the matrix inversion lemma, and is used extensively in the subsequent appendices to remove matrix dimensions as described in Section \ref{sec:preamble}. It is based on an approach in \cite{Honig02Xiao}.

\begin{lemma}
\label{lem:incsig_mhR} %
Let $\mY_N = \mX_N + \vv_N\vu_N^\dag + \vu_N\vv_N^\dag + c_N \vu_N\vu_N^\dag$, where $\vv_N, \vu_N \in \CC^N$, $c_N\in\RR^*$, and $\mX_N=\mM_N-z\mI_N$, where $\mM_N$ is an $N\times N$ Hermitian matrix and $z\in\CC^+$. Denote
\begin{align}
\epsilon_N &= \vu_N^\dag\mX_N^{-1}\vv_N \\
u_N &= \vu_N^\dag\mX_N^{-1}\vu_N \\
v_N &= \vv_N^\dag\mX_N^{-1}\vv_N
\end{align}

Assume that as $N\to\infty$,
\begin{align}
|\epsilon_N| &\asto 0 \label{eq:eps_to_zero}
\end{align}
and
\begin{align}
b \:=\: \inf_N& \abs{\vu_N} > 0 \quad\text{, a.s.,} \label{eq:uN_nz} \\
B \:=\: \sup_N& \max\left\{ \norm{\mX_N},\: \abs{\vv_N},\: \abs{\vu_N},\: \abs{c_N} \right\} <\infty \as \label{eq:vu_bounded}
%\delta &= \min\left\{\inf_N \abs{1 + \epsilon_N},\: \inf_N \abs{1 - u_Nv_N}, \: \inf_N \abs{1 - u_N(v_N-c_N)}\right\} > 0 \label{eq:def_delta}
\end{align}
Then,
\begin{align}
& \abs{\mY_N^{-1}\vu_N - \frac{\mX_N^{-1}(\vu_N -  u_N \vv_N)} {1 - u_N(v_N - c_N)}} \asto 0 \label{eq:leminc_eq1} \\
& \abs{\mY_N^{-1}\vv_N - \frac{\mX_N^{-1} (-v_N \vu_N + (1 +  c_N u_N)\vv_N )}{1 - u_N(v_N - c_N)}} \asto 0 \label{eq:leminc_eq2}
\end{align}
%and  % include?
%\begin{align}
%& \abs{\vu_N^\dag\mY_N^{-1} - \frac{(\vu_N^\dag -  u_N \vv_N^\dag)\mX_N^{-1}} {1 - u_N(v_N - c_N)}} \asto 0 \label{eq:leminc_eq3} \\
%& \abs{\vv_N^\dag\mY_N^{-1} - \frac{(-v_N \vu_N^\dag + (1 +  c_N u_N)\vv_N^\dag )\mX_N^{-1}}{1 - u_N(v_N - c_N)}} \asto 0 \label{eq:leminc_eq4}
%\end{align}
as $N\to\infty$, and
\begin{align}
\delta &= \inf_N \min\left\{ \abs{1 - u_Nv_N},\; \abs{1 - u_N(v_N-c_N)}\right\} \;>\; 0 \label{eq:def_delta}
\end{align} % XXX is the formula we've given for delta the greatest lower bound? maybe not..
almost surely, where $\delta$ depends only on $B$, $b$, and $\imag(z)$.
\end{lemma}

\begin{proof}
First note that from Lemma \ref{lem:uN_in_Cplus}, (\ref{eq:uN_nz}), and (\ref{eq:vu_bounded}), that $u_N \in \CC^+$ almost surely. We therefore consider a realization for which $u_N \in \CC^+$ and (\ref{eq:eps_to_zero})--(\ref{eq:vu_bounded}) holds, and take $N$ sufficiently large such that $\abs{\epsilon_N}\leq 1/2$ and hence
\begin{align}
\abs{1+\epsilon_N} \geq 1/2 . \label{eq:den_Xoinv}
\end{align}

Due to the definition of $\mX$,
\begin{align}
\norm{\mX^{-1}} \leq \imag(z)^{-1}. \label{eq:normXinv}
\end{align}
Now note that $\sup_N \abs{u_N} = B^2\imag(z)^{-1} < \infty$ due to (\ref{eq:vu_bounded}) and (\ref{eq:normXinv}). Also, $\imag(-u_N^{-1})=\imag(u_N)/\abs{u_N}^2$ and $\sup_N \abs{u_N^{-1}} = \imag(u_N)^{-1}$. Additionally, note that $\imag(v_N)\geq 0$, using an identical argument to (\ref{eq:imuN_as_qf}). Using these facts we obtain
\begin{align}
\abs{\frac{1}{1-u_Nv_N}} &= \frac{\abs{-u_N^{-1}}}{\abs{-u_N^{-1}+v_N}} \leq \frac{\abs{u_N^{-1}}}{\imag(-u_N^{-1}+v_N)}  \leq \frac{\abs{u_N^{-1}}}{\imag(-u_N^{-1})} \:\leq\:  \frac{\abs{u_N}^2}{\imag(u_N)^2} \leq  \(\frac{B^2}{\imag(z) b}\)^4 \label{eq:a1muv_ninf}
\end{align}
In the same way, also using $c_N\in\RR^*$, we obtain an identical uniform upper bound on $\abs{1-u_N(v_N-c_N)}^{-1}$, and hence obtain (\ref{eq:def_delta}).

%and therefore we have the following two important facts
%\begin{align}
%\abs{1-u_Nv_N} & > 0 \\
%\abs{1-u_N(v_N-c_N)} & > 0
%\end{align}
%where the latter is shown in the same manner as (\ref{eq:a1muv_ninf}), and also using $c_N\in\RR^*$.

In what follows, we will drop the dependence on $N$ from $u_N$, $v_N$, $c_N$, $\vu_N$, $\vv_N$, $\mX_N$, and $\mY_N$ to clarify the derivations. Define $\mX_1$ and $\mX_2$ according to the following equations
\begin{align}
\mY &=  \mX_2 + c\vu\vu^\dag \\
\mX_2 &= \mX_1 + \vu \vv^\dag \\
\mX_1 &= \mX  + \vv \vu^\dag.
\end{align}
The matrix inversion lemma gives
\begin{align}
\mY^{-1} &=  \mX_2^{-1} - \frac{  c \mX_2^{-1} \vu\vu^\dag \mX_2^{-1} }{1 + c\vu^\dag \mX_2^{-1} \vu }  \label{eq:Y_mil} \\
\mX_2^{-1} &= \mX_1^{-1} - \frac{  \mX_1^{-1} \vu \vv^\dag \mX_1^{-1}} {1 +   \vv^\dag \mX_1^{-1} \vu  } \label{eq:X2_mil} \\
\mX_1^{-1} &= \mX^{-1}-\frac{ \mX^{-1} \vv \vu^\dag \mX^{-1}} {1 +  \vu^\dag \mX^{-1} \vv } \label{eq:X1_mil}.
%&\asymp \mX^{-1}- \mX^{-1} \vv \vu^\dag \mX^{-1}
\end{align}

%The proof proceeds by using (\ref{eq:X1_mil}) and (\ref{eq:X2_mil}) to find $\mX_2^{-1}$ and
%related terms as a function of $\mX^{-1}$, which are in turn combined with (\ref{eq:Y_mil}) to
%establish the result.

%Firstly, note that (\ref{eq:vu_bounded}) and the fact that $\norm{\mX^{-1}}\leq \imag(z)^{-1}$ gives both
%$\max\left\{\abs{u_N},\: \abs{v_N} \right\} \leq B^2 \imag(z)^{-1}$ and
%$\max\left\{\abs{\mX_N^{-1}\vu_N},\: \abs{\mX_N^{-1}\vv_N} \right\} \leq B \imag(z)^{-1}$.

First consider $\mX_1^{-1}$, and note from (\ref{eq:X1_mil}), (\ref{eq:eps_to_zero}), Lemma \ref{lem:ax_by}, (\ref{eq:vu_bounded}), (\ref{eq:den_Xoinv}), and (\ref{eq:normXinv}) that
\begin{align}
\mX_1^{-1} &\asymp \mX^{-1}- \mX^{-1} \vv \vu^\dag \mX^{-1} \label{eq:mX1inv}
\end{align}
In fact, in the remainder of the proof, we shall repeatedly use (\ref{eq:eps_to_zero}), (\ref{eq:vu_bounded}), (\ref{eq:def_delta}), (\ref{eq:den_Xoinv}), and (\ref{eq:normXinv}) in order to apply Lemma \ref{lem:ax_by}, without explicitly stating this, however, it should be clear from the context.

From (\ref{eq:mX1inv}) we obtain
\begin{align}
\mX_1^{-1}\vu &\asymp \mX^{-1}(\vu -  u \vv) \label{eq:mXovu}\\
\mX_1^{-1}\vv &\asymp \mX^{-1}\vv \\
\vv^\dag\mX_1^{-1} &\asymp (\vv^\dag -  v \vu^\dag)\mX^{-1} \label{eq:vvmXo} \\
\vu^\dag\mX_1^{-1} &\asymp \vu^\dag \mX^{-1}
\end{align}
From this, we obtain
\begin{align}
\vu^\dag \mX_1^{-1} \vu &\asymp u  \\
\vv^\dag \mX_1^{-1} \vv &\asymp v  \\
\vv^\dag \mX_1^{-1}\vu &\asymp -  u v \label{eq:vX1u_uv}\\
\vu^\dag \mX_1^{-1}\vv &\asymp 0
%\tr{\mX_1^{-1}} &\asymp \tr{\mX^{-1}}  % HEY
\end{align}

Before we consider $\mX_2^{-1}$, we first analyze the denominator of the second term in (\ref{eq:X2_mil}). Firstly, due to (\ref{eq:vX1u_uv}), we may take $N$ large enough such that $\abs{\vv^\dag\mX_1^{-1}\vu + u v} < \delta/2$, and hence with (\ref{eq:def_delta}) we obtain $\abs{1 + \vv^\dag\mX_1^{-1}\vu} \geq \delta/2$. With this fact, we obtain $(1 +   \vv^\dag \mX_1^{-1} \vu)^{-1} \asymp (1 - u v)^{-1}$, since
\begin{align}
\abs{\frac{1}{1 + \vv^\dag \mX_1^{-1} \vu} - \frac{1}{1 - u v}} &= \frac{\abs{\vv^\dag \mX_1^{-1} \vu + u v}}{\abs{1 + \vv^\dag \mX_1^{-1} \vu}\abs{1 - uv}} \:\leq\: \frac{2}{\delta^2}\abs{\vv^\dag \mX_1^{-1} \vu + u v} \label{eq:den_X2}
\end{align}
%\begin{align}
%\delta \leq \abs{1 - u_N v_N \pm (1 + \vv^\dag\mX_1^{-1}\vu)} \;\leq\; \abs{1 + \vv^\dag\mX_1^{-1}\vu} + \delta/2 \label{eq:1pvX1u_pos}
%\end{align}

Now consider $\mX_2^{-1}$, for which from (\ref{eq:X2_mil}) and the preceding discussion we obtain
\begin{align}
\mX_2^{-1}\vv &= \mX_1^{-1}\vv - \frac{  \mX_1^{-1} \vu \vv^\dag
\mX_1^{-1}\vv}{1 +   \vv^\dag \mX_1^{-1} \vu  } \\
&\asymp \mX_1^{-1}\vv - \frac{  v \mX_1^{-1} \vu } {1 -   u v  } \\
%&\asymp \mX^{-1}\vv - \frac{  v}{1 -   u v}(\mX^{-1}\vu -  u \mX^{-1} \vv) \\
&\asymp \frac{1}{1 - u v}\mX^{-1}(-v\vu + \vv)
\end{align}
Similarly,
\begin{align}
\mX_2^{-1}\vu
% thesis
%&= \mX_1^{-1}\vu - \frac{  \mX_1^{-1} \vu \vv^\dag \mX_1^{-1}\vu}
%{1 +   \vv^\dag \mX_1^{-1} \vu  } \\
%&\asymp \mX_1^{-1}\vu + \frac{  u_1 v_1 \mX_1^{-1} \vu } {1 -   u_1 v_1  } \\
&\asymp \frac{1} {1 - u v} \mX^{-1} (\vu -  u\vv) \label{eq:mX2vu} \\
\vv^\dag\mX_2^{-1} &\asymp \frac{1}{1 - u v}(-v\vu^\dag + \vv^\dag)\mX^{-1} \\
\vu^\dag\mX_2^{-1} &\asymp \frac{1} {1 - u v} (\vu^\dag -  u\vv^\dag)\mX^{-1} \label{eq:vumX2}
\end{align}
and so,
\begin{align}
\vu^\dag \mX_2^{-1}\vv &\asymp \vv^\dag \mX_2^{-1}\vu \:\asymp\: - \frac{  v u}{1 -   u v}  \\
\vu^\dag \mX_2^{-1}\vu &\asymp \frac{u} {1 - u v}  \\
\vv^\dag \mX_2^{-1}\vv &\asymp \frac{v} {1 - u v}  \\
1 + c\vu^\dag \mX_2^{-1}\vu &\asymp \frac{1 - u (v - c)} {1 - u v} \label{eq:uXtv}
%\vu^\dag \mX_2^{-2} \vu &\asymp \frac{u_2 +  u^2 v_2 } {(1 - u v)^2} % HEY
\end{align}
%and
%\begin{align}
%\tr{\mX_2^{-1}} &= \tr{\mX_1^{-1}} - \frac{ \tr{\mX_1^{-1} \vu \vv^\dag
%\mX_1^{-1}}}{1 -   u_1 v_1} \\
%&= \tr{\mX^{-1}} - \frac{ \tr{(\mX^{-1}\vu -  u_1 \mX^{-1} \vv) (\vv^\dag - v_1\vu^\dag)\mX^{-1}}}{1 -   u_1 v_1}\\
%&\asymp \tr{\mX^{-1}} + \frac{u_1 v_2 + v_1 u_2}{1 -   u_1 v_1} .
%\end{align} % HEY

Before considering $\mY^{-1}$, we note that from (\ref{eq:vu_bounded}), (\ref{eq:def_delta}), (\ref{eq:uXtv}), and similar arguments preceding (\ref{eq:den_X2}) that $\abs{1 + c\vu^\dag \mX_2^{-1} \vu}$ has a positive uniform lower bound and $1/(1 + c\vu^\dag \mX_2^{-1}\vu) \asymp (1 -   u v)/(1 -   u (v - c))$.

Considering $\mY^{-1}$ using (\ref{eq:Y_mil}) and the preceding discussion, we obtain
\begin{align}
\mY^{-1}\vu %&=  \mX_2^{-1}\vu - \frac{ c \mX_2^{-1} \vu\vu^\dag \mX_2^{-1}\vu}{1 + c\vu^\dag \mX_2^{-1} \vu }  \\
&=  \frac{ \mX_2^{-1} \vu}{1 + c\vu^\dag \mX_2^{-1} \vu }  \\
% thesis
%&\asymp  \frac{1 -   u v} {1 -   u (v - c)}   \frac{1} {1 -   u v} (\mX^{-1}\vu -  u \mX^{-1} \vv) \\
&\asymp  \frac{1} {1 - u (v - c)} \mX^{-1}(\vu -  u \vv).
\end{align}
Similarly,
\begin{align}
\mY^{-1}\vv &=  \mX_2^{-1}\vv -  c\frac{ \vu^\dag
\mX_2^{-1}\vv}{1 + c\vu^\dag \mX_2^{-1} \vu }\mX_2^{-1} \vu \\
%&\asymp \mX_2^{-1}\vv +  c\frac{1 -   u v}{1 -   u (v - c)} \frac{  v u}{1 -   u v} \mX_2^{-1} \vu \\
&\asymp \mX_2^{-1}\vv +  c \frac{ v u}{1 -   u (v -
c)} \mX_2^{-1} \vu \\
&\asymp \frac{1}{1 -   u (v - c)}\mX^{-1}(-v \vu + (1 +  u c)\vv )
\end{align}
%and (\ref{eq:leminc_eq3})--(\ref{eq:leminc_eq4}) are obtained.

%and
%\begin{align}
%\vu^\dag\mY^{-1} &\asymp  \frac{1} {1 - u (v - c)} (\vu^\dag -  u \vv^\dag)\mX^{-1} \\
%\vv^\dag\mY^{-1} &\asymp \frac{1}{1 - u (v - c)}(-v \vu^\dag + (1 +  u c)\vv^\dag )\mX^{-1}
%\end{align}

%and
%\begin{align}
%\mY^{-1} &=  \mX_2^{-1} - \frac{  c \mX_2^{-1} \vu\vu^\dag
%\mX_2^{-1}}{1 + c\vu^\dag \mX_2^{-1} \vu }  \\
%&\asymp  \mX_2^{-1} -  c \frac{1 -   u v}{1 -   u (v -
%c)}\mX_2^{-1} \vu\vu^\dag \mX_2^{-1}\\
%\tr{\mY^{-1}} &= \tr{\mX^{-1}} + \frac{u v_2 + v u_2}{1 -   u v} -  c \frac{1 - u v}{1 - u
%(v - c)} \frac{u_2 + u^2 v_2 } {(1 - u v)^2} \\
%&= \tr{\mX^{-1}} + \frac{u v_2 + (v-c)u_2}{1 - u(v - c)}.
%\end{align} % HEY

\end{proof}

\newcommand{\uXAXv}{\varepsilon^{(1)}}
\newcommand{\vXAXu}{\varepsilon^{(2)}}
\newcommand{\au}        {\acute{u}}
\newcommand{\av}        {\acute{v}}
\begin{lemma}
\label{lem:trace_incsig_mhR}
Let $\mA_N$ be an $N\times N$ Hermitian matrix, and suppose $A = \sup_N \norm{\mA_N} < \infty$. Using the definitions and assumptions of Lemma \ref{lem:incsig_mhR}, additionally define
\begin{align}
\uXAXv_N &= \vu_N^\dag\mX_N^{-1}\mA_N\mX_N^{-1}\vv_N \label{eq:def_uXAXv} \\
\vXAXu_N &= \vv_N^\dag\mX_N^{-1}\mA_N\mX_N^{-1}\vu_N \label{eq:def_vXAXu} \\
\au_N &= \vu_N^\dag\mX_N^{-1}\mA_N\mX_N^{-1}\vu_N \label{eq:def_auN}  \\
\av_N &= \vv_N^\dag\mX_N^{-1}\mA_N\mX_N^{-1}\vv_N \label{eq:def_avN}
\end{align}
Then,
\begin{align}
\abs{\tr{\mA_N\mY_N^{-1}} - \( \tr{\mA_N\mX_N^{-1}} +  \frac{u_N \av_N + (v_N - c_N)\au_N - \uXAXv_N - \vXAXu_N } {1 - u_N (v_N - c_N)}\) } \asto 0 \label{eq:trace_incsig_mhR}
\end{align}
as $N\to\infty$.
%where $C\in(0,\infty)$ is a deterministic finite constant which depends only on $A$, $B$, $\delta$, and $\imag(z)$.
\end{lemma}
\begin{proof}
The proof continues from the proof of Lemma \ref{lem:incsig_mhR}. Again, we drop the subscript $N$ for convenience. We see from (\ref{eq:eps_to_zero}), (\ref{eq:X1_mil}), (\ref{eq:den_Xoinv}) and (\ref{eq:def_uXAXv}) that
\begin{align}
\tr{\mA\mX_1^{-1}} &\asymp \tr{\mA\mX^{-1}} - \uXAXv
\end{align}
while (\ref{eq:X2_mil}), (\ref{eq:mXovu}), (\ref{eq:vvmXo}), (\ref{eq:vX1u_uv}), and (\ref{eq:def_uXAXv})--(\ref{eq:def_avN}) give
\begin{align}
\tr{\mA\mX_2^{-1}} %&\asymp \tr{\mA\mX_1^{-1}} - \frac{ \vv^\dag \mX_1^{-1}\mA\mX_1^{-1} \vu }{1 -   u_1 v_1} \\
&\asymp \tr{\mA\mX_1^{-1}} - \frac{ (\vv^\dag - v\vu^\dag)\mX^{-1}\mA\mX^{-1}(\vu -  u \vv) }{1 - u v}\\
&= \tr{\mA\mX_1^{-1}} + \frac{u \av + v \au - \vXAXu - uv \uXAXv}{1 - u v} .
\end{align}
Similarly, (\ref{eq:mX2vu}), (\ref{eq:vumX2}) and (\ref{eq:def_uXAXv})--(\ref{eq:def_avN}) give
\begin{align}
\vu^\dag \mX_2^{-1}\mA\mX_2^{-1}\vu &\asymp \frac{(\vu^\dag - u\vv^\dag)\mA\mX^{-1}(\vu - u\vv)}{(1-uv)^2} \\
&= \frac{\au + u^2\av - u(\uXAXv + \vXAXu)} {(1 - u v)^2}
\end{align}
and finally (\ref{eq:Y_mil}) and (\ref{eq:uXtv}) yield
\begin{align}
\tr{\mA\mY^{-1}} &\asymp
\tr{\mA\mX_2^{-1}} -  c \frac{1 - u v}{1 - u (v -c)}\vu^\dag\mX_2^{-1}\mA\mX_2^{-1} \vu.
\end{align}
Combining the above, we obtain (\ref{eq:trace_incsig_mhR}).
\end{proof}

\begin{lemma}\cite[Lemma 2.6]{silverstein95bai}
\label{lem:lem2.6} %
Let $z\in\CC^+$, $\mA$ and $\mB$ $N\times N$ Hermitian, $\tau\in\RR$, and $\vq\in\CC^N$. Then,
\begin{align}
\abs{\tR{\((\mB-z\mI)^{-1}-(\mB+\tau\vq\vq^\dag-z\mI)^{-1}\)\mA}} &\leq \frac{\norm{\mA}}{\imag(z)}.
\end{align}
\end{lemma}

\begin{lemma}
\label{lem:lem3.1} \cite[Lemma 1]{Evans00Tse}
Let $\mC_N$, be an $N\times N$ complex-valued matrix with uniformly bounded spectral radius for all $N$, \ie, $\sup_N \norm{\mC_N} < \infty$, and $\vy = [X_1 \lldots X_N]^\dag/\sqrt{N}$, where the $X_i$'s are \iid complex random variables with mean zero, unit variance, and finite eighth moment. Then
\begin{align}
\Exp{|\vy^\dag \mC \vy - \tr{\mC}|^4} \leq \frac{c}{N^2}
\end{align}
where the constant $c>0$ does not depend on $N$, $\mC$, nor on the distribution of $X_i$.
\end{lemma}
%% Silverstein version
%\begin{lemma}
%\label{lem:lem3.1} \cite[Lemma 3.1]{silverstein95}
%Let $\mC$ be an $N\times N$ complex-valued matrix with $\norm{\mC}\leq 1$, and $\vy = [X_1 \lldots X_N]^\dag$, $\mX_i\in\CC$, where the $X_i$'s are \iid satisfying
%\begin{itemize}
%\item $|X_1|\leq \log N$,
%\item $\Exp{X_1}=0$, $\Exp{|X_1|^2}=1$
%\end{itemize}
%Then,
%\begin{align}
%\Exp{|\vy^\dag \mC \vy - \tr{\mC}|^6} \leq K N^3 \log^{12} N
%\end{align}
%where the constant $K$ does not depend on $N$, $\mC$, nor on the distribution of $X_1$.
%\end{lemma}

\begin{lemma} %
\label{lem:qf_to_tr_iso} %
Let $\mS$ be $K<N$ columns of an $N\times N$ Haar distributed random matrix, and suppose $\vs$ is a column of $\mS$. Let $\mX_{N}$ be an $N\times N$ complex-valued matrix, which is a non-trivial function of all columns of $\mS$ except $\vs$, and $B = \sup_N\norm{\mX_{N}} < \infty$. Then,
\begin{align}
\EXp{ \abs{\vs^\dag \mX_N \vs - \frac{1}{N-K}\tr{\mPi\mX_N} }^4 } &\leq \frac{C}{N^2}
\end{align}
where $\mPi = \mI_N - (\mS\mS^\dag - \vs\vs^\dag)$ and $C$ is a deterministic finite constant which depends only on $B$ and $\alpha=K/N$.
\end{lemma}
\begin{proof}
This result is a straightforward extension of \cite[Proposition 4]{Chaufray03Hachem}.
\end{proof}
% this is not used. XXX
%Using the definitions from Lemma \ref{lem:qf_to_tr_iso}, we note also that without loss of generality \cite[Lemma 2.2]{Petz04Reffy}, $\vs$ can be considered to be obtained from an \iid $N\times 1$ complex Gaussian vector via Gram-Schmidt orthogonalization with respect to the other $K-1$ columns of $\mS$, that is,
%\begin{align}
%\vs &= \frac{\mPi\vx}{|\mPi\vx|} \label{eq:isosig_as_gaussian}
%\end{align}
%where $\vx$ contains \iid proper complex Gaussian elements with $\Exp{\vx}=\mZero$ and $\Exp{\vx\vx^\dag}=\mI_N$.

Throughout the subsequent derivations, we shall use the fact that since we have assumed that the \edf's of $\mA^2$, $\mH\mH^\dag$, and $\mW$ converge in distribution almost surely to compactly supported non-random distributions on $\RR^*$, we have \cite{Chung01}
\begin{align}
&\limK \frac{1}{K}\sum_{k=1}^K \operatorname{f}(P_k) \:=\: \Exp{\operatorname{f}(P)} \label{eq:sum_to_exp_P}\\
&\limN \frac{1}{\beta^* N}\sum_{n=1}^{\beta^* N} \operatorname{f}(d_n^2) \:=\: \Exp{\operatorname{f}(H)} \label{eq:sum_to_exp_H}\\
&\limi \frac{1}{i}\sum_{m=1}^i \operatorname{f}(w_m) \:=\: \Exp{\operatorname{f}(W)} \label{eq:sum_to_exp_W}
\end{align}
almost surely, where $d_n$ is the $n\thh$ singular value of $\mH$, and $f:\RR^*\to\RR^*$ is any fixed bounded continuous function on the support of the \aed of $\mA^2$, the first $\beta^*N$ eigenvalues of $\mH\mH^\dag$, and $\mW$, respectively.

\section{Proof of Theorem \ref{th:unified_mmse}}
\label{ap:mmse_proofs}

The analysis in these appendices is based on removing a single dimension from matrices and vectors, as described in Section \ref{sec:preamble}. The dimension removed will correspond to a particular data stream, transmit/receive dimension, or symbol interval. For example, in what follows, $\mRItn{n}$ represents the matrix $\mR$ with the $n\thh$ transmit dimension removed. The symbol $t_n$ is used in this case since the $n\thh$ \emph{transmit} dimension is removed. We will use $d_k$ when removing the $k\thh$ \emph{data} stream, and $r_m$ for removing the $m\thh$ \emph{received} symbol interval.

We define $\uasymp{k}$ and $\uasymp{n}$ according to Definition \ref{df:uasymp} in Appendix \ref{ap:precursor}, where the maximum is over $k\leq K$ and $n\leq N$, respectively, and the limit is as $(M,N,K)\to\infty$ with $K/N\to\alpha>0$ and $M/N\to\beta>0$ constant, as described in Section \ref{sec:limit}.

\subsection{Definitions}
\label{sec:mmse_var_defns}

Let $\mR = (\mH\mS\mA)^\ddag - z\mI_M$, $z\in\CC^+$. The \Stieltjes transform of the \edf of the eigenvalues of $(\mH\mS\mA)^\ddag$ is given by $G_R^N(z) = \gmNm$, and the MMSE SINR in (\ref{eq:def_sinr}) is given by $P_k\rhomkNm$, where
\begin{align}
\gn{j}^N &= \frac{1}{M}\tr{\mXmse{j}} & \text{ , for }j&=1. \label{eq:def_gn_tr}  \\
%\gn{j,m}^N &= \frac{1}{\sn M} \vn_m^\dag\mXmse{j}\vn_m \hspace{5mm} \text{ , for $j=1$.} \label{eq:def_gn} \\
\rhok{j,k}^N &= \vh_k^\dag\mXmseIuk{j}\vh_k & \text{ , for }j&=1\lldots 4.,& 0&<k\leq K \label{eq:def_rhon}
\end{align}
where $\vh_k = \mH\vs_k$, and
\begin{align}
\mXmse{j} &= \begin{cases}
\mR^{-1} & \text{, $j = 1$,}\\
\mR^{-\dag}\mR^{-1} & \text{, $j = 2$,}\\
\mR^{-\dag}\mH^\ddag\mR^{-1} & \text{, $j = 3$,} \\
\mR^{-\dag}(\mH\mS\mA)^\ddag\mR^{-1} & \text{, $j = 4$.}
\end{cases}
\end{align}
Furthermore, $\mXmseIuk{j}$ is defined by removing the $k\thh$ data stream, $0<k\leq K$, from $\mXmse{j}$, by replacing $\mR$, $\mS$, and $\mA$ by $\mRIuk{k}$, $\mSIuk{k}$, and $\mAIuk{k}$, respectively, where
\begin{align}
\mRIuk{k} &= (\mH\mSIuk{k}\mAIuk{k})^\ddag-z\mI_M ,
\end{align}
$\mSIuk{k}$ is $\mS$ with the $k\thh$ column removed, and $\mAIuk{k}$ is $\mA$ with the $k\thh$ row and column removed. That is, $\mRIuk{k} = \mR - P_k\vh_k\vh_k^\dag$.

The following proposition shows that we may substitute $\mH$ with an equivalent matrix, without lack of generality. This substitution is essential in the analysis which follows.
\begin{proposition}
\label{pr:HisVD} %
For the model (\ref{eq:rcv_sig}), the distribution of both the \Stieltjes transform of the \eed of $(\mH\mS\mA)^\ddag$ and the MMSE SINR are invariant to the substitution of $\mV\mD$ for $\mH$, where $\mV$ is an $M\times M$ Haar-distributed random unitary matrix, $\mD$ is a $M\times N$ diagonal matrix containing the singular values of $\mH$.
\end{proposition}
\begin{proof}
Let $\mT$ be an independent $M\times M$ Haar-distributed random matrix. Now, note that the quantities of interest, namely $\gn{j}^N$ and $\rhok{j,k}^N$, are unchanged by the substitution of $\mT\mH$ for $\mH$. That is,
\begin{align}
\gn{1}^N = \frac{1}{N}\tr{\mR^{-1}} \:=\: \frac{1}{M}\tr{\mT\mT^\dag\mR^{-1}} \:=\: \frac{1}{M}\tr{((\mT\mH\mS\mA)^\ddag-z\mI_M)^{-1}}
\end{align}
Writing $\mT\mH\mS = (\mT\mU_M)\mD(\mU_N^\dag\mS)$, where $\mU_M\mD\mU_N^\dag$ is the singular value decomposition of $\mH$, the unitary invariance of $\mT$ and $\mS$ infers the result for the \Stieltjes transform. A similar treatment of $\rhok{1}^N$ gives the result for the MMSE SINR.
\end{proof}
Therefore, in the remainder of this appendix, we substitute $\mH$ with $\mV\mD$ everywhere.\footnote{We stress that $\mV\mD$ is an equivalent matrix, as defined in Proposition \ref{pr:HisVD}, as opposed to a decomposition of $\mH$.}  We denote the $n\thh$ column of $\mV$ as $\vv_n$, for $0<n\leq M$, and define $\vv_n = \mZero$ for $n>M$. Define $\{d_1\lldots d_{\beta^*N}\}$ as the diagonal elements of $\mD$, note that $\beta^*N = \min(M,N)$, and define $d_n=0$ for $n>\beta^*N$.

We can now define
\begin{align}
\taun{j,n}^N &= \vu_n^\dag\mXmseItn{j}\vu_n & \text{ , for }j&=1,2,3.,& 0&<n\leq N \label{eq:def_taun}
\end{align}
where $\vu_n = \mHItn{n}\mSItn{n}\mA^2\tvs_n$, and recall that $\tvs_n$ denotes the $n\thh$ column of $\mS^\dag$. Also, $\mXmseItn{j}$ denotes $\mXmse{j}$ with the effect of the $n\thh$ transmit dimension removed,  $0<n\leq \beta^*N$, by replacing $\mR$, $\mH$, and $\mS$ with $\mRItn{n}$, $\mHItn{n}$, and $\mSItn{n}$, respectively, where
\begin{align}
\mRItn{n} &= (\mHItn{n}\mSItn{n}\mA)^\ddag - z\mI_M \\
%\mHItn{n}^\ddag &= \mHItn{n}\mHItn{n}^\dag \\
\mHItn{n} &= \mVItn{n}\mDItn{n} \label{eq:HisVD_mmse}
\end{align}
and where $\mVItn{n}$ and $\mSItn{n}$ are $\mV$ and $\mS$ with their $n\thh$ column and row removed, respectively, and $\mDItn{n}$ is $\mD$ with both the $n\thh$ column and row removed.

Returning to (\ref{eq:def_rhon}) and (\ref{eq:def_taun}), note that these quadratic forms are uniformly asymptotically equivalent to the following expressions, derived in Appendix \ref{ap:uniform_conv_mmse}. These will be important in the subsequent analysis.
\begin{align}
\rhok{j,k}^N &\uasymp{k} \rhok{j}^N \:=\: \begin{cases} \label{eq:def_rhon_tr} %
\frac{1}{N} \tr{\mH^\ddag\mXmse{j}} & \iidS \\
\frac{1}{N-K} \tr{\mPi\mH^\dag\mXmse{j}\mH} & \isoS,
\end{cases} \\
\taun{j,n}^N &\uasymp{n} \taun{j}^N \:=\: \begin{cases}
\frac{1}{Ni^2} \tr{(\mH\mS\mA^2)^\ddag\mXmse{j}} & \iidS \\
\frac{1}{N}\sum_{n=1}^N \taun{j,n}^N & \isoS,
\end{cases} \label{eq:def_taun_tr}
\end{align}
where
\begin{align}
\mPi &= \mI_N - \mS\mS^\dag \label{eq:defn_mPi}.
\end{align}

%and where  is conditioned on the application of Lemma \ref{lem:incsig_mhR} to derive (\ref{eq:Rinv_vvn_mmse})--(\ref{eq:supNn_denH_mmse}) below.

Also, note that
\begin{align}
\vs_k^\dag\mH^\dag\mH\vs_k &\uasymp{k} \beta^*\Exp{H} > 0\label{eq:skHHsk-EH}
\end{align}
from Lemma \ref{lem:lem3.1} or Lemma \ref{lem:qf_to_tr_iso}, the Borel-Cantelli lemma, and (\ref{eq:sum_to_exp_H}). The positivity of (\ref{eq:skHHsk-EH}) is implied by $\beta^* > 0$ and $\Exp{H}>0$. Note that $\Exp{H}>0$ is implied by the assumption that the distribution of $H$ has a compact support on $\RR^*$, and does not have all mass at zero.

In addition, letting $c_n = \tvs_n^\dag\mA^2\tvs_n$ and $\EP = \Exp{P}$, we have
\begin{align}
c_n &\uasymp{n} \alpha\EP \label{eq:tvsA2tvs-aEP}
\end{align}
This is shown in a similar manner to (\ref{eq:skHHsk-EH}) using (\ref{eq:sum_to_exp_P}), noting that for isometric $\mS$ it requires $\tvs_n$ written as $\mE_K\vomega_n$, where $\mE_K = [\mI_K, \mZero_{K,N-K}]$, and $\vomega_n^\dag$ is the $n\thh$ row of the $N\times N$ Haar matrix $\mTheta$ from which $\mS$ is taken, \ie, $\mS = \mTheta\mE_K^\dag$.

We now give several bounds on particular matrix and vector norms which are required in order to apply Lemmas \ref{lem:ax_by} and \ref{lem:ax_by_ext} later. Firstly, the assumption that $z\in\CC^+$ gives
\begin{align}
\norm{\mR^{-1}} &\leq \imag(z)^{-1} \label{eq:normRinv}
\end{align}
Secondly, the assumptions on $\mH$, $\mS$, and $\mA$ outlined in Section \ref{sec:sysmodel} imply
\begin{align}
&\sup_N \norm{\mH}^2 < \infty \label{eq:def_Dmax} \\
&\sup_N \norm{\mA}^2 < \infty \label{eq:def_Pmax}\\
& \norm{\mS}^2 \:\asto\: (1 + \sqrt{\alpha})^2 \quad \text{, (\iid $\mS$)}\label{eq:norm_S} \\
& \abs{\vs_k}^2 \:\uasymp{k}\: 1 \label{eq:supKabs_sk} \\
& \abs{\tvs_n}^2 \:\uasymp{n}\: \alpha \label{eq:supNabs_tvsn}
\end{align}
where (\ref{eq:norm_S}) is due to \cite{Yin88Bai}, while (\ref{eq:supKabs_sk}) and (\ref{eq:supNabs_tvsn}) are shown in an identical manner to (\ref{eq:skHHsk-EH}) and (\ref{eq:tvsA2tvs-aEP}), respectively. Of course, $\norm{\mS}=1$ for isometric $\mS$. Moreover, (\ref{eq:normRinv})--(\ref{eq:supNabs_tvsn}) imply
\begin{align}
\sup_N &\max_{n\leq N} \max\{ \abs{\vu_n}, \abs{c_n} \} < \infty \as \label{eq:norm_un_cn_mmse} \\
\sup_N &\max\{ \abs{\rhok{j}^N}, \max_{k\leq K} \abs{\rhok{j,k}^N} \} < \infty  \as \quad, \text{ for } j=1\ldots 4.\label{eq:sup_rho} \\
\sup_N &\max\{ \abs{\taun{j}^N},  \max_{n\leq N} \abs{\taun{j,n}^N} \} < \infty \as \quad, \text{ for } j=1\ldots 3. \label{eq:sup_tau}
\end{align}
and additionally, with the assumption that $\abs{z}<\infty$,
\begin{align}
\sup_N \norm{\mR} < \infty \as \label{eq:sup_normR}
\end{align}

% leave out? XXX
%For example, since
%\begin{align}
%\abs{\rhomNm} &\leq \frac{1}{N}\abs{\tr{\mH^\ddag\mR^{-1}}} \:\leq\: \beta \norm{\mH^\ddag\mR^{-1}} \:\leq\: \beta\Dmax\imag(z)^{-1} \:<\: \infty
%\end{align}
%the norm of $\rhomNm$ is bounded for \iid $\mS$, where we've used $\abs{\tr{\mX}} \leq \norm{\mX}\rank(\mX)$. In a similar manner it can be seen that the same bound holds for isometric $\mS$, additionally using the fact that $\norm{\mPi}=1$ and $\rank(\mPi)=N-K$.

\subsection{Derivations}

We start by using the matrix inversion lemma to extract the $k\thh$ data stream from $\mR$, as described in Section \ref{sec:preamble}.
\begin{align}
\mR^{-1}\vh_k &= \frac{\mRIuk{k}^{-1}\vh_k}{1 + P_k \rhomkNm} \label{eq:mil_mmse}
\end{align}
This may be applied to the following identity to obtain
\begin{align}
1 &= \frac{1}{M}\tr{\mR\mR^{-1}} \:=\: -z\gmNm + \frac{1}{M}\sum_{k=1}^K P_k \vh_k^\dag \mR^{-1} \vh_k \\
&= -z\gmNm + \frac{\alpha}{\beta} \frac{1}{K}\sum_{k=1}^K \frac{P_k \rhomkNm}{1 + P_k \rhomkNm} \label{eq:RRin_dimK_unsimp} \end{align}
Now, we show that
\begin{align}
\frac{P_k \rhomkNm}{1 + P_k \rhomkNm} &\uasymp{k} \frac{P_k \rhomNm}{1 + P_k \rhomNm} \label{eq:RRin_dimK_arg}
\end{align}
First note that (\ref{eq:skHHsk-EH}) and (\ref{eq:sup_normR}) satisfies conditions (\ref{eq:uN_nz_lem}) and (\ref{eq:normXNbound}), respectively, of Lemma \ref{lem:uN_in_Cplus}, and hence $\imag(\rhomkNm)$ is uniformly bounded below over $K = \alpha N$ and $k\leq K$ by some $\delta > 0$.  Now due to (\ref{eq:def_rhon_tr}), we may consider a realization for which $\max_{k\leq K}\abs{\rhomkNm - \rhomNm} \to 0$ holds, and take $N$ sufficiently large such that $\max_{k \leq K}\imag(\rhomkNm - \rhomNm)\leq \delta/2$ so that $\imag(\rhomNm) \geq \delta/2$. Moreover, note that $\abs{1+P_k \rhomNm} \geq \abs{P_k\imag(\rhomNm)} \geq \abs{P_k}\delta/2$ and similarly $\abs{1+P_k \rhomkNm} \geq \abs{P_k} \delta$ so that
\begin{align}
\abs{\frac{P_k \rhomkNm}{1+P_k \rhomkNm} - \frac{P_k \rhomNm}{1+P_k \rhomNm}} &\leq \frac{\abs{P_k}\abs{\rhomkNm-\rhomNm}}{\abs{1+P_k \rhomkNm}} +  \frac{\abs{P_k}^2 \abs{\rhomNm} \abs{\rhomkNm-\rhomNm}}{\abs{1+P_k \rhomkNm}\abs{1+P_k \rhomNm}} \nonumber \\
&\leq \frac{1}{\delta} \(1 + \frac{2\abs{\rhomNm}}{\delta}\) \abs{\rhomkNm-\rhomNm}. \label{eq:denterm_uasymp_unfin}
\end{align}
Taking the maximum over $k$ and using (\ref{eq:sup_rho}) gives (\ref{eq:RRin_dimK_arg}).

From (\ref{eq:RRin_dimK_unsimp}), (\ref{eq:RRin_dimK_arg}), and Lemma \ref{lem:sums_converge} we obtain
\begin{align}
1 + z\gmNm &\asymp \frac{\alpha}{\beta}\rhomNm\sE_{1,1}^N \:=\: \frac{\alpha}{\beta}(1-\sE_{0,1}^N) \label{eq:RRinv_dimK_mmse}
\end{align}
where
\begin{align}
\sE_{m,n}^N &= \frac{1}{K}\sum_{k=1}^K \frac{P_k^m}{(1 + P_k\rhomNm)^n} \label{eq:def_EmnN}
\end{align}

%due to Lemma \ref{lem:ax_by_ext}, where $P_k \rhomkNm$ and $1 + P_k \rhomkNm$ correspond to $a_{N,n}$ and $x_{N,n}$ in the statement of the Lemma, respectively. That is because $P_k < \Pmax$, $\rhomNm$ is uniformly bounded above (as discussed after (\ref{eq:rhomNm_bounded})), and $\abs{1 + P_k\rhomkNm}$ is uniformly bounded below over $N$ and $k\leq \alpha N$. This last fact follows since $\abs{1 + P_k\rhomkNm} \geq 1$ for $P_k=0$ and  $\abs{1 + P_k\rhomkNm} \geq P_k\imag(\rhomkNm)$, for $P_k\neq 0$.  and (\ref{eq:uNgeqimz}) we see that $\abs{1/P_k + \rhomkNm}$ is uniformly bounded below over $N$ and $k\leq \alpha N$.  As $\rhomNm$ is uniformly bounded above, as discussed above, we have

For future reference, note that from (\ref{eq:mil_mmse}), following the proof of (\ref{eq:RRinv_dimK_mmse}) gives
\begin{align}
\frac{1}{N} \tr{(\mH\mS\mA^2)^\ddag\mR^{-1}} &= \frac{1}{N}\sum_{k=1}^K \frac{P_k^2 \rhomkNm}{1 + P_k \rhomkNm} \:\asymp\: \alpha(\EP - \sE_{1,1}^N) \label{eq:um1_iid_mmse_derivation} \\
\frac{1}{N} \tr{(\mH\mS)^\ddag\mR^{-1}} &= \frac{1}{N}\sum_{k=1}^K \frac{\rhomkNm}{1 + P_k \rhomkNm} \:\asymp\: \alpha\rhomNm\sE_{0,1}^N \label{eq:um1_iso_bit_mmse}
\end{align}

To prove the remaining equations in Theorem \ref{th:unified_mmse}, we consider another expansion of the correlation matrix $\mR$, this time to remove the $n\thh$ transmit dimension, $0<n\leq N$, as described in Section \ref{sec:preamble}.
\begin{align}
\mR &= (\mHItn{n}\mSItn{n} + d_n \vv_n \tvs_n^\dag)\mA^2(\mHItn{n}\mSItn{n} + d_n \vv_n
\tvs_n^\dag)^\dag - z\mI_M \\
&= \mRItn{n} + d_n \vu_n \vv_n^\dag + d_n \vv_n \vu_n^\dag + d_n^2 c_n \vv_n\vv_n^\dag \label{eq:Rmse_nexp}
\end{align}
where $\vu_n$ and $c_n$ are defined after (\ref{eq:def_taun}) and above (\ref{eq:tvsA2tvs-aEP}), respectively.

We now apply Lemma \ref{lem:incsig_mhR} to (\ref{eq:Rmse_nexp}), where $\mY_N$, $\mX_N$, $\vv_N$, $\vu_N$, and $c_N$ in the statement of the Lemma correspond to $\mR$, $\mRItn{n}$, $d_n\vu_n$, $\vv_n$, and $c_n$, respectively. We shall now verify that the conditions of the Lemma are satisfied. For any $n\leq N$, since $\mHItn{n}^\dag\vv_n = \mZero$ we have $\mRItn{n}\vv_n = -z\vv_n$ and $\mRItn{n}^\dag\vv_n = -z^*\vv_n$, and moreover
\begin{align}
\vv_n^\dag \mRItn{n}^{-1} \vv_n &= -z^{-1} \label{eq:invsn} \\
\vu_n^\dag \mRItn{n}^{-1}\vv_n &= \vv_n^\dag \mRItn{n}^{-1}\vu_n \:=\: 0 \label{eq:vuRtnvv}
\end{align}
where $\vv_n^\dag \mRItn{n}^{-1} \vv_n$ corresponds to $u_N$ in the Lemma, and (\ref{eq:vuRtnvv}) satisfies condition (\ref{eq:eps_to_zero}) of the Lemma. Since $\abs{\vv_n}=1$, condition (\ref{eq:uN_nz}) is satisfied, and along with (\ref{eq:norm_un_cn_mmse}) and (\ref{eq:sup_normR}) satisfies condition (\ref{eq:vu_bounded}). Note that $\taumnNm$, defined in (\ref{eq:def_taun}), corresponds to $v_N$ in the Lemma.  Therefore,
\begin{align}
\mR^{-1}\vv_n &\asymp \frac{\mRItn{n}^{-1}\( \vv_n + d_n z^{-1}\vu_n \)}{1+d_n^2z^{-1}(\taumnNm-c_n)}  \label{eq:Rinv_vvn_mmse}\\
\mR^{-1}\vu_n &\asymp \frac{\mRItn{n}^{-1}\( -d_n \taumnNm \vv_n + (1 - d_n^2 c_n z^{-1})\vu_n \)}{1+d_n^2z^{-1}(\taumnNm-c_n)} \label{eq:Rinv_vun_mmse} \\
\inf_N &\min_{n\leq N} \abs{1+d_n^2z^{-1}(\taumnNm-c_n)} > 0 \quad\text{, a.s.,} \label{eq:supNn_denH_mmse}
\end{align}
%It's also handy to know that
%\begin{align}
%\mR^{-1}(\vu_n + d_n\alpha\EP\vv_n) &= \frac{\mRItn{n}^{-1}\( \vu_n + d_n(\alpha\EP - \taumm)
%\vv_n \)}{1+d_n^2z^{-1}(\taumm-\alpha\EP)}
%\end{align}
which we shall now use to derive (\ref{eq:rm1_mmse_unified}) and (\ref{eq:tm1_mmse_unified}).

With \iid $\mS$, we see from (\ref{eq:def_taun_tr}) that (\ref{eq:um1_iid_mmse_derivation}) gives an expression for $\taumNm$.  For isometric $\mS$, we use $\mS^\dag\mS = \mI_K$, and
\begin{align}
\frac{1}{N}\tr{(\mH\mS\mA^2)^\ddag\mR^{-1}} &= \frac{1}{N}\tr{(\mH\mS\mA^2\mS^\dag)^\ddag\mR^{-1}}
\:=\: \frac{1}{N}\sum_{n=1}^N \tr{(\mH\mS\mA^2\tvs_n)^\ddag \mR^{-1}}  \\
&= \frac{1}{N}\sum_{n=1}^N \tr{(\vu_n + d_n c_n\vv_n)^\ddag \mR^{-1}} \label{eq:taumm_derive_iso}
\end{align}
where we have used $\mH\mS\mA^2\tvs_n = \vu_n + d_n c_n\vv_n$. Continuing with the preceding application of Lemma \ref{lem:incsig_mhR}, we may use Lemma \ref{lem:trace_incsig_mhR} to determine an equivalent asymptotic representation of the argument in the sum in (\ref{eq:taumm_derive_iso}), where $\mA_N$ in the statement of Lemma \ref{lem:trace_incsig_mhR} corresponds to $(\vu_n + d_n c_n\vv_n)^\ddag$. That is, using  (\ref{eq:vuRtnvv}), (\ref{eq:Rinv_vvn_mmse}), and (\ref{eq:Rinv_vun_mmse}), we note that the terms corresponding to $\uXAXv_N$ and $\vXAXu_N$ are both asymptotically equivalent to $- d_n^2 c_n z^{-1}\taumnNm$, while the terms corresponding to $\au_N$,  $\av_N$, and $\tr{\mA_N\mX_N^{-1}}$ are asymptotically equivalent to $d_n^2 c_n^2 z^{-2}$, $d_n^2 (\taumnNm)^2$, and $\taumnNm - d_n^2 c_n^2 z^{-1}$, respectively. Therefore, after some algebra, we obtain from (\ref{eq:trace_incsig_mhR})
\begin{align}
\tr{(\vu_n + d_n c_n\vv_n)^\ddag \mR^{-1}} &\asymp c_n + \frac{\taumnNm - c_n}{1+d_n^2z^{-1}(\taumnNm-c_n)} \label{eq:taumm_iso_unsimp}
\end{align}
and from Lemma \ref{lem:ax_by_ext}, (\ref{eq:def_taun_tr}), and (\ref{eq:tvsA2tvs-aEP}) we obtain
\begin{align}
c_n + \frac{\taumnNm - c_n}{1+d_n^2z^{-1}(\taumnNm-c_n)} &\uasymp{n} \alpha\EP + \frac{\taumNm - \alpha\EP}{1+d_n^2z^{-1}(\taumNm-\alpha\EP)} \label{eq:taumm_iso_uasymp}
\end{align}
noting that the bounds required for the application of Lemma \ref{lem:ax_by_ext} are satisfied by (\ref{eq:sup_tau}), $\EP<\infty$, and (\ref{eq:supNn_denH_mmse}). We therefore obtain from (\ref{eq:um1_iid_mmse_derivation}), (\ref{eq:taumm_derive_iso}), (\ref{eq:taumm_iso_unsimp}), (\ref{eq:taumm_iso_uasymp}), and Lemma \ref{lem:sums_converge} that
\begin{align}
\alpha(\EP - \sE_{1,1}^N) &\asymp \alpha\EP + \beta^*(\taumNm-\alpha\EP)(\sHm_{0,1}^N + \frac{1}{\beta^*}-1) \label{eq:taum_iso_simp}
\end{align}
or equivalently,
\begin{align}
\taumNm &\asymp \alpha\EP - \frac{\alpha\sE_{1,1}^N}{\beta^*(\sHm_{0,1}^N-1) + 1}
\label{eq:tm1_mmse_unified_ap}
\end{align}
where
\begin{align}
\sHm_{p,1}^N &= \frac{1}{\beta^* N}\sum_{n=1}^{\beta^* N} \frac{d_n^{2p}}{1+d_n^2 z^{-1}(\taumNm-\alpha\EP)} \label{eq:def_HmnN}
\end{align}

%we have the following sequence of relations, which comes from the previous sequence of asymptotic equalities in reverse order, and the identity $\mS^\dag\mS = \mI_K$,
%\begin{align}
%\alpha(\EP - \sE_{1,1}) &\asymp \frac{1}{N}\tr{\mH\mS\mA^4\mS^\dag\mH^\dag\mR^{-1}}
%\label{eq:snASHRHSAsn_unsimp}
%\:=\: \frac{1}{N}\tr{\mH\mS\mA^2\mS^\dag\mS\mA^2\mS^\dag\mH^\dag\mR^{-1}} \\
%&= \frac{1}{N}\sum_{n=1}^N \tvs_n^\dag\mA^2\mS^\dag\mH^\dag\mR^{-1}\mH\mS\mA^2\tvs_n  \\
%&= \frac{1}{N}\sum_{n=1}^N (\vu_n + d_n\alpha\EP\vv_n)^\dag \mR^{-1}(\vu_n + d_n\alpha\EP\vv_n) \\
%&\asymp \frac{1}{N}\sum_{n=1}^N \frac{(\vu_n + d_n\alpha\EP\vv_n)^\dag \mRItn{n}^{-1}(\vu_n +
%d_n(\alpha\EP-\taumm)\vv_n)}{1+d_n^2z^{-1}(\taumm-\alpha\EP)} \\
%&\asymp \frac{1}{N}\sum_{n=1}^N \( \alpha\EP + \frac{\taumm - \alpha\EP}{1+d_n^2z^{-1}(\taumm-\alpha\EP)}\) \\
%&\asymp \alpha\EP + \beta^*(\taumm-\alpha\EP)(\sHm_{0,1} + \frac{1}{\beta^*}-1)
%\end{align}

For \iid $\mS$, using (\ref{eq:def_rhon_tr}) and (\ref{eq:vuRtnvv})--(\ref{eq:Rinv_vvn_mmse}) in the same manner as the derivation of (\ref{eq:taum_iso_simp}), we have that
\begin{align}
\rhomNm &= \frac{1}{N}\sum_{n=1}^{\beta^* N} d_n^2\vv_n^\dag\mR^{-1}\vv_n
\:\asymp\: \frac{1}{N}\sum_{n=1}^{\beta^* N} \frac{-z^{-1} d_n^2}{1+d_n^2z^{-1}(\taumnNm-c_n)} \\
&\asymp -\beta^* z^{-1}\sHm_{1,1}^N \label{eq:rhomNm_derive}
\end{align}
and similarly from (\ref{eq:def_rhon_tr}), (\ref{eq:um1_iso_bit_mmse}), and (\ref{eq:rhomNm_derive}) we obtain for isometric $\mS$
\begin{align}
\rhomm &= \frac{1}{1-\alpha}\( -\beta^* z^{-1}\sHm_{1,1} - \alpha\rhomm\sE_{0,1}\)
\label{eq:rhomi_mmse_derive}
\end{align}

We now simplify the preceding solution by noting that the identity $\frac{1}{M}\tr{\mR\mR^{-1}} = 1$ may also be expanded in the dimension $N$, as opposed to the dimension $K$ in (\ref{eq:RRinv_dimK_mmse}). That is,
\begin{align}
1 + z\gmNm &= \frac{1}{M} \sum_{n=1}^{\beta^*N}d_n \tvs_n^\dag\mA^2(\mSItn{n}^\dag\mHItn{n}^\dag + d_n\tvs_n\vv_n^\dag) \mR^{-1}\vv_n \label{eq:RRinv_dimN_mmse_unsimp}
\end{align}
Applying (\ref{eq:vuRtnvv}) and (\ref{eq:Rinv_vvn_mmse}) to the argument to the above sum gives
\begin{align}
\tvs_n^\dag\mA^2(\mSItn{n}^\dag\mHItn{n}^\dag + d_n\tvs_n\vv_n^\dag)\mR^{-1}\vv_n &\asymp
\frac{(\vu_n^\dag + d_n c_n\vv_n^\dag)\mRItn{n}^{-1}(\vv_n +d_nz^{-1}\vu_n)}{1+d_n^2z^{-1}(\taumnNm-c_n)} \\
&\uasymp{n} 1 - \frac{1}{1+d_n^2z^{-1}(\taumNm-\alpha\EP)} \label{eq:RRinv_dimn_mmse_exp}
\end{align}
and so applying Lemma \ref{lem:sums_converge} to (\ref{eq:RRinv_dimN_mmse_unsimp}) with (\ref{eq:RRinv_dimn_mmse_exp}) gives
\begin{align}
1 + z\gmNm &\asymp \frac{\beta^*}{\beta}(1-\sHm_{0,1}^N) \label{eq:RRinv_dimN_mmse_simp}
\end{align}

We now use (\ref{eq:RRinv_dimN_mmse_simp}) and (\ref{eq:RRinv_dimK_mmse}) to simplify (\ref{eq:tm1_mmse_unified_ap}) and (\ref{eq:rhomi_mmse_derive}) in the case of isometric $\mS$. Combining (\ref{eq:RRinv_dimN_mmse_simp}) with (\ref{eq:RRinv_dimK_mmse}) gives $1+\beta^*(\sHm_{0,1}^N-1) \asymp 1-\beta(1+z\gmNm)$, which combined with (\ref{eq:tm1_mmse_unified_ap}) gives
\begin{align}
\taumNm &\asymp \alpha\EP - \frac{\alpha\sE_{1,1}^N}{1-\beta(1+z\gmNm)} \label{eq:taumNm_mmse_iso}. %\\
\end{align}
Similarly, combining (\ref{eq:RRinv_dimK_mmse}) with (\ref{eq:rhomi_mmse_derive}) gives
\begin{align}
\rhomNm &\asymp \frac{-\beta^* z^{-1}\sHm_{1,1}^N}{1-\beta(1+z\gmNm)} \label{eq:rhomNm_mmse_iso} .
%&= \frac{-z^{-1}\beta^*\sHm_{1,1}}{1-\alpha\rhomm\sE_{1,1}} .
\end{align}

It follows that along a realization for which (\ref{eq:sum_to_exp_P}), (\ref{eq:sum_to_exp_H}), (\ref{eq:RRinv_dimK_mmse}), (\ref{eq:def_EmnN}), (\ref{eq:um1_iid_mmse_derivation}), (\ref{eq:def_HmnN}), (\ref{eq:rhomNm_derive}), (\ref{eq:taumNm_mmse_iso}), and (\ref{eq:rhomNm_mmse_iso}) hold, $\abs{\Gmse^N(z) - \gmm} \to 0$, $\abs{\rhomNm - \rhomm} \to 0$, and $\abs{\taumNm - \taumm} \to 0$, where $\gmm,\rhomm,\taumm\in\CC^+$ are solutions to
(\ref{eq:Gz_mmse_unified})--(\ref{eq:tm1_mmse_unified}).

\section{Proof of (\ref{eq:def_rhon_tr}) and (\ref{eq:def_taun_tr}) in Appendix \ref{ap:mmse_proofs}.}
\label{ap:uniform_conv_mmse}

Here we show $\max_{k\leq K}\abs{\rhok{j,k}^N - \rhok{j}^N}\asto 0$ and $\max_{n\leq N}\abs{\taun{j,n}^N - \taun{j}^N} \asto 0$ for $j=1$ in the limit considered.  The remaining cases $j=2,3,4$ are shown in an identical manner using the same results as outlined below.

Define
\begin{align}
\rhok{1,k}^{N'} &= \begin{cases}
\frac{1}{N} \tr{\mH^\dag\mRIuk{k}^{-1}\mH} & \iidS \\
\frac{1}{N-K} \tr{\mPiIuk{k}\mH^\dag\mRIuk{k}^{-1}\mH} & \isoS,
\end{cases} \\
\rhok{1,k}^{N''} &= \frac{1}{N-K}\tr{\mPi\mH^\dag\mRIuk{k}^{-1}\mH}  \quad \isoS, \\
\mPiIuk{k} &= \mPi + \vs_k\vs_k^\dag
\end{align}
From Lemma \ref{lem:lem3.1}, Lemma \ref{lem:qf_to_tr_iso}, and the Borel-Cantelli lemma, we have $\rhomkNm \uasymp{k} \rho_{1,k}^{N'}$ in the limit considered. For isometric $\mS$, we obtain $\abs{\rho_{1,k}^{N'} - \rho_{1,k}^{N''}} = \frac{1}{N-K}\abs{\rhomkNm}$, which with (\ref{eq:sup_rho}) gives $\rho_{1,k}^{N'} \uasymp{k} \rho_{1,k}^{N''}$. Finally, from Lemma \ref{lem:lem2.6} we have $\abs{\rhomNm - \rho_{1,k}^{N''}} \leq \frac{\Hmax}{\imag(z)N}$ for \iid $\mS$, and $\leq \frac{\Hmax}{\imag(z)(N-K)}$ for isometric $\mS$, and hence $\rho_{1,k}^{N''} \uasymp{k} \rhomNm$. Putting these together, we have $\rho_{1,k}^{N} \uasymp{k} \rho_{1,k}^{N'} \uasymp{k} \rhomNm$ for \iid $\mS$, and $\rho_{1,k}^{N} \uasymp{k} \rho_{1,k}^{N'} \uasymp{k} \rho_{1,k}^{N''} \uasymp{k} \rhomNm$ for isometric $\mS$, as claimed in (\ref{eq:def_rhon_tr}).

Turning our attention to $\taumnNm$, for \iid $\mS$ define
\begin{align}
\tau_{1,n}^{N'} &= \frac{1}{N}\tr{(\mHItn{n}\mSItn{n}\mA^2)^\ddag\mRItn{n}^{-1}} \\
\tau_{1,n}^{N''} &= \frac{1}{N}\tr{(\mHItn{n}\mSItn{n}\mA^2)^\ddag\mR^{-1}}
%\taujN{j} &= \frac{1}{N}\tr{\mA^2\mS^\dag\mH^\dag\mR^{-1}\mH\mS\mA^2}
\end{align}
Firstly, $\taumnNm \uasymp{n} \tau_{1,n}^{N'}$ from Lemma \ref{lem:lem3.1} and the Borel-Cantelli lemma. Now, applying Lemma \ref{lem:trace_incsig_mhR} to
\begin{align}
\abs{\tau_{1,n}^{N'} - \tau_{1,n}^{N''}} &= \frac{1}{N}\abs{\tr{(\mHItn{n}\mSItn{n}\mA^2)^\ddag(\mR^{-1} - \mRItn{n}^{-1})}}
\end{align}
and using (\ref{eq:sup_rho})--(\ref{eq:sup_tau}), and (\ref{eq:supNn_denH_mmse}) it is straightforward to show $\tau_{j,n}^{N'} \uasymp{n} \tau_{j,n}^{N''}$.
Also,
\begin{align}
\abs{\tau_{1,n}^{N''} - \taumNm} &= \abs{\frac{1}{N}\tr{\mR^{-1}((\mHItn{n}\mSItn{n}\mA^2)^\ddag - (\mH\mS\mA^2)^\ddag)} } \nonumber \\
&= \abs{\frac{1}{N}\tr{\mR^{-1}(d_n \vu_n \vv_n^\dag + d_n \vv_n \vu_n^\dag + d_n^2 c_n \vv_n \vv_n^\dag)}} \nonumber \\
%&\leq \frac{1}{N} \( 2\abs{d_n}\abs{\mR^{-1}\vu_n} + \abs{d_n^2} \abs{c_n} \abs{\mR^{-1}\vv_n}\) \label{eq:taud_taudd_mmse}
&\leq \frac{1}{N \imag(z)} \( 2d_n\abs{\vu_n} + d_n^2\abs{c_n}\) \label{eq:taud_taudd_mmse}
\end{align}
where $\vu_n$ and $c_n$ are defined in Appendix \ref{sec:mmse_var_defns}, and we have used $\abs{\vv_n}=1$ and (\ref{eq:normRinv}). It is clear from (\ref{eq:def_Dmax}) and (\ref{eq:norm_un_cn_mmse}) that the terms inside the bracket of (\ref{eq:taud_taudd_mmse}) are uniformly bounded above over $N$ and $n$, so $\tau_{1,n}^{N''} \uasymp{n} \taumNm$. Moreover, as $\taumnNm \uasymp{n} \tau_{1,n}^{N'} \uasymp{n} \tau_{j,n}^{N''} \uasymp{n} \taumNm$ we have (\ref{eq:def_taun_tr}) for \iid $\mS$.

To show (\ref{eq:def_taun_tr}) for isometric $\mS$, define
\begin{align}
%\tau_{1,n}^N &= \vu_n^\dag\mRItn{n}^{-1}\vu_n \\
\tau_{1,n,m}^N &= \vu_{n,m}^\dag\mRItn{n}^{-1}\vu_{n,m} \\
\tau_{1,n,m}^{N'} &= \vu_{n,m}^\dag\mRItn{n,m}^{-1}\vu_{n,m}
%\tau_{1,n,m}^{N''} &= \acute{\vu}_{n,m}^\dag\mRItn{n,m}^{-1}\acute{\vu}_{n,m}
\end{align}
for $m,n\in\{1\lldots N\}$ with $m\neq n$ and where
\begin{align}
\vu_{n,m} &= \mHItn{n,m}\mSItn{n,m}\mA^2\tvs_n \:=\: \vu_n - d_m\tvs_m^\dag\mA^2\tvs_n \vv_m
%\acute{\vu}_{n,m} &= \mHItn{n,m}\mSItn{n,m}\mA^2\mE_K\mTheta^\dag \Psi_{n,m} \ve_n \\
\end{align}
Firstly, note that
\begin{align}
\max_{m,n\leq N,(m\neq n)}\abs{\tvs_{m}^\dag\mA^2\tvs_{n}} &\asto 0 \label{eq:smAjsn} \\
\max_{m,n\leq N,(m\neq n)}\abs{\vu_{m,n}^\dag \mRItn{n,m}^{-1} \vu_{n,m}} &\asto 0  \label{eq:ujmnRnmujnm} \\
\vu_{n,m}^\dag \mRItn{n,m}^{-1} \vv_n \:=\: \vv_n^\dag\mRItn{n,m}^{-1}\vu_{n,m} &= 0, \quad \forall \:m,n\leq N \label{eq:ujnmRvm}
\end{align}
where (\ref{eq:smAjsn}) and (\ref{eq:ujmnRnmujnm}) can be shown using standard arguments after writing $\tvs_{m}$ and $\tvs_{n}$ in the form described in the discussion following (\ref{eq:tvsA2tvs-aEP}). Also, (\ref{eq:ujnmRvm}) is shown in the same way as (\ref{eq:vuRtnvv}). We now focus on a realization for which (\ref{eq:smAjsn}) and (\ref{eq:ujmnRnmujnm}) hold.
Now, $\max_{m,n\leq N,(m\neq n)}\abs{\tau_{1,n}^N - \tau_{1,n,m}^N} \to 0$ since $\abs{\tau_{1,n}^N - \tau_{1,n,m}^N} \leq 2\imag(z)^{-1} \abs{\tvs_m^\dag\mA^2\tvs_n} \abs{d_m}\abs{\vu_n}$, to which we apply (\ref{eq:def_Dmax}), (\ref{eq:norm_un_cn_mmse}), and (\ref{eq:smAjsn}).

Writing $\tau_{1,n,m}^N - \tau_{1,n,m}^{N'} = \tr{\vu_{n,m}^\ddag(\mRItn{n}^{-1}-\mRItn{n,m}^{-1})}$, we have that $\max_{m,n\leq N,(m\neq n)}\abs{\tau_{1,n,m}^N - \tau_{1,n,m}^{N'}} \to 0$ from Lemma \ref{lem:trace_incsig_mhR}, since the terms corresponding to $\au_N$, $\av_N$, $\uXAXv$, and $\vXAXu$ in the statement of the Lemma uniformly converge to zero (independently of $m$ and $n$) due to (\ref{eq:supNn_denH_mmse}), (\ref{eq:ujmnRnmujnm}), (\ref{eq:ujnmRvm}).

Define
\begin{align}
\Psi_{n,m} &= \ve_n\ve_m^\dag + \ve_m\ve_n^\dag + \sum_{\ell\neq m,n}^N \ve_\ell\ve_\ell^\dag
\end{align}
where $\ve_n$ is an $N\times 1$ vector which contains zeros except for a $1$ in the $n\thh$ row. That is, $\mPsi_{n,m}$ is simply the unitary permutation matrix which swaps the $n\thh$ and $m\thh$ entries. Also, recall from the discussion following (\ref{eq:tvsA2tvs-aEP}) that $\mS$ may be written $\mS = \mTheta\mE_K^\dag$, and thus $\tvs_n = \mE_K\mTheta^\dag\ve_n$.
Now, note that
\begin{align}
\EXp{\abs{\tau_{1,n,m}^{N'} - \tau_{1,m,n}^{N'}}^4}
&= \EXp{\abs{\tr{((\mHItn{n,m}\mSItn{n,m}\mA^2\mE_K\mTheta^\dag \ve_n)^\ddag  - (\mHItn{n,m}\mSItn{n,m}\mA^2\mE_K\mTheta^\dag \ve_m)^\ddag)\mRItn{n,m}^{-1}}}^4} \nonumber \\
&= \EXp{\abs{\tr{((\mHItn{n,m}\mSItn{n,m}\mA^2\mE_K\mTheta^\dag\Psi_{n,m} \ve_n)^\ddag  - (\mHItn{n,m}\mSItn{n,m}\mA^2\mE_K\mTheta^\dag \Psi_{n,m}\ve_m)^\ddag)\mRItn{n,m}^{-1}}}^4} \nonumber \\
&= 0
\end{align}
where in the second step, we used the unitary invariance of $\mTheta$ to substitute $\mTheta$ with $\Psi_{n,m}\mTheta$ throughout the previous expression, noting also that this has no effect on $\mSItn{m,n}$ and hence also $\mRItn{n,m}$.  Therefore, $\max_{m,n\leq N,(m\neq n)}\abs{\tau_{1,n,m}^{N'} - \tau_{1,n,m}^{N''}} \to 0$.

Combining the above preceding gives $\max_{m,n\leq N,(m\neq n)}\abs{\tau_{1,n}^N - \tau_{1,m}^N} \to 0$. Moreover,
\begin{align}
\abs{\taumnNm - \taumNm} &\leq \frac{1}{N}\sum_{m=1}^N \abs{\tau_{1,m}^N - \tau_{1,n}^N} \;\leq\; \max_{m\leq N}\abs{\tau_{1,n}^N - \tau_{1,m}^N}
\end{align}
and hence $\taumnNm \uasymp{n} \taumNm$.

\section{Alternate MMSE SINR of Section \ref{sec:alternate_mmse_sinr}}
\label{ap:alternate_mmse_sinr} %

Note from (\ref{eq:mil_mmse}) that the filter $\mRIuk{k}^{-1}\vh_k$ has the same SINR as $\mR^{-1}\vh_k$. The associated signal and interference powers are $P_k\abs{\rhomkNm}^2$ and $\rhomtkNcm + \sn\rhomtkNm$ respectively, as defined in Appendix \ref{sec:mmse_var_defns}. It is easily shown that the latter term simplifies to $\rhomkNmc$ (the complex conjugate of $\rhomkNm$), and hence the MMSE SINR is $P_k\rhomkNm$. Namely,
\begin{align}
\rhomtkNcm + \sn\rhomtkNm &= \vh_k^\dag\mRIuk{k}^{-\dag}(\mRIuk{k}-\sn\mI_M)\mRIuk{k}^{-1}\vh_k +
\sn\vh_k^\dag\mRIuk{k}^{-\dag}\mRIuk{k}^{-1}\vh_k \\
&= \vh_k^\dag\mRIuk{k}^{-\dag}\vh_k \:=\: \rhomkNmc
\end{align}

We now seek expressions for each of the variables which enter the interference power, without using the preceding simplification. Firstly, note that in extension to (\ref{eq:invsn})--(\ref{eq:vuRtnvv}),
\begin{align}
\vv_n^\dag\mRItn{n}^{-\dag}\mRItn{n}^{-1}\vv_n &= \abs{z}^{-2} \label{eq:vRRv_mmse} \\
\vv_n^\dag\mXmseItn{j}\vu_n \:=\: \vu_n^\dag\mXmseItn{j}\vv_n &= 0 \quad,\quad j=1,2,3. \label{eq:vvnXjtnvun}
\end{align}
for $n=1\ldots N$.

Considering terms which arise in $\rhok{j}^N$, defined in (\ref{eq:def_rhon_tr}), we have from (\ref{eq:Rinv_vvn_mmse}) and (\ref{eq:mil_mmse}), and additionally using (\ref{eq:vRRv_mmse})--(\ref{eq:vvnXjtnvun}),
\begin{align}
\frac{1}{N}\tr{\mH^\ddag\mXmse{j}} &= \begin{cases}
\frac{1}{N}\sum_{n=1}^{\beta^* N} d_n^2 \vv_n\mXmse{j}\vv_n &,\quad j=2,3, \\
\frac{1}{N} \sum_{k=1}^K P_k \vh_k^\dag \mR^{-\dag}\mH^\ddag \mR^{-1} \vh_k &,\quad j=4.
\end{cases}\\
&\asymp \begin{cases}
\beta^*\abs{z}^{-2}(\sH_{1,2}^N + \taumtNm\sH_{2,2}^N) &,\quad j=2, \\
\beta^*\abs{z}^{-2}(1 + \taumtNbm)\sH_{2,2}^N &,\quad j=3, \\
\alpha \rhomtNbm \sE_{1,2}^N  &,\quad j=4.
\end{cases} \label{eq:rhokj_derive_iid}
\end{align}
which corresponds to $\rhok{j}^N$ for \iid $\mS$. Additionally note that
\begin{align}
\frac{1}{N}\tr{(\mH\mS)^\ddag\mXmse{j}} &=
\frac{1}{N} \sum_{k=1}^K \vh_k^\dag\mXals{j}\vh_k
\:\asymp\: \begin{cases}
\alpha \rhomtNm \sE_{0,2}^N &,\quad j=2, \\
\alpha \rhomtNbm \sE_{0,2}^N &,\quad j=3, \\
\alpha (\rhomtNcm \sE_{0,2}^N  + \abs{\rhomNm}^2 \sE_{1,2}^N)  &,\quad j=4.
\end{cases} \label{eq:rhokj_derive_iso}
\end{align}
Combining (\ref{eq:rhokj_derive_iid}) and (\ref{eq:rhokj_derive_iso}) according to (\ref{eq:def_rhon_tr}) for isometric $\mS$ gives
\begin{align}
\rhok{j}^N &\asymp \begin{cases}
\displaystyle \frac{\beta^*\abs{z}^{-2}(\sH_{1,2}^N + \taumtNm\sH_{2,2}^N)}{\alpha(\sE_{0,2}^N-1)+1} &,\quad j=2, \\
\displaystyle \frac{\beta^*\abs{z}^{-2}(1 + \taumtNbm)\sH_{2,2}^N}{\alpha(\sE_{0,2}^N-1)+1} &,\quad j=3, \\
\displaystyle \frac{\alpha(\rhomtNbm \sE_{1,2}^N - \abs{\rhomNm}^2\sE_{1,2}^N)}{\alpha(\sE_{0,2}^N-1)+1} &,\quad j=4.
\end{cases}
\end{align}

%\begin{align}
%\rhomtNm &= \frac{1}{N}\sum_{n=1}^{\beta^* N} d_n^2 \vv_n\mR^{-\dag}\mR^{-1}\vv_n  \:\asymp\: \beta^*\abs{z}^{-2}(\sH_{1,2}^N + \taumtNm\sH_{2,2}^N)
%\end{align}
%\begin{align}
%\rhomtNbm &= \frac{1}{N}\sum_{n=1}^{\beta^* N} d_n^2 \vv_n\mR^{-\dag}(\mHItn{n}^\ddag + d_n^2\vv_n\vv_n^\dag)\mR^{-1}\vv_n
%\:\asymp\: \beta^*\abs{z}^{-2}(1 + \taumtNbm)\sH_{2,2}^N
%\end{align}
%\begin{align}
%\rhomtNcm &= %\frac{1}{N}\tr{\mH\mS\mA^2\mS^\dag\mH^\dag\mR^{-1}\mH^\ddag\mR^{-1}} \:\asymp\:
%\frac{1}{N} \sum_{k=1}^{K} P_k \vh_k^\dag \mR^{-\dag}\mH^\ddag \mR^{-1} \vh_k \:\asymp\:
%\alpha \rhomtNbm \sE_{1,2}^N
%\end{align}

%Now, with isometric $\mS$, like (\ref{eq:rhomNm_mmse_iso}), we have
%\begin{align}
%%\rhomtm &\asymp \frac{1}{1-\alpha}( \beta^*z^{-2}(\sH_{1,2} + \taumtm\sH_{2,2}) -
%%\alpha\rhomtm\sE_{0,2})
%\rhomtNm &\asymp \frac{\beta^*\abs{z}^{-2}(\sH_{1,2}^N + \taumtNm\sH_{2,2}^N)}{\alpha(\sE_{0,2}^N-1)+1}
%\end{align}

%whereas with isometric $\mS$, we have that
%\begin{align}
%\alpha \rhomtNbm \sE_{1,2}^N - (1-\alpha)\rhomtNcm &\asymp \alpha\frac{1}{K} \sum_{k=1}^K
%\vh_k^\dag\mR^{-\dag}(\mH\mSIuk{k}\mAIuk{k}^2\mSIuk{k}^\dag\mH^\dag + P_k\vh_k\vh_k^\dag) \mR^{-1} \vh_k \\
%&\asymp \alpha(\rhomtNcm\sE_{0,2}^N + \abs{\rhomNm}^2\sE_{1,2}^N)
%\end{align}
%or equivalently,
%\begin{align}
%\rhomtNcm &\asymp \frac{\alpha(\rhomtNbm \sE_{1,2}^N - \abs{\rhomNm}^2\sE_{1,2}^N)}{\alpha(\sE_{0,2}^N-1)+1} .
%\end{align}
%
%With isometric $\mS$
%\begin{align}
%\rhomtNbm &\asymp \frac{\beta^*\abs{z}^{-2}(1 + \taumtNbm)\sH_{2,2}^N}{\alpha(\sE_{0,2}^N-1)+1} .
%\end{align}

Considering terms which arise in $\taun{j}^N$, defined in (\ref{eq:def_taun_tr}), we have from (\ref{eq:mil_mmse})
\begin{align}
\frac{1}{N} \tr{(\mH\mS\mA^2)^\ddag\mXmse{j}}
&= \frac{1}{N} \sum_{k=1}^K P_k^2 \vh_k^\dag \mXmse{j} \vh_k \:\asymp\: \alpha \rhok{j}^N \sE_{2,2}^N  \label{eq:taumtNm_derive}
\end{align}
for $j=2,3$, which corresponds to $\taun{j}^N$ for \iid $\mS$. Additionally, note that
\begin{align}
\frac{1}{N} \tr{(\mH\mS\mA^2\mS^\dag)^\ddag\mXmse{j}} &= \frac{1}{N}\sum_{n=1}^N (\vu_n + d_n c_n\vv_n)^\dag\mXmse{j}(\vu_n + d_n c_n\vv_n) \\
&\asymp \begin{cases}
\taumtm(\beta^*(\sH_{0,2}^N-1)+1) + \beta^*\abs{z}^{-2}\abs{\alpha\EP-\taumNm}^2\sH_{1,2}^N &,\quad j=2, \\
\taumtNbm(\beta^*(\sH_{0,2}^N-1)+1) + \beta^*\abs{z}^{-2}\abs{\alpha\EP-\taumNm}^2\sH_{2,2}^N &,\quad j=3.
\end{cases} \label{eq:taumtNm_derive_iso}
\end{align}
Now for isometric $\mS$, the left hand side of (\ref{eq:taumtNm_derive}) and (\ref{eq:taumtNm_derive_iso}) are equal since $\mS^\dag\mS=\mI_K$. Hence equating these expressions and solving for $\taun{j}^N$ gives
\begin{align}
\taun{j}^N &\asymp \begin{cases}
\displaystyle \frac{\alpha\rhomtNm\sE_{2,2}^N-\beta^*\abs{z}^{-2}\abs{\alpha\EP-\taumNm}^2\sH_{1,2}^N}{\beta^*(\sHm_{0,2}^N-1) + 1} &,\quad j=2, \\
\displaystyle \frac{\alpha\rhomtNbm\sE_{2,2}^N-\beta^* \abs{z}^{-2}\abs{\alpha\EP-\taumNm}^2\sH_{2,2}^N}{\beta^*(\sH_{0,2}^N-1) + 1} &,\quad j=3.
\end{cases}
\end{align}

It follows from the above and Theorem \ref{th:unified_mmse}, that along a realization for which (\ref{eq:sum_to_exp_P}), (\ref{eq:sum_to_exp_H}), (\ref{eq:rhokj_derive_iid}), (\ref{eq:rhokj_derive_iso}), (\ref{eq:taumtNm_derive}), and (\ref{eq:taumtNm_derive_iso}) hold, $\abs{\rhok{j}^N - \rhok{j}}\to 0$, $j=2,3,4$, and $\abs{\taun{j}^N - \taun{j}}\to 0$, $j=2,3$, where $\rhok{j}$ and $\taun{j}$ are solutions to (\ref{eq:rhomtm_iid})--(\ref{eq:taumtbm_iso}). In addition, since $\rhomtkNcm + \sn\rhomtkNm \uasymp{k} \rhomtNcm + \sn\rhomtNm$ and $\rhomkNm \uasymp{k} \rhomNm$, the asymptotic SINR is given by (\ref{eq:alternate_mmse_sinr}).

\section{Proof of (\ref{eq:dist_W}): Convergence of the \edf of $\mW$ for Exponential Weighting}
\label{ap:dist_W} %

With exponential weighting, the \edf of the $i\times i$ matrix $\mW$ corresponds to the distribution of a random variable, $W_i$, which is uniformly distributed on the set $\{\epsilon^{i(1-j/i)} : j=1 \lldots i\}$, where $\epsilon = (1 - \frac{\eta}{\Leff i})$ is the exponential weighting constant. The corresponding distribution function is $F_{W_i}(w)$. To prove convergence in distribution of the \aed of $\mW$ to $F_W(w)$, we show that $\lim_{i\to\infty} F_{W_i}(w) = F_W(w)$ where $F_W(w)$ is given in (\ref{eq:dist_W}), \ie, pointwise convergence. We have
\begin{align}
F_{W_i}(w) &= \Prob{W_i \leq w} =
\Prob{ \epsilon^{i(1-J_i)} \leq w} \:=\:
\PROB{ J_i \leq  1 - \frac{\ln w}{i\ln \epsilon} } \\
&= \frac{1}{i}\sum_{j=1}^i u\( \frac{j}{i} - 1 + \frac{\ln w}{i\ln \epsilon}\) \label{eq:probWi_lt_w}
\end{align}
for $\epsilon^i \leq w < 1$, where $J_i$ is a discrete random variable uniformly distributed on the set $\{j/i \; : j=1 \lldots i\}$, and where $u(t)$ is the step function, \ie, $u(t)$ is zero for $t<0$ and unity for $t \geq 0$. Now,
\begin{align}
\lim_{i\to\infty} i\ln \epsilon &= \lim_{i\to\infty} \frac{\log(1-\frac{\eta}{i\Leff})}{i^{-1}} \:=\: \frac{\lim_{i\to\infty} \frac{\partial}{\partial i} \log(1-\frac{\eta}{i\Leff})}{\lim_{i\to\infty} \frac{\partial}{\partial i}i^{-1}} \:=\: \frac{\lim_{i\to\infty} \frac{\eta}{\Leff i^2}\frac{\Leff}{\Leff - \frac{\eta}{i}}}{\lim_{i\to\infty} -i^{-2}} \:=\: -\eta/\Leff \nonumber
\end{align}
and similarly, the limit of the lower bound on $w$ simplifies to $\lim_{i\to\infty} \epsilon^i = e^{-\eta/\Leff}$.
Taking the limit of the Riemann sum in (\ref{eq:probWi_lt_w}) then gives
\begin{align}
F_W(w) &= \lim_{i\to\infty} F_{W_i}(w) = \int_{1 + \frac{\Leff}{\eta}\ln w}^1 \text{d}t \:=\: 1 + \frac{\Leff}{\eta}\ln w
\end{align}
where we have also used $\lim_{i\to\infty} F_{W_i}(w=1)=1$. This establishes that as $i\to\infty$, $F_i(w)$ converges in distribution to $F(w) = 1+\frac{\Leff}{\eta}\ln w$ for $e^{-\eta/\Leff}<w<1$.

%\subsection{Old Proof}
%Recall that for $i$ training symbol intervals and $\Leff < \infty$ (\ie, $\mW\neq\mI$),
%$\mW=\diag\(w^{i-1}, w^{i-2} \lldots w, 1\)$ and $w=1-\frac{1}{N\Leff}$. We may consider the powers of
%$w$ in $\mW$ as being uniformly distributed on $[0,i)$, which is true asymptotically. Define $X_N$
%as the scalar r.v.\ with this distribution, \ie, $\Prob{X_N<x}=x/i$ for $0\leq x<i$. Also define
%$W_N=w^{X_N}$, and
%\begin{align}
%\Prob{W_N<y} &= \Prob{w^{X_N}<y} \:=\: \Prob{X_N>\log{y}/\log{w}} \\
%&= 1 - \frac{\log{y}}{i \log(1-\frac{1}{N\Leff})} \;, \hspace{5mm}w^i \leq y < 1
%\label{eq:PrWNlty_unsimp}
%\end{align}
%We seek the \pdf of $W = \lim W_N$, and so examine the denominator of the second term in
%(\ref{eq:PrWNlty_unsimp}) asymptotically, to obtain
%\begin{align}
%\lim \frac{\log(1-\frac{\eta}{i\Leff})}{i^{-1}} &= \frac{\lim \frac{\partial}{\partial i}
%\log(1-\frac{\eta}{i\Leff})}{\lim \frac{\partial}{\partial i}i^{-1}} \:=\: \frac{\lim
%\frac{\eta}{\Leff i^2}\frac{\Leff}{\Leff - \frac{\eta}{i}}}{\lim -i^{-2}} \:=\: -\eta/\Leff
%\end{align}
%Also, $w^i \to e^{-\eta/\Leff}$, and thus
%\begin{align}
%\Prob{W < y} &= \lim \Prob{W_N<y} \:=\: 1+\frac{\Leff}{\eta}\log y\;,\hspace{5mm}
%e^{-\eta/\Leff}<y<1 \label{eq:cdf_W}
%% f_W(y) &= \frac{\Leff}{\eta}\frac{1}{y}\;,\hspace{5mm} e^{-\eta/\Leff}<y<1
%\end{align}
%Differentiating (\ref{eq:cdf_W}) gives the result in (\ref{eq:dist_W}).

\section{Proof of Theorem \ref{th:unified_als}}
\label{ap:proof_unified_als}

As in Appendix \ref{ap:mmse_proofs}, the analysis in this appendix is based on removing a single dimension from matrices and vectors, as described in Section \ref{sec:preamble}. We will use $t_n$ when removing the $n\thh$ \emph{transmit} dimension, $d_k$ when removing the $k\thh$ \emph{data} stream, and $r_m$ for removing the $m\thh$ \emph{received} symbol interval.

We define $\uasymp{k}$, $\uasymp{n}$, and $\uasymp{m}$ according to Definition \ref{df:uasymp} in Appendix \ref{ap:precursor}, where the maximum is over $k\leq K$, $n\leq N$, and $m\leq i$, respectively, and the limit is as $(M,N,K,i)\to\infty$ with $K/N\to\alpha>0$, $i/N\to\eta>0$, and $M/N\to\beta>0$ constant, as described in Section \ref{sec:limit}.

\subsection{Definitions}
\label{sec:als_var_defns}

As in Appendix \ref{ap:mmse_proofs}, throughout this appendix, we substitute $\mH$ with $\mV\mD$ without loss of generality, where $\mV$ is an $M\times M$ Haar-distributed unitary matrix, $\mD$ is a diagonal $M\times N$ matrix containing the singular values of $\mH$, and define $\vv_n$ and $d_n$.  The justification for this substitution in this case will be established later in Proposition \ref{pr:HisVDals}.

Define the following quantities
\begin{align}
\rmn{j,m}^N &= \frac{1}{i}\vr_m^\dag\mXalsIrm{j}\vr_m &\text{ , for }j&=1\lldots 4,& 0&<m\leq i \label{eq:def_rjm} \\
\gmn{j,m}^N &= \tvn_m^\dag\mXalsIrm{j}\tvn_m &\text{ , for }j&=1,\lldots 4.,& 0&<m\leq i \label{eq:def_gmn} \\ % used to be \frac{1}{\sn M}
\omegmn{j,m}^N &= \vomega_m \mXalsIrm{j}\vomega_m &\text{ , for } j&=1\lldots 4,& 0&<m\leq i \label{eq:def_omegmn} \\
\rhomn{j,k}^N &= \vh_k^\dag\mXalsIuk{j}\vh_k &\text{ , for }j&=1\lldots 4,& 0&<k\leq K \label{eq:def_rhomn} \\
\psimn{j,k}^N &= \vq_k^\dag\mXalsIuk{j}\vq_k &\text{ , for }j&=1\lldots 4,& 0&<k\leq K \label{eq:def_psimn} \\
\taumn{j,n}^N &= \vtau_n^\dag \mXalsItn{j}\vtau_n &\text{ , for }j&=1,2,3,& 0&<n\leq N \label{eq:def_taumn} \\
\nmn{j,n}^N &= \vv_n^\dag \mXalsItn{j} \vv_n &\text{ , for }j&=1,2,3,& 0&<n\leq \beta^*N \label{eq:def_nmn}
\end{align}
where
\begin{align}
\tvn_m &= \frac{1}{\sqrt{i}}\vn_m \\
\vh_k &= \mH\vs_k \\
\vq_k &= \frac{1}{i}\sRIuk{k}\mW\vub_k \\
\vtau_n &= \frac{1}{i}\sRItn{n}\mW\mB\mA\tvs_n \\
\vomega_m &= \frac{1}{\sqrt{i}}\mH\mS\mA\vb_m
\end{align}
and
\begin{align}
\mXals{j} &= \begin{cases}
\mhR^{-1} & \text{, $j = 1$,}\\
\mhR^{-\dag}\mhR^{-1} & \text{, $j = 2$,}\\
\mhR^{-\dag}\mH^\ddag\mhR^{-1} & \text{, $j = 3$,} \\
\mhR^{-\dag}(\mH\mS\mA)^\ddag\mhR^{-1} & \text{, $j = 4$}
\end{cases}
%\mXals{1} &= \mhR^{-1} \\
%\mXals{2} &= \mhR^{-2} \\
%\mXals{3} &= \mhR^{-1}\mH^\ddag\mhR^{-1} \\
%\mXals{4} &= \mhR^{-1}\mH\mS\mA^2\mS^\dag\mH^\dag\mhR^{-1}
\end{align}
Further, we define $\mXalsIrm{j}$, $\mXalsIuk{j}$, and $\mXalsItn{j}$ by removing the contribution of the $m\thh$ training symbol, the $k\thh$ data stream, and the $n\thh$ transmit dimension, respectively, from $\mXals{j}$
as follows:
\begin{itemize}
\item To remove the contribution of the $m\thh$ received training symbol interval, for some $0<m\leq i$, replace $\mhR$ and $\sR$ with
$\mhRIrm{m}$ and $\sRIrm{m}$, respectively, where
\begin{align}
\mhRIrm{m} &= \sRIrm{m}\mW_{r_m}\sRIrm{m}^\dag -z\mI_M \label{eq:def_mhRIrm}
\end{align}
and $\sRIrm{m}$ and $\mW_{r_m}$ are $\sR$ and $\mW$ with the $m\thh$ column, and $m\thh$ row and column removed, respectively. That is, $\mhRIrm{m} = \mhR - \frac{w_m}{i}\vr_m^\ddag$.

%Define $\mhRIrm{m}$ as $\mhR$ with the effect of the $m\thh$ received training symbol interval removed, for some $0<m\leq i$. That is, $\mhRIrm{m} = \sRIrm{m}\mW\sRIrm{m}^\dag -z\mI_M$ where $\sRIrm{m}$ is $\sR$ with the $m\thh$ column removed. %

\item To remove the contribution of the $k\thh$ data stream, for some $0<k\leq K$, replace $\mhR$, $\sR$, $\mS$, $\mA$, and $\mB$ with $\mhRIuk{k}$, $\sRIuk{k}$, $\mSIuk{k}$, $\mAIuk{k}$, and $\mBIuk{k}$, respectively, where
\begin{align}
\mhRIuk{k} &= \frac{1}{i}\sRIuk{k}\mW\sRIuk{k}^\dag -z\mI_M \\
\sRIuk{k} &= \mH\mSIuk{k}\mAIuk{k}\mBIuk{k}^\dag + \mN
\end{align}
and $\mSIuk{k}$ and $\mBIuk{k}$ are $\mS$ and $\mB$ with their $k\thh$ columns removed, respectively, and $\mAIuk{k}$ is $\mA$ with the $k\thh$ column and row removed.

%Define  $\mhRIuk{k}$as $\mhR$ with the effect of the $k\thh$ data stream removed, for some $0<k\leq K$. That is, $\mhRIuk{k}=\frac{1}{i}\sRIuk{k}\mW\sRIuk{k}^\dag -z\mI_M$, where $\sRIuk{k}=\mH\mSIuk{k}\mAIuk{k}\mBIuk{k}^\dag + \mN$, while $\mSIuk{k}$ and $\mBIuk{k}$ are $\mS$ and $\mB$ with their $k\thh$ columns removed, respectively, and $\mAIuk{k}$ is $\mA$ with the $k\thh$ column and row removed. %

\item To remove the contribution of the $n\thh$ transmit dimension, for some $0<n\leq \beta^*N$, replace $\mhR$, $\sR$, $\mH$, and $\mS$ with $\mhRItn{n}$, $\sRItn{n}$, $\mHItn{n}$, $\mHItn{n}^\ddag$, and $\mSItn{n}$, respectively, where
\begin{align}
\mhRItn{n} &= \frac{1}{i}\sRItn{n}\mW\sRItn{n}^\dag -z\mI_M \\
\sRItn{n} &= \mHItn{n}\mSItn{n}\mA\mB^\dag + \mN \\
\mHItn{n} &= \mVItn{n}\mDItn{n},
\end{align}
and $\mVItn{n}$ and $\mSItn{n}$ are $\mV$ and $\mS$ with the $n\thh$ column and row removed, respectively. Also, $\mDItn{n}$ is $\mD$ with the $n\thh$ column and row removed.

%Define $\mhRItn{n}$ as $\mhR$ with the effect of the $n\thh$ transmit dimension removed, for some $0<n\leq N$. That is, $\mhRItn{n} = \frac{1}{i}\sRItn{n} \mW \sRItn{n}^\dag - z\mI_M$ where $\sRItn{n} = \mHItn{n}\mSItn{n}\mA\mB^\dag + \mN$, and $\mHItn{n}=\mVItn{n}\mDItn{n}$ (according to Proposition \ref{pr:HisVD}), and $\mVItn{n}$ and $\mSItn{n}$ are $\mV$ and $\mS$ with the $n\thh$ column and row removed, respectively. Also, $\mHItn{n}^\ddag = \mHItn{n}\mHItn{n}^\dag$ and $\mDItn{n}$ is $\mD$ with the $n\thh$ column and row removed.
\end{itemize}

Considering (\ref{eq:def_gmn})--(\ref{eq:def_nmn}) asymptotically, it can be shown using the same steps as in the proof of (\ref{eq:def_rhon_tr}) and (\ref{eq:def_taun_tr}) in Appendix \ref{ap:uniform_conv_mmse} that
\begin{align}
\frac{\eta}{\beta\sn}\gmn{j,m}^N &\uasymp{m} \gmn{j}^N \:=\: \frac{1}{M}\tr{\mXals{j}} \label{eq:def_gmn_tr} \\
\rmn{j,m}^N &\uasymp{m} \rmn{j}^N \:=\: \omegmn{j} + \sn\frac{\beta}{\eta}\gmn{j}^N \label{eq:rdagRhatr_t1_pde} \\
\omegmn{j,m}^N &\uasymp{m} \omegmn{j}^N \:=\: \begin{cases} \label{eq:def_omegmn_tr} %
\frac{1}{i}\tr{(\mH\mS\mA)^\ddag\mXals{j}} & \iidB \\
%\frac{1}{i}\vb_m^\dag\mA^\dag\mS^\dag\mH^\dag\mhRIrm{m}^{-1}\mH\mS\mA\vb_m & \isoBan \\
\frac{1}{i}\sum_{m=1}^i \omegmn{j,m}^N & \isoBan \\
\frac{K}{i(K-i)}\tr{\mUpsilon\mA^\dag\mS^\dag\mH^\dag\mXals{j}\mH\mS\mA} & \isoBna,
\end{cases}
\end{align}
and
\begin{align}
\rhomn{j,k}^N &\uasymp{k} \rhomn{j}^N \:=\: \begin{cases} \label{eq:def_rhomn_tr} %
\frac{1}{N} \tr{\mH^\ddag\mXals{j}} & \iidS \\
\frac{1}{N-K} \tr{\mPi\mH^\dag\mXals{j}\mH} & \isoS,
\end{cases} \\
\psimn{j,k}^N &\uasymp{k} \psimn{j}^N \:=\: \begin{cases} \label{eq:def_psimn_tr} %
\frac{1}{i^2}\tr{(\sR\mW)^\ddag\mXals{j}} & \iidB \\
\frac{1}{i(i-K)}\tr{\mUpsilon\mW\sR^\dag\mXals{j}\sR\mW} & \isoBan \\
\frac{1}{K}\sum_{k=1}^K \psimn{j,k}^N & \isoBna,
%\frac{1}{i^2}\vub_k^\dag\mW\sRIuk{k}^\dag\mhRIuk{k}^{-1}\sRIuk{k}\mW\vub_k &\isoBna
\end{cases}
\end{align}
and
\begin{align}
\taumn{j,n}^N &\uasymp{n} \taumn{j}^N \:=\: \begin{cases}  \label{eq:def_taumn_tr}
\frac{1}{Ni^2} \tr{(\sR \mW \mB\mA)^\ddag \mXals{j}} & \iidS \\
\frac{1}{N}\sum_{n=1}^{N} \taumn{j,n}^N & \isoS,
\end{cases}\\
\nmn{j,n}^N &\uasymp{n} \nmn{j}^N \:=\: \frac{1}{\beta^* N}\sum_{n=1}^{\beta^* N} \nmn{j,n}^N \label{eq:def_nmnjn}
\end{align}
where $\mPi$ is defined in (\ref{eq:defn_mPi}), and
\begin{align}
\mUpsilon &= \begin{cases} %
\mI_i - \frac{1}{i}\mB\mB^\dag & , \; \alpha < \eta, \\
\mI_K - \frac{1}{K}\mB^\dag\mB & , \; \alpha > \eta.
\end{cases}
\end{align}
Although the derivations of the uniform asymptotic equivalence in (\ref{eq:def_rhomn_tr})--(\ref{eq:def_nmnjn}) are not shown, note that they rely on expressions derived later in this appendix, namely (\ref{eq:asympequivLSfilter_blind})--(\ref{eq:infN_dimK_den}) and (\ref{eq:Rinv_vvarrhon})--(\ref{eq:infN_dimN_den}).

% XXX % Also, (\ref{eq:def_rhomn_tr})--(\ref{eq:def_psimn_tr}) are conditioned on the application of Lemma \ref{lem:incsig_mhR} used to derive (\ref{eq:asympequivLSfilter_blind})--(\ref{eq:infN_dimK_den}), while (\ref{eq:def_taumn_tr})--(\ref{eq:def_nmnjn}) are conditioned on the application of Lemma \ref{lem:incsig_mhR} used to derive (\ref{eq:Rinv_vvarrhon})--(\ref{eq:infN_dimN_den}).

%The only exception being that $\rmn{j,m}^N \uasymp{m} \rmn{j}^N$ is shown using $$  % XXX bother to mention?

Note that (\ref{eq:skHHsk-EH}) again applies, and also
\begin{align}
\EW_k &\uasymp{k} \EW \label{eq:EWk_EW} \\
\hat{c}_n &\uasymp{n} \alpha\EW\EP \label{eq:hcn_uasymp_awp} \\
\frac{1}{i}\vr_m^\dag\vr_m & \uasymp{m} \alpha\beta^*\eta^{-1}\EP\Exp{H} + \sn\beta\eta^{-1} \:>\: 0 \label{eq:rmdrm_pos}
%\vomega_m^\dag \vomega_m & \uasymp{m}
\end{align}
for $0<k\leq K$, $0<n\leq N$, and $0<m\leq i$, and where
\begin{align}
\EW_k &= \frac{1}{i}\vub_k^\dag \mW \vub_k \\
\hat{c}_n &= \frac{1}{i}\tvs_n^\dag\mA\mB^\dag\mW\mB\mA\tvs_n
\end{align}
$\EP=\Exp{P}$, and $\EW = \Exp{W}$. (\ref{eq:EWk_EW}) is shown in identical manner to (\ref{eq:skHHsk-EH}) and (\ref{eq:tvsA2tvs-aEP}), in this case using (\ref{eq:sum_to_exp_W}). (\ref{eq:hcn_uasymp_awp}) can be shown in an identical manner to (\ref{eq:tvsA2tvs-aEP}), also using (\ref{eq:EWk_EW}), \ie,
\begin{align}
\hat{c}_n &\uasymp{n} \frac{1}{i N}\tr{\mW\mB\mA^2\mB^\dag} \:=\: \frac{1}{N}\sum_{k=1}^K P_k \EW_k \:\asymp\: \alpha \EW \EP
\end{align}
Finally, (\ref{eq:rmdrm_pos}) can be shown from $\frac{1}{i}\vr_m^\dag\vr_m \uasymp{m} \frac{1}{i}\tr{(\mH\mS\mA)^\ddag+\sn\mI_M}$ using (\ref{eq:sum_to_exp_P}), (\ref{eq:skHHsk-EH}), and $i\abs{\vn_m^\dag\vomega_m}^2 \uasymp{m} \tr{(\mH\mS\mA)^\ddag} \asymp \frac{\alpha\beta^*}{\eta}\Exp{P}\Exp{H}$.

%\begin{lemma} \emph{\citep{Yin88Bai}}
%If $\mS\in\CCNK$ contains \iid complex elements with mean zero, variance $\frac{1}{N}$, and finite fourth-order moment, then
%\begin{align}
%%\lambda_{\text{max}}(\mS^\ddag) &\asto \alpha(1 + 1/\sqrt{\alpha})^2
%\norm{\mS} &\asto 1 + \sqrt{\alpha}
%\end{align}
%as $(N,K)\to\infty$ with $K/N\to\alpha>0$.
%\end{lemma}

We now give several bounds on particular matrix and vector norms which are required in order to apply Lemmas \ref{lem:ax_by} and \ref{lem:ax_by_ext} later. Firstly, the assumption that $z\in\CC^+$ gives
\begin{align}
\norm{\mhR^{-1}} &\leq \imag(z)^{-1} \label{eq:normhRinv}
\end{align}
Recall that assumptions on $\mH$, $\mS$, and $\mA$ outlined in Section \ref{sec:sysmodel} again give (\ref{eq:def_Dmax})--(\ref{eq:supNabs_tvsn}), and additionally the assumptions on $\mW$,  $\mB$ and $\mN$ give
\begin{align}
&\sup_N \norm{\mW} < \infty \\
& \frac{1}{i}\norm{\mB}^2 \:\asto\: (1 + \sqrt{\alpha \eta^{-1}})^2 \quad{\text{, (\iid $\mB$)}}\label{eq:norm_BB} \\
& \frac{1}{i}\norm{\mN}^2 \:\asto\: \sn (1 + \sqrt{\alpha \eta^{-1}})^2 \label{eq:norm_NN} \\
& \frac{1}{i}\abs{\vub_k}^2 \:\uasymp{k}\: 1    \\
& \frac{1}{i}\abs{\vb_m}^2 \:\uasymp{m}\: \alpha \eta^{-1} \label{eq:supNmaxm_absvb} \\
& \frac{1}{i}\abs{\vn_m}^2 \:\uasymp{m}\: \beta \eta^{-1}  \label{eq:norm_nm}
\end{align}
where (\ref{eq:norm_BB})--(\ref{eq:norm_NN}), like (\ref{eq:norm_S}), is due to \cite{Yin88Bai}. Of course, $\frac{1}{N}\norm{\mB}^2 = \eta^*$ for orthogonal $\mB$. Moreover, (\ref{eq:def_Dmax})--(\ref{eq:supNabs_tvsn}), (\ref{eq:normhRinv})--(\ref{eq:norm_nm}) imply
\begin{align}
\sup_N &\frac{1}{i}\norm{\sR}^2 <\infty \as \\
\sup_N &\max_{k\leq K} \max\{ \abs{\vh_k}, \abs{\vq_k}, \EW_k \} < \infty  \as \label{eq:supN_dimKvecs}\\
\sup_N &\max_{n\leq N} \max\{ \abs{\vtau_n}, \abs{\hat{c}_n} \} < \infty  \as \label{eq:supN_dimNvecs}\\
\sup_N &\max\{ \abs{\rmn{j}^N}, \max_{m\leq i} \abs{\rmn{j,m}^N} \} < \infty \as \quad, \text{ for } j=1\ldots 4, \label{eq:supN_rmi}\\
\sup_N &\max\{ \abs{\rhomn{j}^N}, \max_{k\leq K} \abs{\rhomn{j,k}^N} \} < \infty \as \quad, \text{ for } j=1\ldots 4, \label{eq:supN_rhomi}\\
\sup_N &\max\{ \abs{\psimn{j}^N}, \max_{k\leq K} \abs{\psimn{j,k}^N} \} < \infty \as \quad, \text{ for } j=1\ldots 4, \label{eq:supN_psimi} \\
\sup_N &\max\{ \abs{\taumn{j}^N},  \max_{n\leq N} \abs{\taumn{j,n}^N} \} < \infty \as \quad, \text{ for } j=1\ldots 3, \label{eq:supN_taumi}\\
\sup_N &\max\{ \abs{\nmn{j}^N},  \max_{n\leq N} \abs{\nmn{j,n}^N} \} < \infty \as \quad, \text{ for } j=1\ldots 3, \label{eq:supN_nmi}.
\end{align}
With the additional assumption that $\abs{z}<\infty$, we also have
\begin{align}
\sup_N \norm{\mhR} < \infty \as \label{eq:sup_normhR}
\end{align}

\subsection{Derivations}

Note that $\gmi^N = \frac{1}{M}\tr{\mhR^{-1}}$ is the \Stieltjes transform of the \edf of the eigenvalues of $\frac{1}{i}\sR\mW\sR^\dag$. The proof of Theorem \ref{th:unified_als} proceeds as described in Section \ref{sec:preamble}, by applying the matrix inversion lemma to remove the effect of the $m\thh$ training interval from $\mhR$, $0< m\leq i$, and applying Lemma \ref{lem:incsig_mhR} to remove the effect of the $k\thh$ data stream and $n\thh$ transmit dimension from $\mhR$, $0<k\leq K$, $0<n\leq \beta^*N$. The removal of each dimension generates a pair of variables; expressions for which are then derived. Additionally, these results are applied to expansions of the identity $\mhR\mhR^{-1}=\mI$ to yield relationships between $\Gals^N(z)$ and the variables generated.

\subsubsection{Expanding Dimension $i$}
\label{sec:ap_dimi}

%First consider the following expression for $n=1$
%\begin{align}
%\frac{1}{i}\vr_m^\dag\mhRIrm{m}^{-n}\vr_m &= \frac{1}{i}\(\vb_m^\dag\mA^\dag\mS^\dag\mH^\dag
%+ \vn_m^\dag\)\mhRIrm{m}^{-n}\(\mH\mS\mA\vb_m + \vn_m\)\\
%&\asymp \frac{1}{i}\vb_m^\dag\mA^\dag\mS^\dag\mH^\dag\mhRIrm{m}^{-n}\mH\mS\mA\vb_m +
%\frac{1}{i}\vn_m^\dag\mhRIrm{m}^{-n}\vn_m\\
%&\asymp \omegmn{n} + \frac{\beta}{\eta} \sn \hat{\gamma}_{n} \label{eq:rdagRhatr_t1_pde}
%\end{align}

From the matrix inversion lemma, (\ref{eq:def_rjm}), and (\ref{eq:def_mhRIrm}) we have that
\begin{align}
\mhR^{-1}\vr_m &= R_m \mhRIrm{m}^{-1}\vr_m \label{eq:Rrm_mil}
\end{align}
where $R_m = 1/(1 + w_m\rmmNi)$. In addition, since
\begin{align}
\vomega_m^\dag\mhRIrm{m}^{-1}\tvn_m &\uasymp{m} \tvn_m^\dag\mhRIrm{m}^{-1}\vomega_m \:\uasymp{m}\: 0 \label{eq:omegamhRvn}
\end{align}
due to $i \abs{\vomega_m^\dag\mhRIrm{m}^{-1}\tvn_m} \uasymp{m} \sn\omegmti$, we have from the matrix inversion lemma and (\ref{eq:omegamhRvn}) that
\begin{align}
%\mhR^{-1}\vomega_m &\asymp \mhRIrm{m}^{-1}\((1- w_m R_m \omegmmNi)\vomega_m - w_m R_m \omegmmNi \tvn_m\) \\
%\mhR^{-1}\tvn_m &\asymp \mhRIrm{m}^{-1}\( - w_m R_m \sn\beta\eta^{-1}\gmmNi \vomega_m + (1 - w_m R_m \sn\beta\eta^{-1}\gmmNi)\tvn_m\)
\mhR^{-1}\vomega_m &\asymp R_m \mhRIrm{m}^{-1}\((1 + w_m \gmmNi)\vomega_m - w_m \omegmmNi \tvn_m\) \label{eq:hRinv_vomegan}\\
\mhR^{-1}\tvn_m &\asymp R_m \mhRIrm{m}^{-1}\( - w_m \gmmNi \vomega_m + (1 + w_m \omegmmNi)\tvn_m\) \label{eq:hRinv_tvn}
\end{align}

Now, we expand the identity $\mhR^{-1}\mhR=\mI_M$ along dimension $i$ using (\ref{eq:Rrm_mil}).  We have that
\begin{align}
1 &= \frac{1}{M}\tr{\mhR^{-1}\mhR} \:=\: -z\gmNi + \frac{1}{M} \sum_{m=1}^i \frac{w_m \rmmNi}{1 + w_m\rmmNi} \label{eq:RRinv_dimi_unsimp}
\end{align}
In an identical manner to the proof of (\ref{eq:RRin_dimK_arg}),
\begin{align}
\frac{w_m \rmmNi}{1 + w_m\rmmNi} &\uasymp{m} \frac{w_m \rmNi}{1 + w_m\rmNi} \label{eq:mhRmhR_dimi_sumarg}
\end{align}
That is, we use (\ref{eq:rdagRhatr_t1_pde}), and note that (\ref{eq:rmdrm_pos}) and (\ref{eq:sup_normhR}) satisfy conditions (\ref{eq:uN_nz_lem}) and (\ref{eq:normXNbound}) of Lemma \ref{lem:uN_in_Cplus}, which gives a uniform positive lower bound on $\imag(\rmmNi)$ over $i$ and $m\leq i$. Hence $\rmmNi\in\CC^+$ almost surely, and from Lemma \ref{lem:sums_converge},
\begin{align}
1 + z\gmNi &\asymp \begin{cases} %
\frac{\eta}{\beta}(1 - \sW_{0,1}^N)  & \\
\frac{\eta}{\beta}\rmNi\sW_{1,1}^N &
\end{cases} \label{eq:hRhRinv_dimi}
\end{align}
where the expressions in the right hand side of (\ref{eq:hRhRinv_dimi}) are related via (\ref{eq:id_WEHa}), and
\begin{align}
\sW_{p,q}^N &= \frac{1}{i}\sum_{m=1}^i \frac{w_m^p}{(1 + w_m \rmNi)^q} . \label{eq:def_WmnN}
\end{align}

\subsubsection{Expanding Dimension $K$}
\label{ap:als_dimK}

In this case, we write
\begin{align}
\mhR &= \frac{1}{i}(\sRIuk{k} + A_k\vh_k\vub_k^\dag)\mW(\sRIuk{k} + A_k\vh_k\vub_k^\dag)^\dag -z\mI_M \\
%&= \mhRIuk{k} + \frac{1}{i}(A_k\sRIuk{k} \mW \vub_k \vh_k^\dag +  A_k\vh_k
%\vub_k^\dag \mW \sRIuk{k}^\dag + P_k\vh_k \vub_k^\dag \mW \vub_k \vh_k^\dag) \\
&= \mhRIuk{k} + A_k \vq_k \vh_k^\dag + A_k \vh_k \vq_k^\dag + \EW_k P_k \vh_k \vh_k^\dag \label{eq:R_as_RIplus}.
\end{align}
We now apply Lemma \ref{lem:incsig_mhR} to (\ref{eq:R_as_RIplus}), where $\mY_N$, $\mX_N$, $\vu_N$, $\vv_N$, and $c_N$ in the Lemma correspond to $\mhR$, $\mhRIuk{k}$, $\vh_k$, $A_k\vq_k$, and $P_k\EW_k$, respectively. Therefore, $u_N$ and $v_N$ in the Lemma correspond to $\rhomkNi$ and $P_k\psimkNi$, respectively. Note that
\begin{align}
\vh_k^\dag\mhRIuk{k}^{-1}\vq_k &\uasymp{k} \vq_k^\dag\mhRIuk{k}^{-1}\vh_k \:\uasymp{k}\: 0 \label{eq:hRqzero}
\end{align}
due to $N \abs{\vh_k^\dag \mhRIuk{k}^{-1} \vq_k}^2 \uasymp{k} \psimtkNbi$ and (\ref{eq:supN_psimi}) for \iid $\mS$, and similarly with isometric $\mS$, it can be shown $(N-K)\abs{\vh_k^\dag \mhRIuk{k}^{-1} \vq_k}^2$ has a finite uniform upper bound. And so, (\ref{eq:hRqzero}) with (\ref{eq:def_Pmax}) satisfies condition (\ref{eq:eps_to_zero}) of the Lemma.  Additionally, (\ref{eq:skHHsk-EH}), (\ref{eq:supN_dimKvecs}) and (\ref{eq:sup_normhR}) satisfy conditions (\ref{eq:uN_nz}) and (\ref{eq:vu_bounded}) of the Lemma.

%\begin{align}
%\abs{\vh_k^\dag \mhRIuk{k}^{-1} \vq_k}^2 &\uasymp{k}\begin{cases}
%\frac{1}{N}\psimtkNci & \iidS \\
%\frac{1}{N-K}(\psimtkNci - \vq_k^\dag\mhRIuk{k}^{-1}(\mH\mSIuk{k})^\ddag\mhRIuk{k}^{-1}\vq_k) & \isoS
%\end{cases}
%\end{align}

%due to \cite[Corollary 1]{Evans00Tse}\footnote{This is clear for \iid $\mS$ and $\mB$. For isometric $\mS$ it first requires writing $\vs$ as described in (\ref{eq:isosig_as_gaussian}). For isometric $\mB$ with $\alpha<\eta$ it requires writing $\vub_k$ in a similar manner to (\ref{eq:isosig_as_gaussian}), \ie, $\vub_k = \mUpsilon_k\vx/\abs{\mUpsilon_k\vx}$ where $\mUpsilon_k = \mUpsilon + \frac{1}{i}\vub_k\vub_k^\dag$. For isometric $\mB$ with $\alpha>\eta$ it requires writing $\vub_k$ in a similar manner to that described for $\tvs_n$ after (\ref{eq:tvsA2tvs-aEP}) and expressing the resulting expression in a manner similar to (\ref{eq:isosig_as_gaussian}).},

Therefore, from Lemma \ref{lem:incsig_mhR} we have
\begin{align}
\mhR^{-1}\vh_k &\asymp K_k \mhRIuk{k}^{-1}(\vh_k - A_k \rhomkNi \vq_k) \label{eq:asympequivLSfilter_blind} \\
\mhR^{-1}\vq_k &\asymp K_k \mhRIuk{k}^{-1}(- A_k\psimkNi \vh_k + (1+\EW_k P_k\rhomkNi)\vq_k) \label{eq:Rhatvq} \\
\inf_N & \min_{k\leq K} \abs{1-P_k\rhomkNi(\psimkNi-\EW_k)} \:>\: 0 \quad\text{, a.s.} \label{eq:infN_dimK_den}
\end{align}
where $K_k=1/(1-P_k\rhomkNi(\psimkNi-\EW_k))$.

%\begin{align}
%P_k\vh_k^\dag\mhRIuk{k}^{-1}\vh_k &\asymp P_k \rhomi \\
%\vq_k^\dag \mhRIuk{k}^{-1} \vq_k &\asymp \psimi ,
%\end{align}

Now, we expand the identity $\mhR^{-1}\mhR=\mI_M$ along dimension $K$ using (\ref{eq:asympequivLSfilter_blind})--(\ref{eq:Rhatvq}).  We have
\begin{align}
1 + z\gmNi &= \frac{1}{iM}\tr{\mhR^{-1}\mH\mS\mA\mB^\dag\mW\sR^\dag} + \frac{1}{iM}\tr{\mhR^{-1}\mN\mW\sR^\dag} \label{eq:splitKN}
\end{align}

Expanding the first term in (\ref{eq:splitKN}) with respect to dimension $K$ using (\ref{eq:asympequivLSfilter_blind})--(\ref{eq:Rhatvq}), we have
\begin{align}
\frac{1}{iM}\tr{\mhR^{-1}\mH\mS\mA\mB^\dag\mW\sR^\dag} &= \frac{1}{iM}\sum_{k=1}^K A_k \vub_k^\dag\mW\sR^\dag\mhR^{-1}\vh_k \\
% thesis &= \frac{1}{iM}\sum_{k=1}^K A_k \vub_k^\dag\mW(\sRIuk{k}+A_k\vh_k\vub_k^\dag)^\dag\mhR^{-1}\vh_k \\
% thesis &= \frac{1}{iM}\sum_{k=1}^K A_k (A_k\EW_k\vh_k + \vq_k)^\dag \mhR^{-1}\vh_k \\
&\asymp \frac{1}{M}\sum_{k=1}^K A_k K_k (A_k\EW_k\vh_k + \vq_k)^\dag\mhRIuk{k}^{-1}(\vh_k - A_k\rhomkNi\vq_k) \\
&\asymp \begin{cases} %
\frac{\alpha}{\beta}(1 - \sEa_{0,1}^N) & \\
-\frac{\alpha}{\beta}\rhomNi(\psimNi - \EW)\sEa_{1,1}^N &
\end{cases} \label{eq:hRhRinv_dimK_derive}
\end{align}
where we have used (\ref{eq:def_rhomn_tr}), (\ref{eq:def_psimn_tr}), (\ref{eq:EWk_EW}), (\ref{eq:hRqzero}), and (\ref{eq:infN_dimK_den}), and define
\begin{align}
\sEa_{p,q}^N &= \frac{1}{K}\sum_{k=1}^K \frac{P_k^p}{(1-P_k\rhomNi(\psimNi-\EW))^q} \label{eq:def_EamnN}
\end{align}

Now considering the second term in (\ref{eq:splitKN}), using (\ref{eq:Rrm_mil}), (\ref{eq:omegamhRvn}), and (\ref{eq:def_gmn_tr}) we see
\begin{align}
\frac{1}{iM}\tr{\mhR^{-1}\mN\mW\sR^\dag} &= \frac{1}{M}\sum_{m=1}^i \frac{w_m(\vomega_m + \tvn_m)^\dag\mhRIrm{m}^{-1}\tvn_m}{1 + w_m\rmmNi} \:\asymp\: \sn\gmNi\sW_{1,1}^N \label{eq:hRhR_inv_npart}
\end{align}

Combining (\ref{eq:splitKN}), (\ref{eq:hRhRinv_dimK_derive}), and (\ref{eq:hRhR_inv_npart}) gives
\begin{align}
\beta(1 - \gmNi/\nmNi) &\asymp \alpha(1 - \sEa_{0,1}^N) \:=\: -\alpha\rhomNi(\psimNi - \EW)\sEa_{1,1}^N \label{eq:RRinv_dimK_derive}
\end{align}

Note that up until this point, none of the analysis has relied on the fact that we have made the substitution of $\mH$ with $\mV\mD$. Therefore, we are in a position to state the following proposition, which shows that this substitution is valid. Note that, the analysis in the remainder of this appendix relies on this substitution having been made throughout.
\begin{proposition}
\label{pr:HisVDals}
For the model (\ref{eq:rcv_sig}), the distribution of both the \Stieltjes transform of the \eed of $\frac{1}{i}\sR\mW\sR^\dag$ and the asymptotically equivalent ALS SINR given in (\ref{eq:asympSINR}) are invariant to the substitution of $\mV\mD$ for $\mH$, where $\mV$ is an $M\times M$ Haar-distributed random unitary matrix, and $\mD$ is a $M\times N$ diagonal matrix containing the singular values of $\mH$.
\end{proposition}
\begin{proof}
Using (\ref{eq:asympequivLSfilter_blind})--(\ref{eq:Rhatvq}), we may derive the asymptotically equivalent form of the ALS SINR given in (\ref{eq:asympSINR}) (see Appendix \ref{ap:derivationSINR}), which depends on $\rhomn{j}^N$ and $\psimn{j}^N$, $j=1\lldots 4$. Now note that the distributions of $\gmi^N$, $\rhomn{j}^N$ and $\psimn{j}^N$, $j=1\lldots 4$ are unchanged by the substitution of $\mH$ by $\mV\mD$. That is, let $\mT$ be an independent $M\times M$ Haar-distributed random matrix. Then,
\begin{align}
\gmi^N = \frac{1}{M}\tr{\mhR^{-1}} \:=\: \frac{1}{N}\tr{\mT\mT^\dag\mR^{-1}} \:=\: \frac{1}{M}\tr{((\mT\sR)\mW(\mT\sR)^\dag-z\mI_M)^{-1}}
\end{align}
Note that $\mT\sR = \mT\mH\mS\mA\mB^\dag+\mT\mN$, and so, writing $\mH\mS = (\mT\mU_M)\mD(\mU_N^\dag\mS)$, where $\mU_M\mD\mU_N^\dag$ is the singular value decomposition of $\mH$, the unitary invariance of $\mT$, $\mS$, and $\mN$ implies the result for the \Stieltjes transform of $\mhR$. A similar treatment of $\rhomn{j}^N$ and $\psimn{j}^N$ gives the result for the asymptotically equivalent form of the ALS SINR given in the right-hand side of (\ref{eq:asympSINR}).
\end{proof}

\subsubsection{Expanding Dimension $N$}
\label{ap:als_dimN}

%We can obtain equivalent asymptotic representations for $\mhR^{-1}\vv_n$ and $\mhR^{-1}\vtau_n$ using Lemma \ref{lem:incsig_mhR}. We have
Writing
\begin{align}
\sR &= \sum_{n=1}^N d_n \vv_n (\mB\mA\tvs_n)^\dag + \mN \label{eq:sR_sumrhotau}
\end{align}
and substituting this into (\ref{eq:def_mhR}), in a similar manner to (\ref{eq:R_as_RIplus}), we have
\begin{align}
\mhR &= \mhRItn{n} + d_n \vv_n \vtau_n^\dag + d_n \vtau_n \vv_n^\dag +d_n^2\hat{c}_n \vv_n \vv_n^\dag \label{eq:R_RIplus_H} .
\end{align}
We now apply Lemma \ref{lem:incsig_mhR} to (\ref{eq:R_RIplus_H}), where $\mY_N$, $\mX_N$, $\vu_N$, $\vv_N$, and $c_N$ in the Lemma correspond to $\mhR$, $\mhRItn{n}$, $\vv_n$, $d_n\vtau_n$, and $d_n^2\hat{c}_n$, respectively. Therefore, $u_N$ and $v_N$ in the Lemma correspond to $\nmnNi$ and $d_n^2\taumnNi$, respectively. We have (\ref{eq:supN_dimNvecs}), (\ref{eq:sup_normhR}) and $\abs{\vv_n}=1$, which satisfy conditions (\ref{eq:uN_nz}) and (\ref{eq:vu_bounded}) of the Lemma. We now show
\begin{align}
\vv_n^\dag \mhRItn{n}^{-1} \vtau_n &\uasymp{n} \vtau_n^\dag \mhRItn{n}^{-1} \vv_n \:\uasymp{n}\: 0 \label{eq:vvnmhRItntvaun_zero}
\end{align}
which, with (\ref{eq:def_Dmax}) satisfies condition (\ref{eq:eps_to_zero}) of the Lemma.  To see this, note that $\vv_n^\dag\vtau_n \uasymp{n} 0$, since
\begin{align}
N \abs{\vv_n^\dag\vtau_n}^2 &= \frac{N}{i^2}(\vv_n^\dag\mN\mW\mB\mA\tvs_n)^\ddag \uasymp{n} \frac{1}{i^2 M} \tr{(\mN\mW\mB\mA)^\ddag}
%\asymp \frac{\sn}{i^2} \tr{(\mW\mB\mA)^\ddag} &\asymp \frac{\sn}{i^2} \sum_{k=1}^K P_k \vub_k^\dag\mW^2\vub_k
\asymp \frac{\sn \alpha}{\eta} \Exp{P}\Exp{W^2} \label{eq:vvntaun_zero}
\end{align}
Also, since $\mHItn{n}^\dag\vv_n = \mZero$, we have
\begin{align}
%\vtau_n &= \frac{1}{i}\sRItn{n}\mW\mB\mA\tvs_n \\
\mhRItn{n}\vv_n &= \frac{1}{i}\sRItn{n}\mW\mN^\dag\vv_n - z\vv_n \label{eq:hRItn_vvn}
\end{align}
From (\ref{eq:hRItn_vvn}), we find
\begin{align}
z\vtau_n^\dag\mhRItn{n}^{-1}\vv_n &\uasymp{n} \frac{1}{i}\vtau_n^\dag \mhRItn{n}^{-1}\sRItn{n}\mW\mN^\dag\vv_n \:=\: \frac{1}{i} \sum_{m=1}^i w_m \vn_m^\dag\vv_n\vtau_n^\dag \mhRItn{n}^{-1}\vr_m \label{eq:tauRinvvn_unsimp}
\end{align}
Considering the argument in the preceding sum, note that
\begin{align} % XXX have we made all the neccesary assumptions regarding HSAb and n and gamma, omega, r, in order for this derivation to be correct?
w_m\vn_m^\dag\vv_n\vtau_n^\dag \mhRItn{n}^{-1}\vr_m
%thesis
%&= \frac{w_m \vn_m^\dag\vv_n\vtau_n^\dag \mhR_{t_n,r_m}^{-1}\vr_m}{1 + \frac{w_m}{i}\vr_{m,t_n}^\dag\mhR_{t_n,r_m}^{-1}\vr_{m,t_n}}
\:\uasymp{m}\: \frac{w_m \sn \vtau_n^\dag \mhRItn{n}^{-1}\vv_n}{1 + w_m\rmn{1,t_n}^N} \label{eq:tauRv_sumarg}
\end{align}
where we define $\gmn{1,t_n}^N$, $\rmn{1,t_n}^N$, and $\omegmn{1,t_n}^N$ from $\gmNi$, $\rmNi$, and $\omegmNi$ in (\ref{eq:def_gmn_tr}), (\ref{eq:rdagRhatr_t1_pde}), and (\ref{eq:def_omegmn_tr}), respectively, with the contribution of transmit dimension $n$ removed, as explained in Appendix \ref{sec:als_var_defns}.
%thesis
%where $\mhR_{t_n,r_m} = \mhRItn{n} - \frac{w_m}{i}\vr_{m,t_n}\vr_{m,t_n}^\dag$, and $\vr_{m,t_n} = \mHItn{n}\mSItn{n}\mA\vb_m + \vn_m$, and we have used $\vtau_n^\dag\mhR_{t_n,r_m}^{-1}\vv_n \uasymp{m} \vtau_n^\dag\mhRItn{n}^{-1}\vv_n$, which can be shown using Lemma \ref{lem:lem2.6} and $\vv_n^\dag\vtau_n \uasymp{n} 0$.
Returning to (\ref{eq:tauRinvvn_unsimp}), and additionally using $1 + z\gmn{1,t_n}^N \asymp \rmn{1,t_n}^N \frac{1}{M}\sum_{m=1}^i \frac{w_m}{1 + w_m \rmn{1,t_n}^N}$, which is shown in an identical manner to (\ref{eq:hRhRinv_dimi}), it follows that
\begin{align}
%\vtau_n^\dag\mhRItn{n}^{-1}\vv_n\(z - \sn \sW_{1,1}\) &\uasymp{n} 0
\vtau_n^\dag\mhRItn{n}^{-1}\vv_n \frac{z \omegmn{1,t_n}^N}{\rmn{1,t_n}^N} &\uasymp{n} 0
\end{align}
Now, since $z$, $\omegmn{1,t_n}^N$, and $\rmn{1,t_n}^N$ are in $\CC^+$ almost surely (through an application of Lemma \ref{lem:uN_in_Cplus}), and also since $\abs{\rmn{1,t_n}^N}$ is uniformly bounded above over $N$ and $n$ (which is shown identically to (\ref{eq:supN_rmi})), we have (\ref{eq:vvnmhRItntvaun_zero}).

%\frac{1}{i}\tvs_n^\dag\mA\mB^\dag\mW\sRItn{n}^\dag

So we may apply Lemma \ref{lem:incsig_mhR} to (\ref{eq:R_RIplus_H}), to obtain
\begin{align}
\mhR^{-1}\vv_n &\asymp V_n \mhRItn{n}^{-1} (\vv_n - d_n \nmnNi\vtau_n) \label{eq:Rinv_vvarrhon} \\
\mhR^{-1}\vtau_n &\asymp V_n \mhRItn{n}^{-1} (-d_n\taumnNi \vv_n + (1 + d_n^2\hat{c}_n\nmi)\vtau_n) \label{eq:Rinv_vtaun} \\
\inf_N & \min_{n\leq N} \abs{1-d_n^2\nmnNi(\taumnNi-\hat{c}_n)} \:>\: 0 \quad\text{, a.s.}\label{eq:infN_dimN_den}
\end{align}
where $V_n = 1/(1-d_n^2\nmnNi(\taumnNi-\hat{c}_n))$.

%\subsubsection{Aside: \iid $\mH$}
%
%$\nmi = \gmi$ for \iid $\mH$
%
%$\nmti = \gmti$ for \iid $\mH$.

Now, we expand the identity $\mhR^{-1}\mhR=\mI_M$ along dimension $N$ using (\ref{eq:Rinv_vvarrhon})--(\ref{eq:Rinv_vtaun}). Continuing from (\ref{eq:splitKN}), we may expand the first term along dimension $N$ to obtain
\begin{align}
\frac{1}{iM}\tr{\mhR^{-1}\mH\mS\mA\mB^\dag\mW\sR^\dag} &= \frac{1}{iM} \sum_{n=1}^N d_n\tvs_n^\dag\mA\mB^\dag\mW\sR^\dag\mhR^{-1}\vv_n \\
% thesis &= \frac{1}{iM} \sum_{n=1}^N d_n \tvs_n^\dag\mA\mB^\dag\mW(\sRItn{n} + d_n\vv_n(\mB\mA\tvs_n)^\dag)^\dag\mhR^{-1}\vv_n \\
&\asymp \frac{1}{M} \sum_{n=1}^N d_n V_n(d_n\hat{c}_n\vv_n + \vtau_n )^\dag\mhRItn{n}^{-1}(\vv_n - d_n\nmnNi\vtau_n) \\
&\asymp \frac{1}{M} \sum_{n=1}^N d_n^2 V_n \nmnNi(\hat{c}_n - \taumnNi) \\
&\asymp \begin{cases} %
-\frac{\beta^*}{\beta}(\sHa_{0,1}^N-1) & \\
-\nmi(\taumNi - \alpha\EP\EW)\frac{\beta^*}{\beta}\sHa_{1,1}^N &
\end{cases} \label{eq:hRhRinv_dimN_derive}
\end{align}
where
\begin{align}
\sHa_{p,q}^N &= \frac{1}{\beta^*N}\sum_{n=1}^{\beta^*N}\frac{d_n^{2p}} {(1 - d_n^2 \nmNi(\taumNi-\alpha\EP\EW) )^q} \label{eq:def_HamnN}
\end{align}
and we have also used (\ref{eq:def_taumn_tr}), (\ref{eq:def_nmnjn}), (\ref{eq:hcn_uasymp_awp}), and (\ref{eq:infN_dimN_den}).

Combining (\ref{eq:splitKN}), (\ref{eq:hRhRinv_dimN_derive}), and (\ref{eq:hRhR_inv_npart}) gives
\begin{align}
\beta(1 - \gmNi/\nmNi) &= \beta^*(1-\sHa_{0,1}^N) \:=\: -\nmNi(\taumNi - \alpha\EP\EW)\beta^*\sHa_{1,1}^N \label{eq:RRinv_dimN_derive}
\end{align}

%\begin{align}
%\beta(1+z\gmi) &= \eta(1-\sW_{0,1}) \\
%\beta(1 - \gmi/\nmi) &= \eta \omegmi\sW_{1,1} \\
%&= \alpha(1 - \sEa_{0,1}) \:=\: -\alpha\rhomi(\psimi - \EW)\sEa_{1,1} \\
%&= \beta^*(1-\sHa_{0,1}) \:=\: -\nmi(\taumi - \alpha\EP\EW)\beta^*\sHa_{1,1}
%\end{align}

\subsubsection{Identities via moment definitions}

In order to find expressions for $\omegmNi$, $\rhomNi$, $\psimNi$, $\taumNi$, and $\nmNi$, we now apply (\ref{eq:Rrm_mil}), (\ref{eq:hRinv_vomegan})--(\ref{eq:hRinv_tvn}), (\ref{eq:asympequivLSfilter_blind})--(\ref{eq:Rhatvq}), and (\ref{eq:Rinv_vvarrhon})--(\ref{eq:Rinv_vtaun}) to expansions of the definitions of these variables, or expressions related to them. After applying (\ref{eq:Rrm_mil}) or (\ref{eq:hRinv_vomegan})--(\ref{eq:hRinv_tvn}), we shall use (\ref{eq:def_gmn_tr}), (\ref{eq:rdagRhatr_t1_pde}), (\ref{eq:def_omegmn_tr}), (\ref{eq:omegamhRvn}), and (\ref{eq:mhRmhR_dimi_sumarg}) in conjunction with Lemmas \ref{lem:ax_by_ext} and \ref{lem:sums_converge} to further simplify the resulting expression.  Similarly, after applying (\ref{eq:asympequivLSfilter_blind})--(\ref{eq:Rhatvq}), we shall use (\ref{eq:def_rhomn_tr}), (\ref{eq:def_psimn_tr}), (\ref{eq:supN_rhomi}), (\ref{eq:supN_psimi}), (\ref{eq:EWk_EW}), (\ref{eq:hRqzero}), and (\ref{eq:infN_dimK_den}) in conjunction with Lemmas \ref{lem:ax_by_ext} and \ref{lem:sums_converge} to further simplify the resulting expression. Finally, after applying (\ref{eq:Rinv_vvarrhon})--(\ref{eq:Rinv_vtaun}), we shall use (\ref{eq:def_taumn_tr}), (\ref{eq:def_nmnjn}), (\ref{eq:supN_taumi}), (\ref{eq:supN_nmi}), (\ref{eq:hcn_uasymp_awp}), (\ref{eq:vvnmhRItntvaun_zero}), and (\ref{eq:infN_dimN_den}) in conjunction with Lemmas \ref{lem:ax_by_ext} and \ref{lem:sums_converge} to further simplify the resulting expression.

Firstly, consider $\omegmNi$ in (\ref{eq:def_omegmn_tr}). From (\ref{eq:asympequivLSfilter_blind})--(\ref{eq:Rhatvq}) we obtain
\begin{align}
\frac{1}{i}\tr{(\mH\mS\mA)^\ddag\mhR^{-1}}
%&= \frac{1}{i}\sum_{k=1}^{K} P_k \vh_k^\dag\mhR^{-1}\vh_k\\
&\asymp \frac{1}{i} \sum_{k=1}^{K}  P_k K_k \vh_k^\dag\mhRIuk{k}^{-1}\(\vh_k - A_k \rhomNi \vq_k\)
\:\asymp\: \frac{\alpha}{\eta}\rhomNi\sEa_{1,1}^N \label{eq:rdagRhatr_t2_pde}
\end{align}
which corresponds to $\omegmNi$ for \iid $\mB$. For orthogonal $\mB$ with $\alpha<\eta$, we use (\ref{eq:rdagRhatr_t2_pde}) and $\frac{1}{i}\mB^\dag\mB=\mI_K$ to obtain
\begin{align}
\frac{\alpha}{\eta}\rhomNi\sEa_{1,1}^N &\asymp \frac{1}{i^2}\tr{(\mH\mS\mA\mB^\dag)^\ddag\mhR^{-1}}
\:\asymp\: \frac{1}{i}\sum_{m=1}^i \vomega_m^\dag \mhR^{-1}\vomega_m \\
&\asymp \omegmNi(1-\omegmNi\sW_{1,1}^N) \label{eq:omegmi_derive_iso}
\end{align}
which is proven in an identical manner to (\ref{eq:hRhRinv_dimi}), also using (\ref{eq:hRinv_vomegan}), and may be simplified using (\ref{eq:hRhRinv_dimi}) to give
\begin{align}
\omegmNi &\asymp \frac{\frac{\alpha}{\eta}\rhomNi\sEa_{1,1}^N}{1-\frac{\beta}{\eta}(1 - \gmNi/\nmNi)} \label{eq:omegmi_derive_isoae}
\end{align}
Now if $\alpha>\eta$, we have directly from (\ref{eq:rdagRhatr_t2_pde}) and (\ref{eq:omegmi_derive_iso})
\begin{align}
\omegmNi &= \frac{\alpha}{\eta}\frac{1}{K-i}\tr{\mUpsilon\mA^\dag\mS^\dag\mH^\dag\mhR^{-1}\mH\mS\mA} \\
&\asymp \frac{\alpha}{\alpha-\eta} \( \frac{\alpha}{\eta}\rhomNi\sEa_{1,1}^N -\frac{\eta}{\alpha} \omegmNi(1-\omegmNi\sW_{1,1}^N) \)
\end{align}
Combining this with (\ref{eq:hRhRinv_dimi}) gives
\begin{align}
\omegmNi &= \frac{\frac{\alpha}{\eta}\rhomNi\sEa_{1,1}^N}{1-\frac{\beta}{\alpha}(1 - \gmNi/\nmNi)} \label{eq:omegmi_derive_isoea}.
\end{align}

Now we consider $\taumNi$, defined in (\ref{eq:def_taumn_tr}), using (\ref{eq:asympequivLSfilter_blind})--(\ref{eq:Rhatvq}). Firstly, note that $\frac{1}{i}\sR \mW \vub_k = \frac{1}{i}(\sRIuk{k} + A_k \vh_k\vub_k^\dag)\mW \vub_k = \vq_k + A_k \EW_k \vh_k$, and hence
\begin{align}
\frac{1}{Ni^2} \tr{(\sR \mW \mB\mA)^\ddag\mhR^{-1}}
&= \frac{1}{Ni^2} \sum_{k=1}^K P_k \vub_k^\dag \mW\sR^\dag \mhR^{-1} \sR \mW\vub_k \\
&= \frac{1}{N} \sum_{k=1}^K P_k (\vq_k + A_k\EW_k \vh_k)^\dag \mhR^{-1} (\vq_k + A_k\EW_k \vh_k) \label{eq:tauRIinvtau} \\
&\asymp \alpha \frac{1}{K} \sum_{k=1}^K P_k K_k (\psimNi - \EW P_k \rhomNi (\psimNi-\EW)) \\
%&\asymp \alpha \EXp{ P \frac{\psimi-\EW + \EW(1 - P\rhomi
%(\psimi-\EW))}{1-P\rhomi(\psimi-\EW)} } \\
%&\asymp \alpha (\psimi-\EW) \sEa_{1,1} + \alpha \EW\EP \label{eq:def_taumi} \\
&\asymp \alpha (\EW\EP + (\psimNi-\EW)\sEa_{1,1}^N) \label{eq:def_taumi}
\end{align}
which corresponds to $\taumNi$ for \iid $\mS$. To find an expression for $\taumNi$ with isometric $\mS$, we use $\mS^\dag\mS=\mI_K$ and (\ref{eq:def_taumi}) in reverse, \ie,
\begin{align}
\alpha (\EW\EP + (\psimNi-\EW)\sEa_{1,1}^N)
&\asymp \frac{1}{Ni^2} \tr{(\sR\mW\mB\mA\mS^\dag)^\ddag\mhR^{-1}} \\
% thesis &= \frac{1}{N} \sum_{n=1}^N (\vtau_n + d_n\hat{c}_n\vv_n)^\dag\mhR^{-1}(\vtau_n + d_n\hat{c}_n\vv_n) \\
&\asymp \frac{1}{N} \sum_{n=1}^N V_n (\vtau_n + d_n\hat{c}_n\vv_n)^\dag\mhRItn{n}^{-1} (\vtau_n + d_n(\hat{c}_n-\taumnNi) \vv_n) \\
&\asymp \frac{1}{N} \sum_{n=1}^N \( \hat{c}_n + (\taumnNi - \hat{c}_n)V_n \) \\
&\asymp \alpha\EP\EW + (\taumNi - \alpha\EP\EW)(\beta^*(\sHa_{0,1}^N-1) + 1) \label{eq:taumi_derive}
\end{align}
where we have used $\frac{1}{i}\sR\mW\mB\mA\tvs_n = \frac{1}{i}(\sRItn{n}+d_n\vv_n\tvs_n)\mW\mB\mA\tvs_n = \vtau_n + d_n\hat{c}_n\vv_n$. Moreover, using (\ref{eq:RRinv_dimN_derive}) to simplify (\ref{eq:taumi_derive}) we obtain
\begin{align}
\taumNi &\asymp \alpha\EP\EW + \frac{\alpha(\psimNi-\EW)\sEa_{1,1}^N}{1-\beta(1-\gmNi/\nmNi)} \label{eq:def_taumi_iso}
\end{align}

Now we consider $\rhomNi$ defined in (\ref{eq:def_rhomn_tr}), using (\ref{eq:Rinv_vvarrhon})--(\ref{eq:Rinv_vtaun}). Firstly,
\begin{align}
\frac{1}{N}\tr{\mH^\ddag\mR^{-1}} &= \frac{1}{N}\sum_{n=1}^{\beta^*N} d_n^2 \vv_n^\dag\mhR^{-1}\vv_n \:\asymp\: \frac{1}{N}\sum_{n=1}^{\beta^*N} d_n^2 V_n \vv_n^\dag\mhRItn{n}^{-1}\vv_n \\
&\asymp \nmNi \beta^*\sHa_{1,1}^N \label{eq:rhomi_derive}
\end{align}
which corresponds to $\rhomNi$ with \iid $\mS$. Note that $\frac{1}{N}\tr{(\mH\mS)^\ddag\mhR^{-1}} \asymp \alpha\rhomNi\sEa_{0,1}^N$ follows from an analogous derivation to (\ref{eq:rdagRhatr_t2_pde}). Therefore, combining this with (\ref{eq:rhomi_derive}), we obtain for isometric $\mS$,
\begin{align}
\rhomNi &= \frac{1}{1-\alpha}\(\frac{1}{N}\tr{\mH^\ddag\mhR^{-1}}-\frac{1}{N}\tr{(\mH\mS)^\ddag\mhR^{-1}}\) \\
&\asymp \frac{1}{1-\alpha}(\beta^*\nmNi \sHa_{1,1}^N - \alpha\rhomNi\sEa_{0,1}^N)
\label{eq:rhomi_iso_derive}
\end{align}
Moreover, combining with (\ref{eq:RRinv_dimK_derive}) gives
\begin{align}
\rhomNi &= \frac{\beta^*\nmNi \sHa_{1,1}^N}{1-\beta(1-\gmNi/\nmNi)} \label{eq:rhomi_iso_derive_fin}
\end{align}

Now we consider $\psimNi$, as defined in (\ref{eq:def_psimn_tr}), using (\ref{eq:Rrm_mil}). Firstly,
\begin{align}
\frac{1}{i^2}\tr{(\sR\mW)^\ddag\mhR^{-1}} &= \frac{1}{i}\sum_{m=1}^i \frac{w_m^2 \rmmNi}{1+w_m\rmmNi}  \:\asymp\: \EW - \sW_{1,1}^N \label{eq:psimi_derive}
\end{align}
which corresponds to $\psimNi$ for \iid $\mB$. Note that $\frac{1}{N i^2}\tr{(\sR\mW\mB)^\ddag\mhR^{-1}} \asymp \alpha(\EW + (\psimNi-\EW)\sEa_{0,1}^N)$ from an analogous derivation to (\ref{eq:def_taumi}). Therefore, combining this with (\ref{eq:psimi_derive}), we obtain for orthogonal $\mB$ and $\alpha<\eta$,
\begin{align}
\psimNi &= \frac{\eta}{\eta-\alpha}\( \frac{1}{i^2}\tr{(\sR\mW)^\ddag\mhR^{-1}} - \frac{1}{i^3}\tr{(\sR\mW\mB)^\ddag\mhR^{-1}}\) \\
&\asymp \frac{\eta}{\eta-\alpha} (\EW - \sW_{1,1}^N - \frac{\alpha}{\eta}(\EW + (\psimNi-\EW)\sEa_{0,1}^N))
\end{align}
Moreover, combining with (\ref{eq:RRinv_dimK_derive}) gives
\begin{align}
\psimNi &= \EW - \frac{\sW_{1,1}^N}{1 - \frac{\beta}{\eta}(1 - \gmNi/\nmNi)} \label{eq:psimi_derive_iso_ae}
\end{align}
Finally, for orthogonal $\mB$ and $\alpha>\eta$, from (\ref{eq:psimi_derive}) and $\frac{1}{K}\mB^\dag\mB=\mI_i$, we have
\begin{align}
\EW - \sW_{1,1}^N &\asymp \frac{1}{K i^2}\tr{(\sR\mW\mB)^\ddag\mhR^{-1}} \\
&\asymp \EW + (\psimNi-\EW)\sEa_{0,1}^N
\end{align}
following an analogous derivation to (\ref{eq:def_taumi}). Moreover, combining with (\ref{eq:RRinv_dimK_derive}) gives
\begin{align}
\psimNi &\asymp \EW - \frac{\sW_{1,1}^N}{1-\frac{\beta}{\alpha}(1-\gmNi/\nmNi)} . \label{eq:psimi_derive_iso_ea}
\end{align}

To derive an expression for $\nmi$, we start with (\ref{eq:hRItn_vvn}) and obtain
\begin{align}
1 + z\nmnNi &= \frac{1}{i} \sum_{m=1}^i w_m \vn_m^\dag \vv_n \vv_n^\dag\mhRItn{n}^{-1}\vr_m \label{eq:nmi_unsimp}
\end{align}
for which, like (\ref{eq:tauRv_sumarg}), we have
\begin{align}
w_m\vn_m^\dag\vv_n\vv_n^\dag \mhRItn{n}^{-1}\vr_m &\uasymp{m} \frac{w_m \sn \nmnNi}{1 + w_m\rmNi} \label{eq:nmi_derive_arg}
\end{align}
where we have additionally used $\rmn{1,m,t_n}^N \uasymp{n} \rmn{1,m}^N \uasymp{m} \rmNi$, which follows from Lemma \ref{lem:trace_incsig_mhR}, continuing on from the application of Lemma \ref{lem:incsig_mhR} in Appendix \ref{ap:als_dimN}. Therefore, from (\ref{eq:nmi_unsimp}) and (\ref{eq:nmi_derive_arg}), we obtain
\begin{align}
1 + z\nmNi &\asymp \sn\nmNi\sW_{1,1}^N \label{eq:nmi_derive}
\end{align}

It follows that along a realization for which (\ref{eq:sum_to_exp_P})--(\ref{eq:sum_to_exp_W}), (\ref{eq:rdagRhatr_t1_pde}), (\ref{eq:hRhRinv_dimK_derive}), (\ref{eq:rdagRhatr_t2_pde}), (\ref{eq:omegmi_derive_isoae}), (\ref{eq:omegmi_derive_isoea}), (\ref{eq:def_taumi}), (\ref{eq:def_taumi_iso}), (\ref{eq:rhomi_derive}), (\ref{eq:rhomi_iso_derive_fin}), (\ref{eq:psimi_derive}), (\ref{eq:psimi_derive_iso_ae}), (\ref{eq:psimi_derive_iso_ea}), and (\ref{eq:nmi_derive}) hold, $\abs{\rhomNi - \rhomi} \asto 0$, $\abs{\taumNi - \taumi} \asto 0$, $\abs{\psimNi - \psimi} \asto 0$, $\abs{\omegmNi - \omegmi} \asto 0$, $\abs{\nmNi - \nmi} \asto 0$, and $\abs{\rmNi - \rmi} \asto 0$, where $\gmi$, $\rhomi$, $\taumi$, $\psimi$, $\omegmi$, $\nmi$, and $\rmi\in\CC^+$ are solutions to (\ref{eq:gmi_gen_als})--(\ref{eq:nmi_gen}).

\section{Proof of Theorem \ref{th:sinr_als}: Asymptotic SINR}
\label{ap:derivationSINR}

This proof continues on from the proof of Theorem \ref{th:unified_als} in Appendix \ref{ap:proof_unified_als}. Firstly, note that the steering vector (\ref{eq:ALS_shat}) can be written as
\begin{align}
\hat{\vs}_k &= \begin{cases} %
\vq_k + A_k \EW_k \vh_k & \text{, ALS with training} \label{eq:als_steering} \\
\vh_k & \text{, semi-blind ALS.}
\end{cases}
\end{align}
That is, (\ref{eq:asympequivLSfilter_blind}) is the equivalent asymptotic form for the semi-blind LS filter given in (\ref{eq:ALS_filter}). Also, using (\ref{eq:als_steering}), (\ref{eq:asympequivLSfilter_blind})--(\ref{eq:Rhatvq}) we obtain the equivalent asymptotic form for the ALS filter with training. We can express the asymptotic form for both receivers as
\begin{align}
\vc_k &\asymp K_k \mhRIuk{k}^{-1}(a_{k,1}^N \vh_k + a_{k,2}^N \vq_k) \label{eq:asympequivLSfilter_gen}
\end{align}
where
\begin{align}
(a_{k,1}^N,\; a_{k,2}^N) &= \begin{cases}
(1,\;  -A_k\rhomkNi) & \text{, semi-blind LS} \\
(A_k(\EW_k-\psimkNi),\; 1) & \text{, LS with training} ,
\end{cases} \label{eq:def_a1a2N}
\end{align}

%$a_{k,1}$ and $a_{k,2}$ are given in (\ref{eq:def_a1a2}).

We now compute the large-system SINR for the filter (\ref{eq:asympequivLSfilter_gen}) for stream $k$, and symbol interval $m>i$. For notational simplicity, we drop the subscript $m$ in this appendix.  Note therefore that in this appendix $\vbIuk{k}$ denotes $\vb_m$ with the $k\thh$ element removed.  Since $K_k$ cancels in the ratio when calculating the SINR, we ignore this constant. The signal component is
\begin{align}
A_k \vc_k^\dag \vh_k \vb(k) &\asymp A_k \(a_{k,1}^N\vh_k + a_{k,2}^N\vq_k\)^\dag\mhRIuk{k}^{-\dag}\vh_k \vb(k) \\
&\asymp A_k a_{k,1}^{N*} \vb(k) \vh_k^\dag\mhRIuk{k}^{-\dag}\vh_k \label{eq:sigcomp_als}
\end{align}
using (\ref{eq:def_Pmax}), (\ref{eq:supN_rhomi}), and (\ref{eq:hRqzero}). The interference component is
\begin{align}
\vc_k^\dag \(\mH \mSIuk{k} \mAIuk{k} \vbIuk{k} + \vn\) &\asymp \(a_{k,1}^N\vh_k + a_{k,2}^N\vq_k\)^\dag \mhRIuk{k}^{-\dag} \(\mH \mSIuk{k} \mAIuk{k} \vbIuk{k} + \vn\) \\
&\asymp a_{k,1}^{N*}\( \vh_k^\dag \mhRIuk{k}^{-\dag}\mH \mSIuk{k} \mAIuk{k} \vbIuk{k} + \vh_k^\dag \mhRIuk{k}^{-\dag}\vn\) \nonumber \\
& \quad + a_{k,2}^{N*} \( \vq_k^\dag \mhRIuk{k}^{-\dag}\mH \mSIuk{k} \mAIuk{k} \vbIuk{k} + \vq_k^\dag \mhRIuk{k}^{-\dag}\vn\) . \nonumber
\end{align}
Therefore, the signal power is asymptotically equivalent to $P_k \abs{a_{k,1}}^2 \abs{\rhomkNi}^2$. Also, the interference power, averaged over the data symbols and noise is asymptotically equal to
\begin{align}
\abs{a_{k,1}^N}^2(\rhomtkNci + \sn\rhomtkNi) + \abs{a_{k,2}^N}^2(\psimtkNci + \sn\psimtkNi) .
\end{align}
Moreover, from (\ref{eq:def_rhomn}), (\ref{eq:def_psimn}), (\ref{eq:def_rhomn_tr}), (\ref{eq:def_psimn_tr}), we have
\begin{align}
\SINR^{\text{ALS}}_{k,N} \uasymp{k} \frac{P_k \abs{a_{k,1}^N}^2 \abs{\rhomi}^2}{\abs{a_{k,1}^N}^2(\rhomtNci + \sn\rhomtNi) + \abs{a_{k,2}^N}^2(\psimtNci + \sn\psimtNi)} \label{eq:als_sinr_derive} .
\end{align}
% is this ok to use uasymp{k} here?

%%%%% THIS IS THE OMITTED PROOF %%%%%
%\input{Drafts/proof_lemma1}

\section{Steady-State ALS SINR with Windowing}
\label{ap:unlimited_training} %

%In each case above, the appropriate $\eta$-scaling in the limit is chosen to give sensible expressions in the limit as $\eta\to\infty$.

By inspection of Theorem \ref{th:unified_als} and Lemma \ref{lem:extra_moments}, we note the following limits as $\eta\to\infty$:
\begin{align}
\tgmn{j} &= \lim_{\eta\to\infty} \begin{cases}
\gmi/\eta & \text{, $j=1$,} \\
\gmn{j}/\eta^2 & \text{, $j=2,3,4$.}
\end{cases} \label{eq:def_tgmi} \\
\tomegmn{j} &= \lim_{\eta\to\infty} \begin{cases}
\omegmi & \text{, $j=1$,} \\
\omegmn{j}/\eta & \text{, $j=2,3,4$.}
\end{cases} \\
\trmn{j} &= \lim_{\eta\to\infty}
\begin{cases}
\rmn{1} & \text{, $j=1$,} \\
\rmn{j}/\eta & \text{, $j=2,3,4$.}
\end{cases}
\end{align}
and
\begin{align}
\trhomn{j} &= \lim_{\eta\to\infty} \begin{cases}
\rhomi/\eta & \text{, $j=1$,} \\
\rhomn{j}/\eta^2 & \text{, $j=2,3,4$.}
\end{cases} \label{eq:def_trhomti} \\
\tpsimn{j} &= \lim_{\eta\to\infty} \begin{cases}
\eta\psimi & \text{, $j=1$,} \\
\psimn{j} & \text{, $j=2,3,4$.}
\end{cases}
\end{align}
and
\begin{align}
\tnmn{j} &= \lim_{\eta\to\infty} \begin{cases}
\nmi/\eta & \text{, $j=1$,} \\
\nmn{j}/\eta^2 & \text{, $j=2,3$.}
\end{cases} \\
\ttaumn{j} &= \lim_{\eta\to\infty} \begin{cases}
\eta\taumi & \text{, $j=1$,} \\
\taumn{j} & \text{, $j=2,3$.}
\end{cases} \label{eq:def_tnmtbi}
\end{align}
and
\begin{align}
\tEW &= \lim_{\eta\to\infty} \eta\EW \\
\tsW_{m,n} &= \lim_{\eta\to\infty} \eta\sW_{m,n}  \\
\tsEa_{m,n} &= \lim_{\eta\to\infty} \sEa_{m,n} \\
\tsHa_{m,n} &= \lim_{\eta\to\infty} \sHa_{m,n}
\end{align}
Moreover, the variables $\tgmn{j}$, $\trhomn{j}$, $\ttaumn{j}$, $\tpsimn{j}$, $\tomegmn{j}$, $\tnmn{j}$, $\trmn{j}$, $\tsEa_{m,n}$, $\tsHa_{m,n}$, and $\tsW_{m,n}$ satisfy the same set of equations as $\gmn{j}$, $\rhomn{j}$, $\taumn{j}$, $\psimn{j}$, $\omegmn{j}$, $\nmn{j}$, $\rmn{j}$, $\sEa_{m,n}$, $\sHa_{m,n}$, and $\sW_{m,n}$, respectively (for appropriate values of $j$, $m$, and $n$), if we set $z=-\mu$, $\eta=1$, and assume that $\mB$ is \iid.

\section{Sketch of Proof of Theorem \ref{th:relationship_mmse_als}} %
\label{ap:relationship_mmse_als} %

%Next, by inspecting the equations of Theorem \ref{th:unified_mmse} and Theorem \ref{th:unified_als}
%with $\mu=0$, we note that for both \iid and isometric $\mS$, the values of
%$\frac{\sW_{1,1}}{D_\mB}\rhomi$, $\sW_{1,1}\gmi$, and $\frac{D_\mB}{\sW_{1,1}}(\taumi -
%\alpha\EW\EP)$ from Theorem \ref{th:unified_als} satisfy the same set of equations as $\rhomm$,
%$\gmm$ and $(\taumm - \alpha\EP)$ in Theorem \ref{th:unified_mmse}, respectively, where $D_\mB = 1
%- \frac{\beta}{\eta^*}(1 - \sn\sW_{1,1}\gmi)$ for orthogonal $\mB$, and unity otherwise. This is
%reasonably clear for \iid $\mB$, however for orthogonal $\mB$ it helps to expand $\sHa_{1,1}$ in
%(\ref{eq:rhomi_gen}) and (\ref{eq:rhomi_gen}), $\sEa_{1,1}$ in (\ref{eq:taumi_gen}) and
%(\ref{eq:taumi_gen_iso}), $\sH_{1,1}$ in (\ref{eq:rm1_mmse_unified}), and $\sE_{1,1}$ in
%(\ref{eq:tm1_mmse_unified}) using the identity in (\ref{eq:id_WEH}). That is, we have
%\begin{align}
%\rhomi &= \frac{D_\mB}{\sW_{1,1}} \rhomm \\
%\gmi &= \frac{1}{\sW_{1,1}} \gmm \label{eq:gmi_gmm_transient} \\
%\taumi &= \frac{\sW_{1,1}}{D_\mB}(\taumm - \alpha\EP) + \alpha\EW\EP \\
%D_\mB &= \begin{cases} 1 & \iidB \\
%1 - \frac{\alpha}{\eta^*}\rhomm\sE_{1,1} & \isoB
%\end{cases}
%\end{align}
%where we have simplified $D_{\mB}$ using (\ref{eq:RRinv_mmse_dimK}) and
%(\ref{eq:gmi_gmm_transient3}).
%
%%, where $D_B$ is unity for \iid $\mB$, and $1/(1 - \frac{\beta}{\eta^*}(1 - \sn\sW_{1,1}\gmi))$ for
%%orthogonal $\mB$

We provide a sketch of the proof for \iid $\mS$ only, mainly due to the fact that the equations are simpler to manipulate. However, the same approach is valid for isometric $\mS$.

Firstly, according to the remark made after Theorem \ref{th:unified_als}, with \iid $\mB$ and $\mu=0$ we have $\sW_{1,1}\rhomi = \rhomm$. That is, the term $\rhomi$ in the numerator of the asymptotic SINR in (\ref{eq:asympSINR}) can be written as $\frac{1}{P_k\sW_{1,1}}\SINR^{\text{MMSE}}_k$.

Consider the denominator of the alternate MMSE SINR expression in Section
\ref{sec:alternate_mmse_sinr} with \iid $\mS$. Solving
(\ref{eq:rhomtm_iid})--(\ref{eq:taumtbm_iid}) for $\rhomtcm + \sn\rhomtm$ gives
\begin{align}
\rhomtcm + \sn\rhomtm &= \frac{\beta^*(\alpha\sE_{1,2}\sH_{2,2} - z\sH_{1,2})}{z^2 -
\alpha\beta^*\sE_{2,2}\sH_{2,2}} ,
\end{align}
which from Appendix \ref{ap:alternate_mmse_sinr} equals $\rhomm$.

The next step is to simplify the $\rhomtci+\sn\rhomti$ term in the denominator of the ALS SINR
(\ref{eq:asympSINR}).  To do this, we solve (\ref{eq:rhomnj_iid}), (\ref{eq:taumnj_iid}),
(\ref{eq:psimnj_iid}), and (\ref{eq:nmnj}) to find $\rhomtbi$, $\taumtbi$, $\psimtci$, $\nmtbi$
in terms of $\rhomtci$ and $\rhomti$, and then substitute into (\ref{eq:rhomnj_iid}). We then
solve (\ref{eq:gmnj_eq}), (\ref{eq:taumnj_iid}), and (\ref{eq:psimnj_iid}) for $\gmti$
$\taumti$, and $\psimti$ in terms of $\rhomtci$ and $\rhomti$, and
substitute into (\ref{eq:rhomnj_iid}).  Combining these results gives
\begin{align}
\rhomtci+\sn\rhomti &= \frac{\frac{\rhomi}{\sW_{1,2}}(1-\alpha\frac{\sW_{2,2}}{\sW_{1,1}}\rhomi\sEa_{1,1})}
{1-\alpha\frac{\sW_{2,2}}{\sW_{1,1}}\rhomi(\sEa_{1,2}+\rhomi\sW_{1,1}\sEa_{2,2})}
\end{align}
Now we use the identity (\ref{eq:id_WEHb}) to obtain $\sEa_{1,1} = \sEa_{1,2}+\rhomi\sW_{1,1}\sEa_{2,2}$,
and so we have that
\begin{align}
\rhomtci+\sn\rhomti &= \frac{\rhomi}{\sW_{1,2}} \:=\: \frac{1}{P_k\sW_{1,1}\sW_{1,2}}\SINR^{\text{MMSE}}_k . \label{eq:kappati_sn_rhomti}
\end{align}

Next we simplify the
$\psimtci + \sn\psimti$ term in the denominator of the ALS SINR
(\ref{eq:asympSINR}).
To do this, we solve (\ref{eq:gmnj_eq}), (\ref{eq:omegmnj_iid}),
(\ref{eq:psimnj_iid}), (\ref{eq:psimnj_iid}) and (\ref{eq:omegmnj_iid}) for $\gmti$,
$\omegmti$, $\psimti$, $\psimtci$, and $\omegmtci$, and form the sum
\begin{align}
\psimtci + \sn\psimti &= \frac{\sW_{2,2}((\rhomtci+\sn\rhomti)\sEa_{1,2} + \rhomi^2\sEa_{2,2})\sW_{1,2}\alpha +
\sn\beta\gmi}{\sW_{1,2}(\eta-\alpha\rhomi^2\sEa_{2,2}\sW_{2,2})} .
\end{align}
Substituting (\ref{eq:kappati_sn_rhomti}) and $\rmi = \frac{1}{\eta}(\alpha\rhomi\sEa_{1,1} + \beta\sn\gmi)$, and simplifying with (\ref{eq:id_WEHb}) gives
\begin{align}
\psimtci + \sn\psimti &= \frac{\sW_{1,1}}{\sW_{1,2}}-1 \label{eq:psimtci_sn_psimti}
\end{align}

Substituting (\ref{eq:kappati_sn_rhomti}) and
(\ref{eq:psimtci_sn_psimti}) into (\ref{eq:asympSINR}) along with
the expressions for $\hat{a}_{k,1}$ and $\hat{a}_{k,2}$
from (\ref{eq:def_a1a2}), and simplifying gives
(\ref{eq:sinr_als_from_mmse_train}) and (\ref{eq:sinr_als_from_mmse_blind}).

\end{appendices}

\bibliographystyle{ieeetr}
\bibliography{matt}

\begin{thebibliography}{10}

\bibitem{Tulino04Verdu}
A.~M. Tulino and S.~Verd\'u, ``Random matrix theory and wireless
  communications,'' {\em Foundations and Trends in Communications and
  Information Theory}, vol.~1, no.~1, pp.~1--182, 2004.

\bibitem{silverstein95bai}
J.~W. Silverstein and Z.~D. Bai, ``On the empirical distribution of eigenvalues
  of a class of large dimensional random matrices,'' {\em Journal of
  Multivariate Analysis}, vol.~54, no.~2, pp.~175--192, 1995.

\bibitem{Girko90}
V.~L. Girko, {\em Theory of Random Determinants}.
\newblock Kluwer Academic, 1990.

\bibitem{Hiai00Petz}
F.~Hiai and D.~Petz, {\em The semicircle law, free random variables and
  entropy}.
\newblock American Mathematics Society, Mathematical Surveys and Monographs,
  Vol.\ 77, 2000.

\bibitem{Li04Tulino}
L.~Li, A.~M. Tulino, and S.~Verd\'u, ``Design of reduced-rank {MMSE} multiuser
  detectors using random matrix methods,'' {\em IEEE Trans.\ on Information
  Theory}, vol.~50, pp.~986--1008, June 2004.

\bibitem{Debbah03Hachem}
M.~Debbah, W.~Hachem, P.~Loubaton, and M.~de~Courville, ``{MMSE} analysis of
  certain large isometric random precoded systems,'' {\em IEEE Trans.\ on
  Information Theory}, vol.~49, pp.~1293--1311, May 2003.

\bibitem{Chaufray03Hachem}
J.~M. Chaufray, W.~Hachem, and P.~Loubaton, ``Asymptotic analysis of optimum
  and suboptimum {CDMA} downlink {MMSE} receivers,'' {\em IEEE Trans.\ on
  Information Theory}, vol.~50, pp.~2620--2638, Nov. 2004.

\bibitem{Haykin96}
S.~Haykin, {\em Adaptive Filter Theory}.
\newblock Prentice Hall, 3rd~ed., 1996.

\bibitem{Poor97Wang}
H.~V. Poor and X.~Wang, ``Code-aided interference suppression for {DS}/{CDMA}
  communications— {P}art {II}: Parallel blind adaptive implementations,'' {\em
  IEEE Trans.\ on Communications}, vol.~45, pp.~1112--1122, Sept. 1997.

\bibitem{Ling84Proakis}
F.~Ling and J.~Proakis, ``Nonstationary learning characteristics of least
  squares adaptive estimation algorithms,'' in {\em IEEE ICASSP}, vol.~9,
  pp.~118--121, Mar. 1984.

\bibitem{Eleftheriou86Falconer}
E.~Eleftheriou and D.~Falconer, ``Tracking properties and steady-state
  performance of {RLS} adaptive filter algorithms,'' {\em IEEE Trans.\ on
  Acoustics, Speech, and Signal Processing}, vol.~34, pp.~1097--1110, Oct.
  1986.

\bibitem{Honig95}
M.~L. Honig, U.~Madhow, and S.~Verd\'u, ``Blind adaptive multiuser detection,''
  {\em IEEE Trans.\ on Information Theory}, vol.~41, pp.~944--960, July 1995.

\bibitem{Caire00}
G.~Caire, ``Two-stage nondata-aided adaptive linear receivers for
  {DS}/{CDMA},'' {\em IEEE Trans.\ on Communications}, vol.~48, pp.~1712--1724,
  Oct. 2000.

\bibitem{Honig98Poor}
M.~L. Honig and H.~V. Poor, {\em {\em``Adaptive Interference Mitigation'' in}
  Wireless Communications: A Signal Processing Perspective}, ch.~2,
  pp.~64--128.
\newblock Englewood Cliffs, NJ: Prentice-Hall, 1998.
\newblock H. V. Poor and G. Wornell, eds.

\bibitem{Xiao00Honig}
W.~Xiao and M.~L. Honig, ``Convergence analysis of adaptive reduced-rank linear
  filters for {DS}-{CDMA},'' in {\em Conference on Information Sciences and
  Systems}, (Princeton University), pp.~WP2--6 -- WP2--11, Mar. 2000.

\bibitem{Honig02Xiao}
W.~Xiao and M.~L. Honig, ``Large system transient analysis of adaptive least
  squares filtering,'' {\em IEEE Trans.\ on Information Theory}, vol.~51,
  pp.~2447--2474, July 2005.

\bibitem{Zhang02Wang}
J.~Zhang and X.~Wang, ``Large-system performance analysis of blind and
  group-blind multiuser receivers,'' {\em IEEE Trans.\ on Information Theory},
  vol.~48, pp.~2507--2523, Sept. 2002.

\bibitem{Xu04Wang}
Z.~Xu and X.~Wang, ``Large-sample performance of blind and group-blind
  multiuser detectors: a perturbation perspective,'' {\em IEEE Trans.\ on
  Information Theory}, vol.~50, pp.~2389--2401, Oct. 2004.

\bibitem{Host-Madsen04Wang}
A.~Host-Madsen, X.~Wang, and S.~Bahng, ``Asymptotic analysis of blind multiuser
  detection with blind channel estimation,'' {\em IEEE Trans.\ on Signal
  Processing}, vol.~52, pp.~1722--1738, June 2004.

\bibitem{Muller02}
R.~R. M{\"u}ller, ``A random matrix model of communication via antenna
  arrays,'' {\em IEEE Trans.\ on Information Theory}, vol.~48, pp.~2495--2506,
  Sept. 2002.

\bibitem{verdu88MultiuserDetection}
S.~Verd\'u, {\em Multiuser Detection}.
\newblock Cambridge University Press, 1998.

\bibitem{PeacockCollingsMCCDMAJournal}
M.~J.~M. Peacock, I.~B. Collings, and M.~L. Honig, ``Asymptotic analysis of
  {MMSE} multiuser receivers for multi-signature multicarrier {CDMA} in
  {R}ayleigh fading,'' {\em IEEE Trans.\ on Communications}, vol.~52,
  pp.~964--972, June 2004.

\bibitem{Chuah02Tse}
C.~Chuah, D.~N.~C. Tse, J.~M. Kahn, and R.~A. Valenzuela, ``Capacity scaling in
  {MIMO} wireless systems under correlated fading,'' {\em IEEE Trans.\ on
  Information Theory}, vol.~48, pp.~637--650, Mar. 2002.

\bibitem{Vasilchuk01}
V.~Vasilchuk, ``On the law of multiplication of random matrices,'' {\em
  Mathematical Physics, Analysis and Geometry}, vol.~4, no.~1, pp.~1 -- 36,
  2001.

\bibitem{Peacock05}
M.~J.~M. Peacock.
\newblock PhD thesis, The University of Sydney, Aug. 2005.

\bibitem{Proakis01}
J.~Proakis, {\em Digital Communications}.
\newblock McGraw Hill, 4th~ed., 2001.

\bibitem{Hassibi03Hochwald}
B.~Hassibi and B.~M. Hochwald, ``How much training is needed in
  multiple-antenna wireless links?,'' {\em IEEE Trans.\ on Information Theory},
  vol.~49, pp.~951 -- 963, Apr. 2003.

\bibitem{VikaloHassibi}
H.~Vikalo, B.~Hassibi, B.~M. Hochwald, and T.~Kailath, ``On the capacity of
  frequency-selective channels in training-based transmission schemes,'' {\em
  Submitted to IEEE Trans.\ on Information Theory}, 2004.

\bibitem{Sun03}
Y.~Sun, {\em Transmitter and Receiver Technqiues for Wireless Fading Channels}.
\newblock PhD thesis, Northwestern University, June 2003.

\bibitem{Girko01}
V.~L. Girko, {\em Theory of stochastic canonical equations}.
\newblock Kluwer Academic, 2001.

\bibitem{Evans00Tse}
J.~Evans and D.~N.~C. Tse, ``Large system performance of linear multiuser
  receivers in multipath fading channels,'' {\em IEEE Trans.\ on Information
  Theory}, vol.~46, pp.~2059--2078, Sept. 2000.

\bibitem{Chung01}
K.~L. Chung, {\em A Course in Probability Theory}.
\newblock Academic Press, third~ed., 2001.

\bibitem{Yin88Bai}
Y.~Q. Yin, Z.~D. Bai, and P.~R. Krishnaiah, ``On limit of the largest
  eigenvalue of the large dimensional sample covariance matrix,'' {\em
  Probability Theory Related Fields}, vol.~78, pp.~509--521, 1988.

\end{thebibliography}

\end{document}